\documentclass[a4paper,fleqn,usenatbib]{mnras}

\usepackage{newtxtext,newtxmath}

\usepackage{ae,aecompl}

\usepackage{graphicx}	
\usepackage{natbib}
\usepackage{soul}
\usepackage{amsmath}
\usepackage[utf8]{inputenc}
\usepackage[default,regular,black]{sourceserifpro}

\usepackage{txfonts}

\bibpunct{(}{)}{;}{a}{}{,}
\usepackage{graphicx}
\usepackage{multirow}

\usepackage{graphicx}
\usepackage{txfonts}
\usepackage{natbib}
\usepackage{xcolor}
\usepackage{xspace}
\usepackage{dcolumn}
\usepackage{longtable}
\usepackage{lscape}
\usepackage{soul}        

\usepackage{rotating}
\usepackage{afterpage}

\definecolor{amethyst}{rgb}{0.6, 0.4, 0.8}

\title[The Ophiuchus DIsc Survey Employing ALMA --- III]{The Ophiuchus DIsc Survey Employing ALMA (ODISEA)--III: the evolution of substructures in massive discs at 3-5 au resolution}

\author[L.A. Cieza et al.]{Lucas A. Cieza$^{1}$\thanks{E-mail: lucas.cieza@mail.udp.cl}, 
Camilo Gonz\'alez-Ruilova$^{1}$, 
Antonio S. Hales$^{2,3}$,
Paola Pinilla$^{4}$,
\newauthor 
Dary Ru\'iz-Rodr\'iguez$^{3}$,
Alice Zurlo$^{1,5}$,
Sim\'on Casassus$^{6}$,
Sebasti\'an P\'erez$^{8}$,
\newauthor
Hector C\'anovas$^{7}$,
Carla Arce-Tord$^{6}$,
Mario Flock$^{3}$,
Nicolas Kurtovic$^{3}$,
\newauthor
Sebastian Marino$^{9}$,
Pedro H. Nogueira$^{1}$,
Laura Perez$^{6}$,
Daniel J. Price$^{10}$,
\newauthor
David A. Principe$^{11}$,
Jonathan P. Williams$^{12}$
\\ \\
$^{1}$N\'ucleo de Astronom\'ia, Facultad de Ingenier\'ia, Universidad Diego Portales, Av. Ejercito 441, Santiago, Chile\\
$^{2}$Joint ALMA Observatory, Alonso de Cordova 3107, Vitacura 763-0355, Santiago, Chile\\
$^{3}$National Radio Astronomy Observatory, 520 Edgemont Road, Charlottesville, VA 22903-2475, USA\\
$^{4}$Max Planck Institute for Astronomy, Königstuhl 17, 69117 Heidelberg, Germany\\
$^{5}$Escuela de Ingenier\'ia Industrial, Facultad de Ingenier\'ia y Ciencias, Universidad Diego Portales, Av. Ejercito 441, Santiago, Chile\\
$^{6}$Universidad de Chile, Camino el Observatorio 1515, Santiago, Chile\\
$^{7}$Aurora Technology for ESA/ESAC, Camino bajo del Castillo s/n, Urbanización Villafranca del Castillo, Villanueva de la Ca\~nada, 28692 Madrid, Spain\\
$^{8}$Departamento de F\'isica, Universidad de Santiago de Chile, Av. Ecuador 3493, Estaci\'on Central, Santiago, Chile \\
$^{9}$Institute of Astronomy, University of Cambridge, Madingley Road, Cambridge CB3 0HA, UK\\
$^{10}$School of Physics and Astronomy, Monash University, Clayton VIC 3800, Australia\\
$^{11}$MIT Kavli Institute for Astrophysics and Space Research, Cambridge, MA, USA\\
$^{12}$Institute for Astronomy, University of Hawaii at Manoa, Honolulu, HI, 96822, USA
}
\date{ }
\pubyear{2021}

\begin{document}
\label{firstpage}
\pagerange{\pageref{firstpage}--\pageref{lastpage}}
\maketitle

\begin{abstract}

We present 1.3 mm continuum ALMA long-baseline observations at 3-5 au resolution of 10 of the brightest discs from the Ophiuchus DIsc Survey Employing ALMA (ODISEA) project. We identify a total of 26 narrow rings and gaps distributed in 8 sources and 3 discs with small dust cavities (r $<$10 au). We find that two discs around embedded protostars lack the clear gaps and rings that are ubiquitous in more evolved sources with Class II SEDs.  Our sample includes 5 objects with previously known large dust cavities (r $>$20 au). We find that the 1.3 mm  radial profiles of these objects are in good agreement with those produced by numerical simulations of dust evolution and planet-disc interactions, which predict the accumulation of mm-sized grains at the edges of  planet-induced cavities. Our long-baseline observations resulted in the largest sample of discs observed at $\sim$3-5 au resolution in any given star-forming region (15 objects when combined with Ophiuchus objects in the DSHARP Large Program) and allow for  a demographic study of the brightest $\sim5\%$ of the discs in Ophiuchus (i.e. the most likely formation sites of giant planets in the cloud). We use this unique sample to propose an evolutionary sequence and discuss a scenario in which the substructures observed in massive protoplanetary discs are mainly the result of planet formation and dust evolution. If this scenario is correct, the detailed study of disc substructures might provide a window to investigate a population of planets that remains mostly undetectable by other techniques.

\end{abstract}





\section{Introduction}

Understanding how the diverse populations of protoplanetary discs in young stellar regions results in the range of exoplanet types and architectures found in the Galaxy is one of the major goals of planet-formation theory.
This is an extremely challenging task due in part to the limited observational constraints available. The Atacama Large Millimeter/submillimeter Array (ALMA) is providing truly transformational images of protoplanetary discs with unprecedented sensitivity and resolution \citep{andrews2020}. However,  millimeter wavelength images reveal the locations of small dust grains but provide little information on the presence of larger particles, beyond centimeter scales.  Gas giant planets are mostly made of hydrogen and helium, which ALMA cannot directly observe; therefore, the information on the gas content relies on the observations of less abundant molecules, such as CO and its isotopologues, that are subjected to uncertain depletion processes in the gas-phase (e.g., \citealt{miotello2016}).
Planets might be detectable by ALMA, although indirectly, by the effects they have on the gas and/or dust in the disc. When planets become massive enough, they can carve gaps (e.g., \citealt{rice2006,zhu2012,pinilla2012}) and disturb the  dynamics of the gas (\citealt{Teague2018,pinte2019,casassusperez2019}). 
The minimum gap-opening mass depends on the viscosity and scale-height of the disc (\citealt{crida2006,duffellmacfayden2013}), but mini-Neptune-mass \citep{perez2019} or even Earth-mass planets (\citealt{rosotti2016,dongfung2017}) could produce detectable gaps. 
Gaps consistent with fully-formed planets have been imaged by ALMA in discs with estimated ages ranging from $<$ 1 Myr (HL Tau and Elias 2-24; \citealt{alma2015,cieza2017}) to $\sim$10 Myr (TW Hydra; \citealt{andrews2016}). However, the origin of these gaps still remains to be established and several alternative explanations have been proposed, including the effect of snow-lines on the dust/gas evolution of different volatiles (\citealp{Zhang2015}), magneto-hydrodynamic effects (\citealt{flock2015}), secular gravitational instability (e.g., \citealt{youdin2011,takahashi2014}), and viscous ring-instabilities \citep{dullemondpenzlin2018}.  Each one of the proposed mechanisms has their merits and shortcomings, and it is possible that different mechanisms operate together or dominate in different objects or in the same object at at different times.  For a recent review on disc (sub)structures, see \citet{andrews2020}.
 Substructures are also expected to be ubiquitous in protoplanetary discs from a theoretical point of view. Without substructures to halt the migration of mm-size grains at large radii, dust particles should migrate toward the innermost part of the disc in timescales shorter than 0.1 Myr (e.g., \citealt{brauer2007}), which is inconsistent with the observations showing significant mm emission at large radii ($\gtrsim$10 au) at much older ages. Understanding the origin and evolution of substructures in protoplanetary discs and their implications for planet formation is currently one the main challenges in the field.
To better understand the incidence and properties of disc substructures in any given molecular cloud, here we present  1.3 mm/230 GHz continuum ALMA long-baseline observations at 3-5 au resolution of the 10 brightest  targets of the "Ophiuchus DIsc Survey Employing  ALMA" (ODISEA) project \citep{cieza2019} that were not included in  "The disc Substructures at High Angular Resolution Project" (DSHARP) ALMA Cycle-4 Large Program \citep{Andrews2018}.
Our new observations result in the largest sample of disc images at $\sim$3-5 au resolution in any star-forming region observed so far at mm wavelengths (15 objects when combined with the brightest Ophiuchus objects in DSHARP). 
In Section~2, we discuss the sample selection, the long-baseline observations, and the data reduction.
In Section~3, we characterize the observed substructures, including gaps, rings, inner discs, and cavities.   
In Section~4, we discuss individual objects and use the full sample of 15 bright Ophiuchus discs observed at high-resolution to construct a tentative evolutionary sequence in which the observed substructures are mostly driven by dust evolution and the formation of giant planets. We also discuss possible connections between the substructures observed in primordial discs and those seen in more evolved debris disc systems.   
A summary of our results and conclusions is presented in Section~5. 
\section{Sample selection, observations and data reduction}\label{s:obs}

\subsection{Sample Selection}\label{s:sample}

All our long-baseline ALMA Cycle-6 targets were selected from the original ``Ophiuchus Disc Survey Employing ALMA” (ODISEA) sample (\citealt{cieza2019,williams2019}).
ODISEA started as a Band-6  (1.3mm/225 GHz) continuum and CO line imaging survey in ALMA Cycles 4 and 5 (PIDs = 2016.1.00545.S; 2017.1.00007.S), aiming to study both the gas and dust components of the entire population of discs identified by the \emph{Spitzer} Legacy Project “Cores to Discs" \citep{evans2009} in the Ophiuchus star-forming region.  
With almost 300 \emph{Spitzer}-identified  Young Stellar Object (YSOs) candidates, Ophiuchus contains the largest disc population of all nearby (d $<$ 200 pc) star-forming regions. 
Ophiuchus is also a particularly interesting region to study disc evolution because it presents objects in a wide range of evolutionary stages, from Class I to Class III (see Sec.~\ref{s:sed} for a detailed discussion on YSO classification).
Some of the main results of ODISEA so far are that most protoplanetary discs in Ophiuchus have much lower dust masses (median mass  Class II sources $\sim$1 M$_{\oplus}$; \citealt{williams2019}) and smaller dust disc radii  (r $\lesssim$ 15 au; \citealt{cieza2019}) than previously thought, and that visual binaries (a $\sim$10-200 au) mostly affect the upper end in the size and mass distributions, leaving typical discs largely unaffected \citep{zurlo2020}.

\begin{figure}
\begin{center}
\includegraphics[trim=20mm 25mm 5mm 35mm,clip,width=9.0cm]{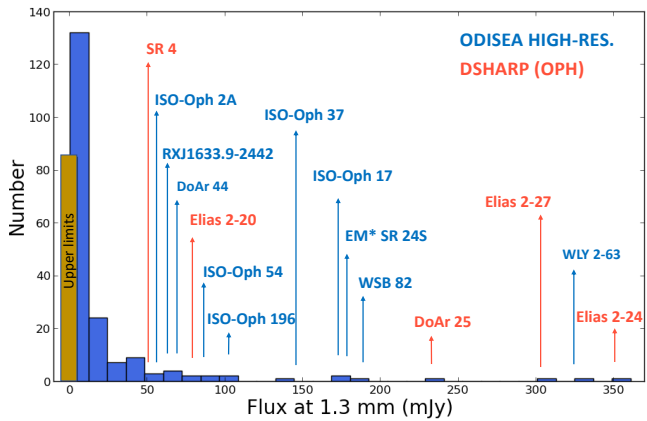}
\caption{Histogram of 1.3 mm fluxes of the 289 objects in the full ODISEA sample (\citealt{cieza2019,williams2019}) observed at at 0.2$''$ (28 au) to 0.6$''$ (84 au) resolution, which corresponds to all YSO candidates identified by \emph{Spitzer} in the Ophiuchus molecular cloud \citep{evans2009}. The 10 ``ODISEA long-baseline targets" observed at 0.02$''$ (3 au) to 0.035$''$ (5 au) resolution are at the upper end of the flux distribution. We combine our long-baseline sample with the 5 Ophiuchus objects brighter than 70 mJy that were observed by DSHARP at 5 au resolution \citep{Andrews2018} to create a flux-limited sample containing the $\sim$5$\%$  brightest  discs in the cloud, which we refer to as the ``Ophiuchus long-baseline sample".  
}
\label{f:histo_flux}
\end{center}
\end{figure}

\begin{table*}
\begin{minipage}{1\textwidth}
\centering
\caption{{Ophiuchus long-baseline targets ordered by descending $\alpha_{IR}$ value. }}
\label{t:sample}
\begin{tabular}{llclcccrrrll}
\hline
\hline
Name &  c2d designation & $\alpha_{IR}$ &SED Class& Gaia Ra &  Gaia Dec & Gaia Dist. &  Gaia $\mu_{\alpha}$ & Gaia $\mu_{\delta}$ & F$_{1.3mm}$ & Project \\
 &  (SSTc2d J+) &   &  & (deg) &(deg)&(pc) &  (mas y$^{-1}$) & (mas y$^{-1}$) & (mJy) &  &\\
  (1)      &  (2) & (3) & (4) & (5) & (6) &  (7) & (8) & (9) & (10) & (11) \\
\hline
ISO-Oph 54 &	 162640.5-242714   	& 0.45   &  I &...& ... & ... & ...	&  ... &	97 	& ODISEA\\
WLY 2-63    &	 163135.6-240129    & 0.14	 &	FS &...&...&  ... & ... & ... & 335  & ODISEA \\
ISO-Oph 37 &	 162623.6-242439  	& -0.01  &  FS &...&...& ...	& ...   &	... &  139  & ODISEA	\\
ISO-Oph 17 &	 162610.3-242054 	& -0.46  &  II/Full &... & ...& ... & ... & ...  &	 173 &  ODISEA	\\
EM* SR4    &     162556.1-242048    & -0.60  &  II/PTD &246.4839894& -24.3468477&  134$\pm$2.0 &  -7.48 & -26.61 &  70 &  DSHARP    \\  
DoAr 44    &	 163133.4-242737    & -0.61	 &	II/PTD  &247.8894038& -24.4604259&  146$\pm$1.0 &  -6.10 & -24.21 &  85	& ODISEA 	\\
Elias 2-27 &     162645.0-242308    & -0.64  &  II/Full &246.6875951& -24.3856199&  116$\pm$15  &  -8.21 & -27.28 & 313 & DSHARP    \\ 
Elias 2-24 &     162624.1-241613    & -0.71  &  II/Full &246.6003316& -24.2705088&  136$\pm$3.0 &  -8.83 & -24.20 & 361 & DSHARP    \\  
Elias 2-20 &     162618.9-242820    & -0.73  &  II/Full &246.5786169& -24.4722560&  138$\pm$5.0 &  -8.99 & -27.34 & 96   & DSHARP    \\ 
WSB 82     &	 163945.4-240204    & -0.73  &  II/Full &249.9393177&  -24.0345083&  155$\pm$2.4 & -6.04 & -22.28 &  191	 & ODISEA \\
ISO-Oph 2  &	 162538.1-242236	& -0.79  &  II/PTD &246.4088384 & -24.3768303 & 144$\pm$9.5  & -5.50 & -25.18 & 72   & ODISEA		 \\
ISO-Oph 196 &	 162816.5-243658    & -0.80	 &	II/Full & 247.0687710 &  -24.616238&  137$\pm$2.4 &  -7.14 &   -25.83  &	98 	 & ODISEA \\
EM* SR 24S  &	 162658.5-244537    & -0.92	 &  II/PTD & 246.7437779 & -24.7603309 &  115$\pm$4.5 & 3.50	&  -31.82  & 180 & ODISEA	 	\\
DoAr 25     &    162623.7-244314   & -1.12  &   II/Full & 246.5986758 & -24.7206373 &  138$\pm$3.0 & -7.83   &  -26.29   &  239  & DSHARP    \\  
RX J1633.9-2442&163355.6-244205     & -1.22	 &	II/TD  &248.4817058 & -24.7014913 & 141$\pm$1.4  &-4.89 & -23.37  & 80	& ODISEA	\\
\hline 
\end{tabular}
\vspace{0.25cm} \\
\noindent (4): SED classes as defined in Sec.~\ref{s:sed}.
(7): a distance of 140 pc is adopted for all targets without a Gaia distance. 
(10): fluxes from Cieza et al. (2019).
\end{minipage}
\end{table*}

Fig.~\ref{f:histo_flux} shows the histogram of 1.3 mm fluxes of all ODISEA objects.  The median flux in the distribution is only $\sim$2 mJy, but there is a long tail of bright discs extending to $\sim$350 mJy. In particular, there are 16 objects brighter than 70 mJy. Five of them (SR4, Elias 2-20,  DoAr 25 and Elias 2-27 and Elias 2-24) are part of the  DSHARP ALMA Cycle-4 program \citep{Andrews2018} and were already observed at $\sim$5 au resolution.  One additional source (ISO-Oph 167) is a binary system with two  compact discs that were barely resolved by our 0.2$''$ resolution observations (ODISEA 105 target shown in Fig~5 in \citealt{cieza2019}) and was excluded from our long-baseline observations. We note that archival ALMA Band-6 data of this system at 0.1$''$ (14 au) resolution (PID = 2016.1.001042.S) indicate that the primary and secondary sources are only 0.14$''$ and 0.2$''$ in size (FWHM) and  the discs lack any clear substructures, but we otherwise exclude ISO-Oph 167 from the rest of the paper as it lacks data at 3-5 au resolution. 

The other 10 ODISEA objects brighter than 70 mJy that are not part of DSHARP 
constitute the ``ODISEA  long-baseline sample". 
The DSHARP project observed a total of 20 discs in several star forming regions, but their sample was restricted to Class II sources and also excluded  “transition'' discs with known dust cavities. 
When discussing the demographics of massive discs in Ophiuchus (Sec.~\ref{s:dis}), we focus on the properties of a flux-limited sample (with the exception of ISO-Oph 167 noted above) and refer to the ``Ophiuchus long-baseline sample", which is the 
combination of  the 10 objects in the  ``ODISEA  long-baseline sample" and the 5 Ophiuchus objects in DSHARP  listed above. 
The 15 objects in the ``Ophiuchus long-baseline sample"  are listed in Table~\ref{t:sample}.

\begin{figure*}
\includegraphics[trim=5mm 30mm 5mm 35mm,clip,width=18.0cm]{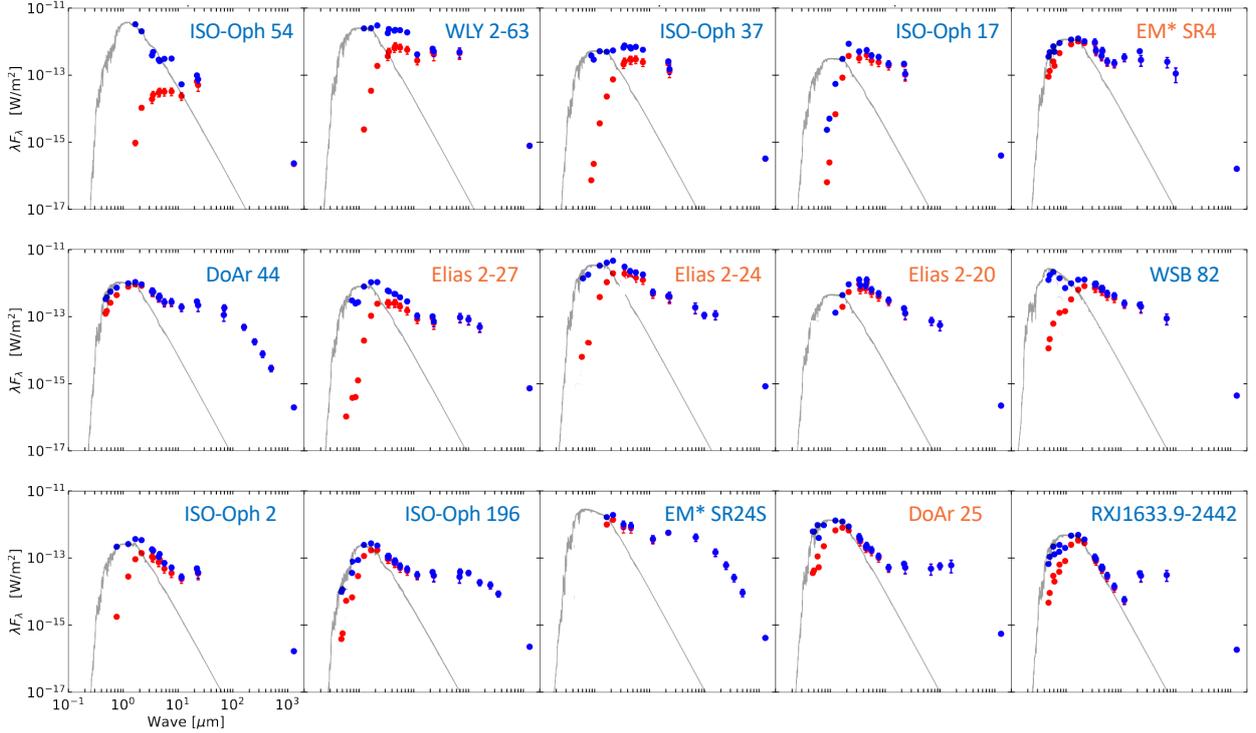}
\caption{Optical to mm spectral energy distribution (SED) of the Ophiuchus long-baseline targets ordered by decreasing $\alpha_{IR}$ value from the top-left corner. Red points correspond to observed fluxes in Gaia, Pan-STARRS, WISE, 2MASS, \emph{Spitzer}, \emph{Herschel} and ALMA. Blue points indicate the extinction-corrected fluxes. 
The solid lines show the model photospheres from \citep{allard2012} normalized to the J-band extinction-corrected values  and corresponding to the effective temperatures listed in Table~\ref{t:basic_prop}.
The visual extinction (A$_V$) ranges from $\sim$1 mag in DoAr~44 to $\sim$50 mag in ISO-Oph 54 (see Section~\ref{s:stellar}.).}
\label{f:seds}
\end{figure*}

\subsection{Spectral Energy Distributions}\label{s:sed}

Since the advent of  space-based mid-IR photometry with \emph{IRAS} and \emph{ISO}, YSOs have been classified based  on the slope ($\alpha_{IR}$) of the spectral energy distribution (SED) between $\sim$2 and 20 $\mu$m \citep{lada1987}. Several classifications schemes exists, but we adopt the following definitions from \citet{williamscieza2011}:

\begin{itemize}
    \item Class I~: $\alpha_{IR}$ $>$ +0.3
    \item Flat Spectrum (FS): +0.3  $>\alpha_{IR} >$ -0.3 
    \item Class II: -0.3  $>\alpha_{IR} >$ -1.6 
    \item Class III: -1.6  $>\alpha_{IR}$
\end{itemize}

These IR SED classes are well correlated with evolutionary stages, where Class I sources typically are very young ($<$ 0.5 Myr, \citealt{evans2009}) protostars still embedded in their natal envelopes and Class II sources are mostly optically revealed YSOs with significant mid-IR emission arising from their discs.  Flat Spectrum sources are intermediate cases between the two.  However, we note that foreground extinction and disc  inclination can change the SED of a YSO,  rendering the SED classification an imperfect 
indicator of evolutionary stage. 

 YSO surveys have shown a large diversity of IR SEDs and the need for expanding the original classification scheme \citep{espaillat2014}. In particular, Class II objects with identical $\alpha_{IR}$ values might show very different SED morphologies at intermediate wavelengths in between 2 and 20 $\mu$m. Since such a diversity is clearly present in our small sample of Class~II targets, we further classify them as follows:

\begin{itemize}

    \item Full discs:  SED between 2 - 24 $\mu$m remains completely within the quartiles defined for Class II sources by \citet{ribas2017}.

     \item Pre-transition discs (PTD):  Class II sources with significant IR excess at $\lambda$ $\lesssim$ 4.5 $\mu$m ([3.6]-[4.5] $>$ 0.25) from an inner disc, but with a mid-IR SED falling below the the quartiles defined for Class II sources, indicating the presence of a large gap. 
    
    \item Transition discs (TD):  Class II sources without significant IR excess at $\lambda$ $\lesssim$ 4.5 $\mu$m (i.e.  [3.6]-[4.5] $<$ 0.25, \citealt{cieza2010}), indicating depleted inner dust discs. 
   
\end{itemize}

The $\alpha_{IR}$ values taken from \citet{evans2009} and the SED classifications (as discussed above) are also listed in Table~\ref{t:sample}.
While these classifications are purely observational, they do provide relevant information regarding the evolutionary status of the systems (see Sec.~\ref{s:evol-sequence}). We note that there are no Class III sources in the ODISEA long-baseline sample, implying that massive and large/cold discs with very weak mid-IR excesses are rare.  
The SEDs of the 15 objects in the Ophiuchus long-baseline sample (from the optical to the mm) are shown in Fig.~\ref{f:seds}. The SED include data from Gaia \citep{GaiaDR2}, Pan-STARRS \citep{chambers2016}, 2MASS \citep{cutri2003}, WISE \citep{wise2010}, \emph{Spitzer} \citep{evans2009}, \emph{Herschel} \citep{rebollido2015,ribas2017} and ALMA \citep{cieza2019}. The SEDs include the observed and extinction-corrected fluxes with extinctions calculated as discussed in Section~\ref{s:stellar}.

\begin{table*}
\caption{{ODISEA long-baseline observing long}} 
\label{t:obs_long}
\centering
\begin{tabular}{lccccccccc}
\hline
\hline
Name    & Execution Block & N. Ant. & Date & ToS   & Avg. Elev. & Mean PWV & Baselines  \\
        &                 &         &      & (sec) &   (deg)    & (mm)     &  (m)       \\
(1)     &  (2)            &   (3)   & (4)  & (5)   & (6)        & (7)  & (8)\\        
\hline
ISO-Oph 54  & uid://A002/Xdeb725/X96a0 & 41	& 2019-07-12 & 3183 &	  51.7 &	1.2 &	111.2-12644.7 \\
WLY 2-63    & uid://A002/Xde0eb4/Xf81  & 48 	& 2019-06-24 & 3058 &	  62.6 &	0.4 & 	83.1-16196.3  \\
ISO-Oph 37  & uid://A002/Xde9c3e/Xd1f  & 44	& 2019-07-08 & 3223 &	  69.4 &	1.2 & 	149.1-13894.4  \\
ISO-Oph 17  & uid://A002/Xddc5da/Xf88  & 44	& 2019-06-19 & 3069 &     55.9 & 	1.1 &	83.1-16196.3   \\
DoAr 44     & uid://A002/Xdeb725/Xc94  & 44	& 2019-07-11 & 3258 &	  65.4 &	1.5 &	111.2-12644.7 \\
DoAr 44     & uid://A002/Xdeb725/X9dd5 & 43	& 2019-07-13 & 3256 &	  78.9 &	1.1 &	111.2-12644.7 \\
WSB 82      & uid://A002/Xdd3de2/X1c59 & 45	& 2019-06-05 & 3162 &	  79.1 &	1.0 &	83.1-15238.4   \\
ISO-Oph 2      & uid://A002/Xdd7b18/X8be5 &45	& 2019-06-12 & 2119 &     68.5 &	1.2 &	83.1-16196.3   \\
ISO-Oph 2      & uid://A002/Xdde745/Xa12 & 46	& 2019-06-21 & 3089 &	  72.4 &	0.9 &	83.1-16196.3  \\
ISO-Oph 196 & uid://A002/Xdeb725/Xf1e  & 44	& 2019-07-11 & 3204 &     52.3 &	1.4 &	111.2-12644.7  \\
SR 24S      & uid://A002/Xdeb725/X98e1 & 43	& 2019-07-12 & 3199 &     64.5 &	1.2 &   111.2-12644.7 \\
RXJ1633.9-2442 & uid://A002/Xde0eb4/Xd43 & 48	& 2019-06-24 & 3142 &	  75.1 & 	0.4 & 	83.1-16196.3 \\
\hline
\end{tabular}
\vspace{0.25cm} \\
\noindent (5) Time on Source. (7) Mean Precipitable Water Vapour.
\end{table*}

\subsection{Observations}

The ODISEA long-baseline targets were observed during ALMA Cycle~6 under program 2018.1.00028.S,  in Band-6 (1.3 mm/230 GHz) and with maximum baselines ranging from  12.6 to 16.2 km. Most of the objects were observed in only one epoch, but ISO-Oph~2 and DoAr~44 were observed in two different epochs.  The observing log of the ODISEA long-baseline observations, including total number of antennas, date,  total time on source (ToS), target average elevation, mean precipitable water vapor column (PWV) in the atmosphere, and minimum and maximum baseline lengths is shown in Table~\ref{t:obs_long}.  

With a continuum bandwidth of 7.5 GHz, the correlator setup was chosen to maximize continuum sensitivity. Three spectral windows overlap the continuum frequency from the Cycle~4 ODISEA observations (see Sec.~\ref{s:sample}) at 217, 219, and 233~GHz and were configured in Time Division Mode, with spectral resolution of 43 ~km~s$^{-1}$. A fourth spectral window was centered in the $^{12}$CO J = 2-1 line (230.538 GHz) with a modest resolution of 1.5 
km~s$^{-1}$, but the $^{12}$CO data are not discussed in this paper.

\begin{table*}
\caption{{Properties of the ODISEA long-baseline images}} 
\label{t:images}
\centering
\begin{tabular}{lcccccrcccc}
\hline
\hline
Name    &   rms & Res.$_{ANG}$ &  Res.$_{PHY}$ & MRS  & Peak Flux   & Total Flux \\
& (mJy/beam)&  (mas $\times$ mas) & (au $\times$ au)& (arcsec) & (mJy/beam)&  (mJy)  \\
(1) & (2) & (3) & (4) & (5) & (6) & (7) &  \\
\hline
 ISO-Oph 54    & 2.9x10$^{-2}$	& 38 $\times$ 58  &  5.3 $\times$ 8.1  &  0.6 &   1.0 & 61\\ 
 WLY 2-63     & 1.8x10$^{-2}$	& 21 $\times$ 22 &  2.9 $\times$ 3.1  &  0.4 & 4.1 & 288\\ 
 ISO-Oph 37    &  2.9x10$^{-2}$ &  24 $\times$ 27   &  3.4 $\times$ 3.8 & 0.5     & 4.4  & 116 \\
 ISO-Oph 17    & 2.7x10$^{-2}$ 	&  23 $\times$ 25 &  3.2 $\times$ 3.5  &  0.3 & 3.4  & 164 \\ 
 DoAr 44    & 2.6x10$^{-2}$ 	&  25 $\times$ 34  & 3.6 $\times$ 5.0   &  0.6 &  0.4 &  60 \\ 
 WSB 82    & 2.7x10$^{-2}$ & 24 $\times$ 31  & 3.7 $\times$ 4.8  &  0.4 & 0.6   & 171  \\
 ISO-Oph 2    & 1.7x10$^{-2}$  &  21 $\times$ 30 & 3.0 $\times$ 4.3  &  0.3 &   0.2 &  63    \\
 ISO-Oph 196  & 3.2x10$^{-2}$ 	&  22 $\times$ 45 & 3.0 $\times$ 6.2  &  0.5 &  1.9 & 90 \\ 
 EM* SR24S     &   2.7x10$^{-2}$  	&  33 $\times$ 37 &  3.8 $\times$ 4.3  &  0.6  &  1.3 & 182 \\
 RXJ1633.9-2442    &  2.1x10$^{-2}$  &   20 $\times$ 21 &  2.8 $\times$ 3.0  & 0.4  & 0.4  & 71 \\
\hline
\end{tabular}
\vspace{0.25cm} \\
\noindent 
(5) Maximum Recoverable Scale.  
(7) the total flux for WSB~82 corresponds to the data set combining the 0.03$''$ resolution data with previous 0.2$''$ resolution observations.
\end{table*}

\subsection{Data reduction}

In general, the ODISEA long-baseline data sets were processed by themselves because, in oder to maximize the observing efficiency, no short-baseline observations were taken as part of the Cycle-6 program.
We have also produced images combining the long-baseline data with the snapshot ODISEA observations at 0.2$''$ resolution from Cycle-4 (PID = 2016.1.00545.S). However, we find that this degrades the resolution and does not improve the image quality or provides additional results. The only exception is WSB~82, for which the combined data shows an additional ring in the outer disc, extending up to $\sim$360 au from the star (see Sec.~\ref{s:images}).

The long-baseline data was calibrated by ALMA staff using the ALMA Calibration Pipeline in CASA version 5.4.0-70. The standard visibility calibration procedure includes  Water Vapor Radiometer, system temperature and antenna position correction, in addition to bandpass, phase, and amplitude calibration.  

The continuum  imaging was performed using the {\sc TCLEAN} task in CASA v.5.6.1  \citep{2007ASPC..376..127M}, with Briggs weighting and robust parameter of 0.5. 
The  resulting images have synthesized beams ranging tom of  0.02$''$ $\times$ 0.02$''$ to 0.04$''$ $\times$ 0.06$''$,  centered at 225 GHz.
Manual masks around each source were defined during the CLEANing process. Two iterations of phase-only self-calibration were conducted on each source (and each epoch). The self-calibration improved the peak signal-to-noise ratio of the final images by 10 to 30$\%$. 

The properties of the images (continuum rms, angular and physical resolution, maximum recoverable scale, peak flux and total flux) are listed in Table~\ref{t:images}. We find that the total flux recovered by the long-baseline observations are typically lower but within $\sim$15$\%$ of the fluxes measured for the objects from the previous ODISEA data at 0.2$''$ resolution (Table~\ref{t:sample}). The differences in flux are only slightly above the calibration uncertainties and indicate that the long-baseline observations  recover most of the flux. Even though some of the discs are larger than the maximum recoverable angular scale of the long-baseline observations, their flux distributions are far from smooth across the disc and little emission is filtered out. 
The exception is the Class I source ISO-Oph 54, which is $\sim$50$\%$ brighter in the low-resolution data, probably due to contamination from envelope emission.   

\begin{table*}
\caption{{Basic stellar properties for Ophiuchus long-baseline sample}} 
\label{t:basic_prop}
\centering
\begin{tabular}{lcccccccrl}
\hline
\hline
Name& SpT  & $M_{acc}$ & Ref. & T$_{eff}$   &  A$_{V}$ & L$_{\star}$ & M$_{\star}$ & Age  & Project  \\
   &   & (M$_{\odot}$yr$^{-1}$) &   & (K)&  (mag) &  (L$_{\odot}$) &  (M$_{\odot}$) & (Myr) &    \\
    (1)      &  (2) & (3) & (4) & (5) & (6) &  (7) & (8) & (9)  & (10) \\     
\hline
ISO-Oph 54 &      ...    & ...	& ...          & 3300 & 53 & 1.7 & ... & $\lesssim$  0.5 & ODISEA  \\
WLY 2-63    &     ...    &...&	 ...           & 4200 & 28 & 3.8 & ... & $\lesssim$  1.0 & ODISEA\\
ISO-Oph 37 &      ...    &...&   ...           & 3900 & 20 & 0.8 & ... & $\lesssim$  1.0 & ODISEA\\
ISO-Oph 17 &	  M0 	 & ... & 	1          & 3800 &  8.4 & 0.2  & 0.5 & 2.0          & ODISEA\\
EM* SR4    &      K7     &  10$^{-6.9}$   &     2          & 4100 &  1.3  &  1.2    &  0.7   &     1.0  & DSHARP \\
DoAr 44    &	  K2	 & 10$^{-8.2}$ &	3  & 4760 & 0.9	& 1.8 & 1.4 & 2.0            & ODISEA \\ 
Elias 2-27 &      M0     & 10$^{-7.2}$   &     2          & 3800 &   15 &   0.9  &   0.5   &   1.0       & DSHARP \\  
Elias 2-24 &      K5     & 10$^{-6.4}$    &    2           & 4200 &    8.7&  6.0    &  0.8   &     $\lesssim$ 1.0     & DSHARP \\  
Elias 2-20 &      M0     & 10$^{-6.9}$ &   2            & 3800 & 14     &  2.2   &     0.5   &       $\lesssim$ 1.0     & DSHARP \\  
WSB 82     &	  K0     & ...     &	4      & 5000 & 5.0 & 5.1 &1.5& 2.0             & ODISEA \\
ISO-Oph 2A &	  M0	 & 10$^{-8.7}$&	5	   & 3800 & 9.0  & 0.7 &  0.5 & 1.0        & ODISEA  \\
ISO-Oph 196 &	  M5     & 10$^{-8.9}$ &3	   & 2900 & 3.1  & 0.2 &   0.2 & 1.0        & ODISEA \\
EM* SR 24S  &	  K2     &10$^{-7.2}$&6    & 4800 & 3.3  &2.0 &  1.4 &  2.0 	     & ODISEA \\
DoAr 25     &     K5     &10$^{-8.3}$          & 2               & 4200 &  2.7  &  1.0    &  0.9   &  2.0                & DSHARP \\  
RX J1633.9-2442&  K5	 &10$^{-10}$&	7	   & 4200 & 2.5  & 1.0 &  0.8               &  2.0       & ODISEA \\
\hline
\end{tabular}
\vspace{0.25cm} \\
\noindent 
(4): Reference for spectral types and accretion rates:  
1 = \citet{ricci2010}; 
2 = \citet{Andrews2018}
3 = \citet{manara2014};
4 Ruiz-Rodriguez et al., in prep; 
5 = \citet{gatti2006};
6 = \citet{natta2006}; 
7 = \citet{cieza2012}. 
(10): All stellar properties  for DSHARP objects are taken from \citet{Andrews2018}. 
\end{table*}

\begin{figure*}
\includegraphics[trim=15mm 90mm 15mm 5mm,clip,width=18.0cm]{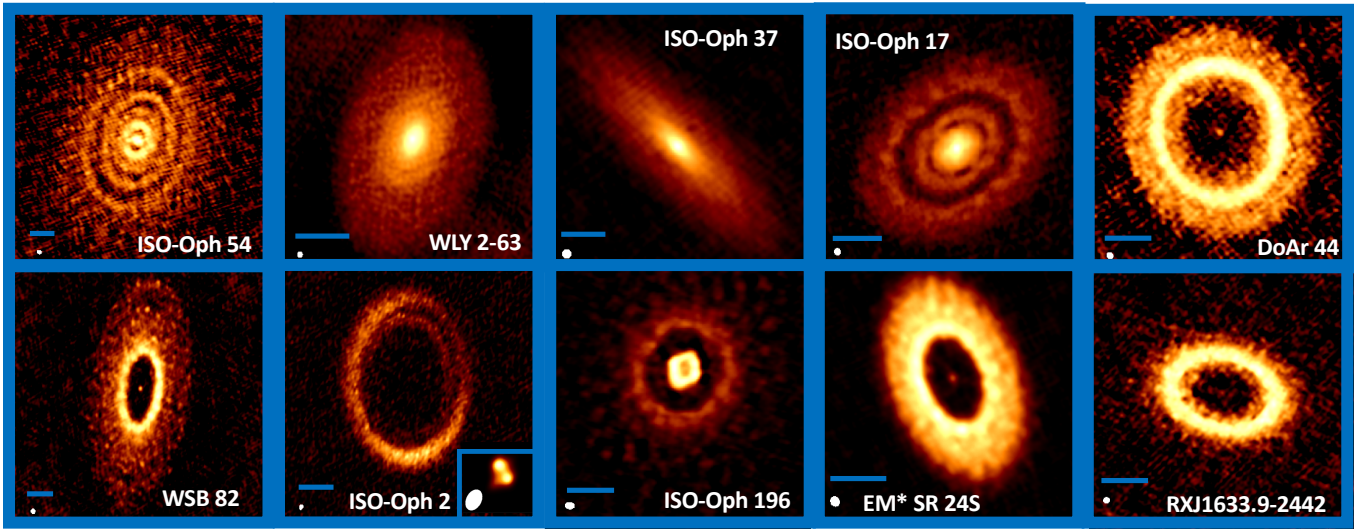}
\caption{The 1.3 mm ALMA images of the ODISEA long-baseline targets ordered by decreasing $\alpha_{IR}$ value from the top-left corner.
The images use the Hot~Metal~2 color map within CASA and scales cycles between -1 and -2 
to bring out the faint features close to the bright peaks. The peak flux of each image is listed in Table~\ref{t:images}.
The white ellipses correspond to the synthesized ALMA beams, the sizes of which are also listed in Table~\ref{t:images} in both mas and au. 
The horizontal bars above the beams are 30 au in length.
The inset in the ISO-Oph 2 image corresponds to the secondary object located at 1.7$''$ to the South of the primary.}
\label{f:all_images}
\end{figure*}

\begin{figure}
\begin{center}
\includegraphics[trim=35mm 5mm 35mm 5mm,clip,width=8.0cm]{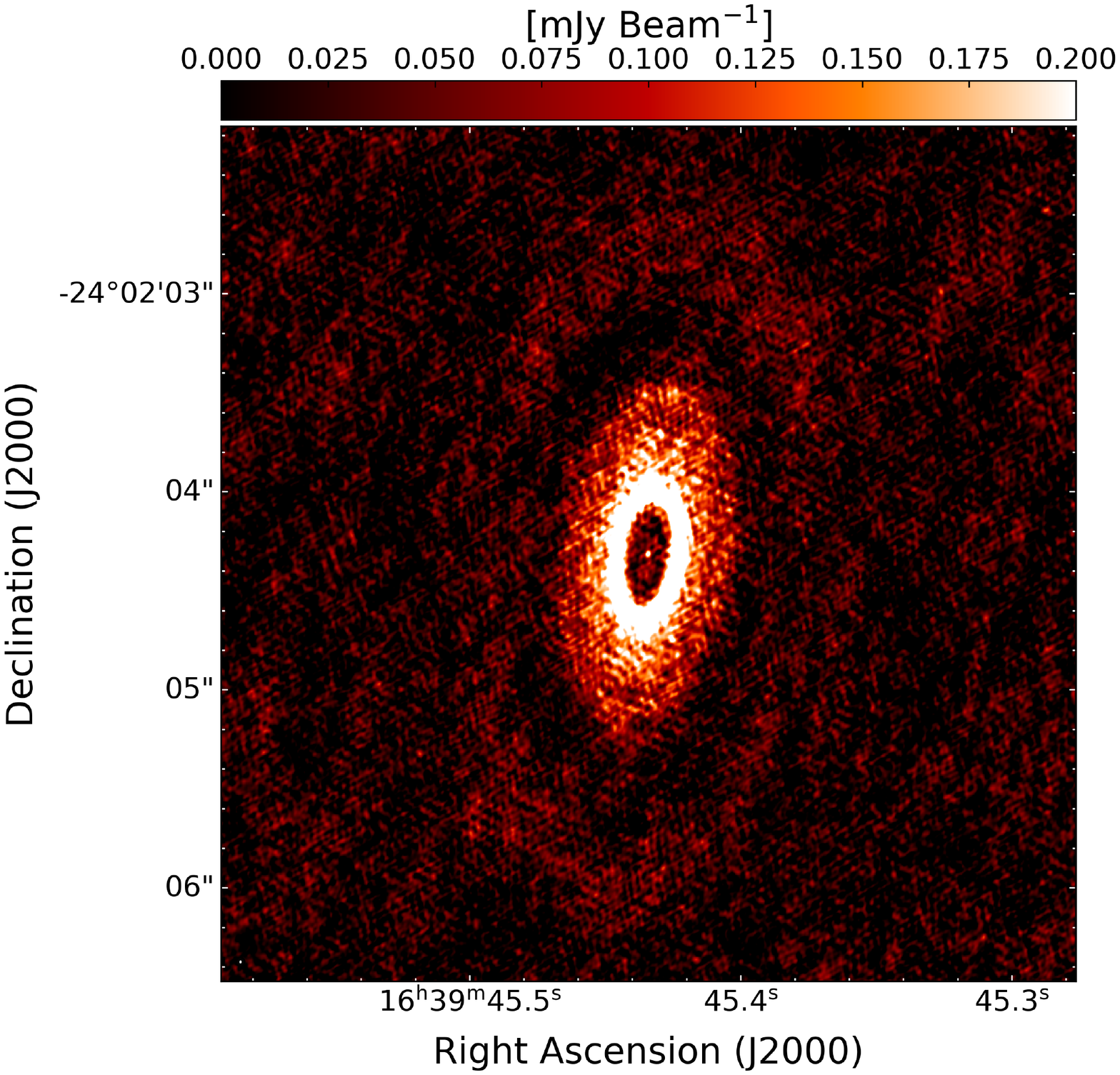}
\caption{The 1.3 mm image of WSB~82 combining the long-baseline data and previous observations at 0.2$''$ resolution. A very faint outer ring is seen extending from 1.2$''$ to 2.3$''$ from the star.}
\label{f:WSB82}
\end{center}
\end{figure}

\section{Results}\label{s:res}

\subsection{Stellar properties}\label{s:stellar}

For Class II sources, we adopt the spectral types listed in Table~\ref{t:basic_prop} and the temperature scale from \citet{pecautmamajek2013}. We derive the luminosities by applying the bolometric corrections from \citep{allard2012}, appropriate for the given effective temperature, to the extinction-corrected values in the J-band.
The A$_V$ values were calculated using the extinction law from \citet{fitzpatrick1999}, including the NIR empirical corrections by \citet{indebetouw2005}.
We adopt the  A$_{V}$ that provides the best match between the extinction-corrected optical SED and the corresponding model photospheres from \citet{allard2012}. 
The resulting A$_{V}$ and luminosities are also listed in Table~\ref{t:basic_prop}. 
The 3 most embedded sources in the sample (ISO-Oph 54, WLY 2-63, and ISO-Oph 37) do not have optical spectral types in the literature, but we adopt the effective  temperatures and luminosities derived by \citet{doppmann2005} based  on high-resolution near-IR spectroscopy. 
For these objects,  we estimate the extinction by dereddening the near-IR colors to  match the expected photospheric color for the given effective temperature.  This approach results in A$_V$ values ranging from 20 to 53 mag, confirming that these targets are deeply embedded objects. 

For Class II sources, we estimate stellar masses and ages by comparing the effective temperatures and luminosities to the predictions of the BT-Settl evolutionary models from \citet{allard2012}. 
Since the photospheric temperatures and luminosities of the embedded sources are very uncertain, we adopt ages for them based on the statistical duration  of each stage provided by \citet{evans2009}:  $\lesssim$ 1 Myr for Flat Spectrum sources and $\lesssim$ 0.5 Myr for Class I sources.  
The masses and ages of low-mass pre-main-sequence stars are 
model-dependent and notoriously difficult to derive (\citealt{baraffe2002,soderblomPPVI}). While the relative values reported in Table~\ref{t:basic_prop} are most robust, the absolute masses and ages should be interpreted with prudence. 

For completeness, in Table~\ref{t:basic_prop} we also include the stellar properties of the DSHARP objects that are part of the  ``Ophiuchus long-baseline sample" (as defined in Sec.~\ref{s:sample}). For these DSHARP objects, we adopt the stellar properties (measured and derived) listed by \citet{Andrews2018}. The derived properties (A$_V$, L$_{\star}$, M$_{\star}$, and age) are in good agreement with the values obtained by our own procedures, within the uncertainties discussed above.

\subsection{ALMA continuum images}\label{s:images}

In Fig.~\ref{f:all_images}, we show all images from the ODISEA long-baseline data. These images display a stunning diversity of substructures, including: concentric gaps and rings (ISO-Oph 54, ISO-Oph 17, ISO-Oph 196), axisymmetric rings (RXJ1633.9-2442), rings with azimuthal asymmetries (ISO-Oph 2), disc with large cavities with bright inner edges and unresolved inner discs (DoAr 44, WSB~82, EM* SR24S), and discs with bright cores and mostly featureless outer discs (WLY 2-63, ISO-Oph 37).
ISO-Oph~2 is a 240 au separation binary system in which we detect the discs around both components.  
As discussed in the previous section, the only object that exhibits additional features when the long-baseline data is combined with the previous ODISEA observations at 0.2$''$ resolution is WSB~82. An image produced with the combined data sets is shown in Figure~\ref{f:WSB82}. The feature is faint, but an additional outer ring can be seen extending from 1.2$"$ (190 au) to 2.3$''$ (360 au). This outer ring is also detectable in the 0.2$''$ resolution images alone \citep{cieza2019}.

\begin{figure*}
\includegraphics[trim=19mm 95mm 19mm 20mm,clip,width=18.0cm]{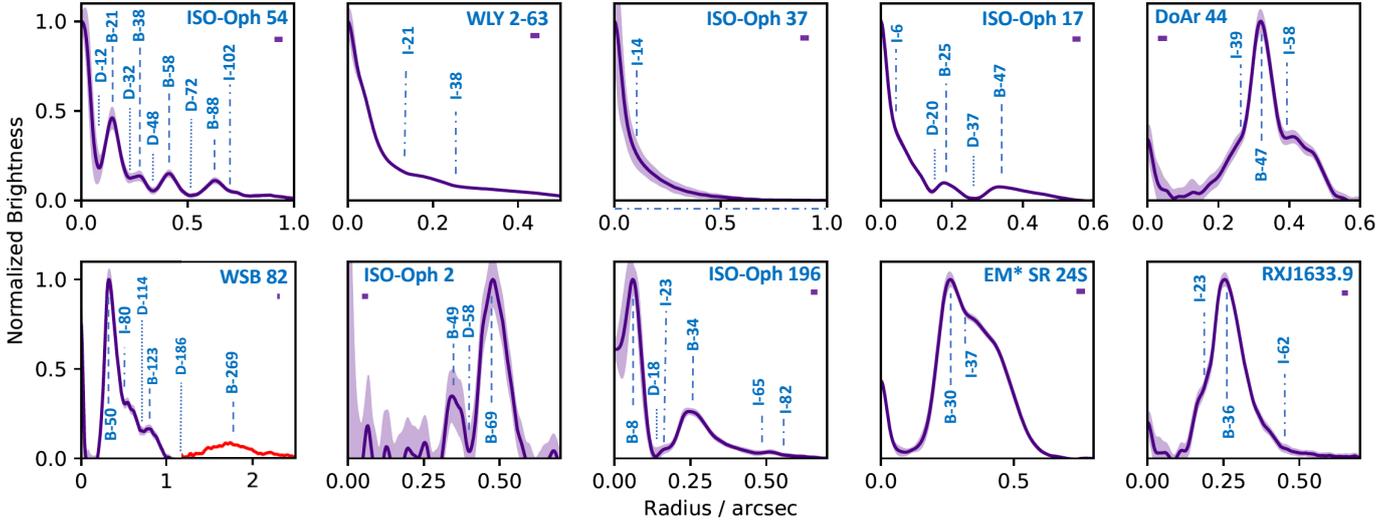}
\caption{Deprojected radial profiles of the "ODISEA long-baseline sample" normalized to the peak flux. The small bar below the name of each source indicates the size of the beam. Substructures are labeled with a prefix ("D" for gaps, "B" for rings, and "I" for inflection points), and a number indicating their location in au. The blue profile corresponds to the long-baseline  data alone, while the red profile in WSB~82 includes the observations at 0.2$''$ resolution. The shaded regions indicate the 3-$\sigma$ error around the mean of the profiles.
}
\label{f:profiles}
\end{figure*}

\begin{table*}
\caption{{Basic disc properties for ODISEA long-baseline sample}} 
\label{t:basic_disc_prop}
\centering
\begin{tabular}{lccccrr}
\hline
\hline
Name&   Ra$_{disc}$ & Dec$_{disc}$&  $i$   & $PA$    &  R$_{90\%}$ & M$_{dust}$\\
    &   (deg) & (deg) & (deg)  &  (deg)  &   (au) & (M$_{\oplus}$)     \\
    (1)      &  (2) & (3) & (4) & (5) & (6) &  (7)  \\ 
\hline
ISO-Oph 54  & 246.6686071    & -24.45417917 & 32.5$\pm$0.4  & 160$\pm$2.5& 119 & 56 \\
WLY 2-63    & 247.8985657  &-24.02500088 & 46.2$\pm$1.2  & 149$\pm$5.0 & 67 & 194 \\
ISO-Oph 37  &  246.5982233 &-24.41112958 & 72.4$\pm$0.2  & 49$\pm$0.6 & 91 & 81 \\
ISO-Oph 17  &   246.5430326 &  -24.34870389 & 42.4$\pm$0.7  & 131$\pm$0.7& 63 & 100\\
DoAr 44     &  247.8893970  &    -24.46045289               & 21.8$\pm$0.9  &  60$\pm$2.7 & 60 & 54\\
WSB 82      &249.9393110	 &    -24.03453309      & 61.2$\pm$0.5  & 173$\pm$1.0 & 256  &  136\\
ISO-Oph 2A &246.4088324	  &    -24.37685830 & 37.6$\pm$0.8  & 0.4$\pm$1.4 &72  & 44  \\
ISO-Oph 196 &	247.0687631   &  -24.61626712 & 22.3$\pm$1.4  & 132$\pm$7.1 &69 & 54 \\
EM* SR 24S  &	 246.7437652   &    -24.76022836       & 47.3$\pm$3.1  & 28$\pm$1.2  & 58 & 70	\\
RX J1633.9-2442&  248.4817004  &    -24.70151736   & 47.9$\pm$1.1  & 77$\pm$1.4  & 53 & 47 \\
\hline
\end{tabular}
\vspace{0.25cm} \\
\noindent 
(2) and (3) coordinates of the disc centers used to produce the radial profiles shown in Fig.~\ref{f:profiles}.
\end{table*}

\subsection{Basic disc properties}

In Table~\ref{t:basic_disc_prop}, we list the  positions (Ra$_{disc}$ and Dec$_{disc}$) of the centers of the discs, together with their Position Angles ($PA$) and inclinations ($i$). 
Following \citet{isella2019}, the  Ra$_{disc}$ and Dec$_{disc}$ values were obtained by   minimizing the imaginary part of the continuum visibilities, which in turn minimizes the asymmetry of the continuum emission  relative  to the  phase  center.
After fixing the Ra$_{disc}$ and Dec$_{disc}$ values, the $PA$ and $i$ were
calculated by minimizing the dispersion of the deprojected visibilities on
circular annuli in the Fourier space. 
The Ra$_{disc}$ and Dec$_{disc}$ values are consistent with the positions and proper motions from Gaia listed in Table~\ref{t:sample}.
%
%
In Table~\ref{t:basic_disc_prop}, we also list the radii of the discs containing 90$\%$ of the flux which we call R$_{90\%}$. These values were calculated by measuring the flux in concentric ellipses with aspect ratios corresponding to the inclination of each disc. 
Finally, we calculate the dust mass of the discs as in the first paper of the ODISEA series \cite{cieza2019}: 

\begin{equation}
M_{dust}~$=$~0.58 \times \frac{F_{1.3mm}}{\mathrm{mJy}}  \left( \frac{ distance }{(140~pc)}\right)^2 M_{\oplus}
\end{equation} \\

This equation assumes a 1.3 mm dust opacity, $\kappa_{1.3mm}$, of 2.3 cm$^2$ g$^{-1}$, and a dust temperature of 20 K. 
The adopted opacity value is taken from \citet{1990AJ.....99..924B} and is still widely used in the 
field \citep{williamscieza2011}.  However, dust opacities at mm wavelengths depend on grain size  distribution, composition, and structure and still remain highly uncertain.  See \citet{2018ApJ...869L..45B} for a recent discussion.
We adopt the 1.3 mm fluxes from the low-resolution  ODISEA data listed in Table~\ref{t:sample} as the long-baseline flux is  subjected to higher calibration uncertainties and potential flux losses by the incomplete sampling of the $u-v$ plane. 
For objects without distances from Gaia, we adopt 140 pc, which is very close to the mean Gaia distance obtained by \citet{canovas2019} for Ophiuchus members (139 pc).

\begin{figure*}
\includegraphics[trim=15mm 115mm 10mm 20mm,clip,width=18.0cm]{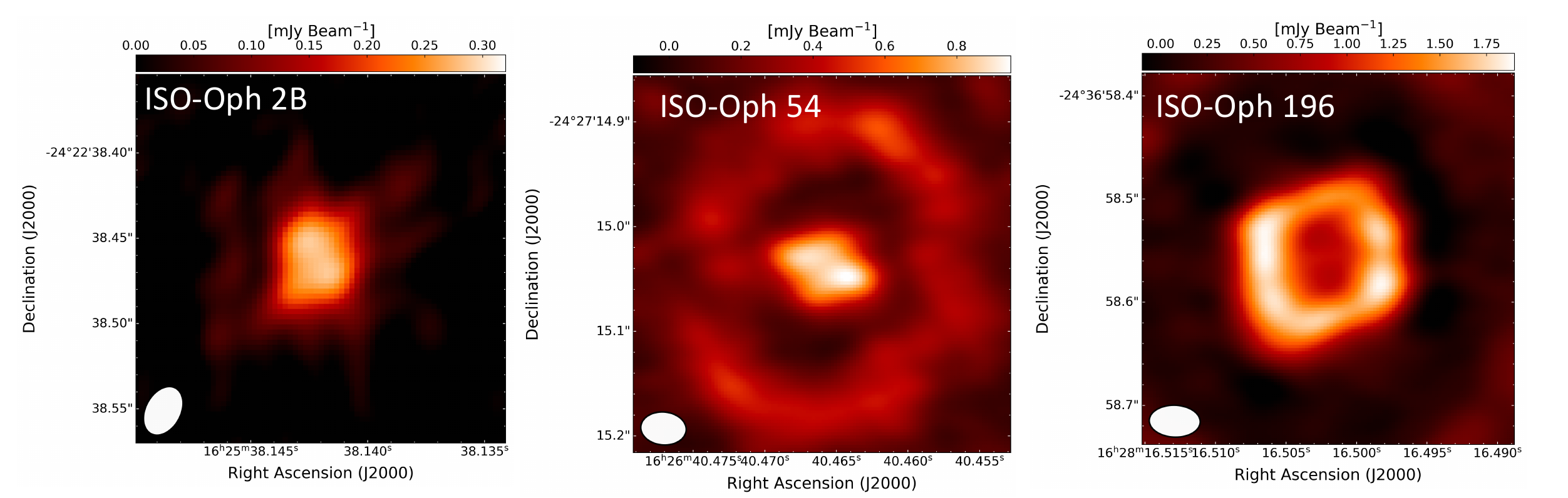}
\caption{Zoomed-in 1.3 mm images of the ODISEA long-baseline targets with small (r $<$ 10 au) cavities ordered by increasing cavity size.  
The white ellipses correspond to the synthesized ALMA beams.
The estimated cavity sizes for ISO-Oph 2B, ISO-Oph 54, and ISO-Oph 196 are 2.2 au, 2.5 au and 8.5 au, respectively. 
}
\label{f:small_cav}
\end{figure*}

\begin{table}
\caption{{Properties of Gaps (Dark) and Rings (Bright)}} 
\label{t:gaps}
\begin{tabular}{llrcccccc}
\hline
\hline
Target & Gap/Ring &  r$_{o}$ &   Depth & Width     \\
       &     &   (au) &    & (au)     \\
(1)    &    (2)  &   (3)      &  (4)       &    (5)      \\
\hline
 ISO-Oph  54 & D-12  &  12 $\pm$ 4  & 0.39$\pm$0.02 &  10$\pm$ 3 \\ 
 ISO-Oph  54 & B-21  &   21 $\pm$ 5 & - &  9$\pm$ 3 \\ 
 ISO-Oph  54 & D-32  &  32$\pm$ 7 & 0.9$\pm$0.05 &  11$\pm$3 \\ 
 ISO-Oph  54 & B-38  &   38 $\pm$ 8 & - &  7$\pm$ 3\\
 ISO-Oph  54 & D-48  &  48 $\pm$ 10 & 0.35$\pm$0.01 &  10$\pm$3 \\ 
 ISO-Oph  54 & B-58  &  58$\pm$ 12 & - &  11$\pm$3  \\ 
 ISO-Oph  54 & D-72 &  72 $\pm$ 15 & 0.24$\pm$0.01 &  17$\pm$4 \\ 
 ISO-Oph  54 & B-88 &  88 $\pm$ 18  & - &  16$\pm$4  \\
 ISO-Oph  17 & D-20    &  20 $\pm$ 4  & 0.49$\pm$0.02	&  18 $\pm$4 \\ 
 ISO-Oph  17 & B-25  &   25 $\pm$ 5 & - &  7$\pm$ 2  \\ 
 ISO-Oph  17 & D-37  &  37 $\pm$ 8  &  0.14$\pm$ 0.01 &   12$\pm$3 \\ 
 ISO-Oph  17 & B-47  &  47 $\pm$ 9 & - &  20$\pm$ 4 \\ 
DoAr 44      & B-47  &  47 $\pm$ 2  & - & 13$\pm$2 \\
 WSB 82 & B-50   &  50 $\pm$ 2  &  -  & 26$\pm$ 2\\ 
 WSB 82 & D-114   &  114 $\pm$ 3 &   0.7$\pm$ 0.02 & 52$\pm$2\\ 
 WSB 82 & B-123   & 123 $\pm$ 3  &  -   & 25$\pm$2 \\ 
 WSB 82 & D-186   & 186 $\pm$ 4 &   0.2$\pm$ 0.05 & 103$\pm$ 2\\ 
 WSB 82 & B-269   & 269 $\pm$ 5 &    - &  78$\pm$2 \\ 
 ISO-Oph  2A & B-49  & 49 $\pm$ 4 &	-&  11$\pm$ 2 \\ 
 ISO-Oph  2A & D-58  & 57 $\pm$ 4 & 0.03$\pm$0.005 & 7 $\pm$ 2 \\ 
 ISO-Oph  2A & B-69  & 69 $\pm$ 5 &	-&   13$\pm$2 \\ 
 ISO-Oph 196& B-8 & 8 $\pm$ 2  & -	&  8 $\pm$ 2\\
 ISO-Oph 196& D-18 & 18 $\pm$ 2 & 0.04$\pm$0.005	&  16 $\pm$2\\
 ISO-Oph 196& B-34 &  34 $\pm$ 2  & -	&  16 $\pm$ 2\\
EM*  SR ~24S & B-30 & 30 $\pm$ 2 & -& 30$\pm$2\\
RXJ1633.9-2442 & B-36 & 36 $\pm$ 1 & -& 18$\pm$1\\
\hline
\end{tabular}
\vspace{0.25cm} \\
\noindent 
(2) Gaps are labeled with he prefix "D" (dark/local minimum) followed by the location in au.  Similarly, rings are labeled with a prefix "B" (bright/local maximum). 
(3) The location of the local maximum or minimum.
(4) Gap depths are defined from the ratio of the local maxima and local minima in gap-ring pairs, $I_{min}$/$I_{max}$.  
( 5) Measured at the mean intensity  between these adjacent pairs of gaps and rings, 0.5 $\times$ ($I_{min}$ + $I_{max}$). 
For rings adjacent to a cavity, the width of the ring is measured at 50$\%$ of the peak intensity.
\end{table}

\begin{table*}
\caption{{Properties of dust cavities}} 
\label{t:cav}
\begin{tabular}{lccccccccc}
\hline
\hline
Large dust cavities & & & & & \\
 (r $>$ 20 au) & \\
\hline
\hline
Target &  R$_{Cav.10\%}$ & R$_{Cav.50\%}$  &  R$_{Cav.90\%}$  &  R$_{Cav}$ &  F$_{CB}$  & M$_{dust,CB}$        \\
       &   (au)      &   (au)      &    (au)  &   (au)  & (mJy)   &   (M$_{\oplus}$) \\
(1) & (2) & (3) & (4) & (5) & (6) & (7)       \\
\hline
 DoAr44           &  26 $\pm$ 2  & 41 $\pm$2	& 45 $\pm$2  &47 $\pm$ 2 & 6$\times$ 10$^{-2}$ & 4$\times$ 10$^{-2}$\\
 WSB~82            &  33 $\pm$ 2& 41	$\pm$2& 47  $\pm$  2&50 $\pm$ 2 & 2$\times$ 10$^{-1}$& 1$\times$ 10$^{-1}$\\ 
 ISO-Oph~2A       &  45 $\pm$3  &46 $\pm$3	&  48 $\pm$4 & 49 $\pm$ 4& $<$ 2$\times$ 10$^{-2}$ & $<$ 1$\times$ 10$^{-2}$ \\ 
 EM*~SR24S         &  18 $\pm$2  &24 $\pm$2	&   27 $\pm$ 2& 30 $\pm$ 2& 2$\times$ 10$^{-1}$& 8$\times$ 10$^{-2}$\\ 
 RXJ1633.9-2442   &  16 $\pm$ 1 &24 $\pm$1	&  27 $\pm$ 1 & 36 $\pm$ 1  & $<$ 4$\times$ 10$^{-2}$ & $<$ 2$\times$ 10$^{-2}$\\ 

\hline
\hline
Small dust cavities & & & & & \\
(r $<$ 10 au) & \\
\hline
\hline
Target      &   R$_{Cav.}$ & & &  \\
            &  (au) & & & \\
            & (8) &&& \\
\hline
 ISO-Oph~54     &   2.5 $\pm$ 2.5   &	 &    \\
 ISO-Oph~2B   &   2.2   $\pm$ 2 &	 &    \\ 
 ISO-Oph~196      &   8.5  $\pm$ 2  &	 &    \\ 
\hline
\end{tabular}
\end{table*}

\begin{table}
\caption{\textbf{Inflections points}}\label{t:inflections}
\begin{tabular}{llr}
\hline
\hline
Target & Inflection point  &     r$_{inflec.}$     \\
                  &    & (au)               \\ 
(1)               &    (2) & (3)\\
\hline
ISO-Oph-54 & I-102 & 102 $\pm$ 21\\
WLY 2-63 & I-21 & 21 $\pm$ 4\\
WLY 2-63 & I-38 & 38 $\pm$ 8\\
ISO-Oph 37 & I-14 & 14 $\pm$ 3\\
ISO-Oph~17 & I-6    & 6 $\pm$ 2\\
DoAr~44    & I-39 & 39 $\pm$ 2\\
DoAr~44    & I-58 &  58 $\pm$ 2\\
WSB~82     & I-80 & 80  $\pm$ 2\\
ISO-Oph~196& I-23 & 23 $\pm$ 2\\
ISO-Oph~196& I-65 & 65 $\pm$ 2\\
ISO-Oph~196& I-82 & 82 $\pm$ 3\\
EM*~SR~24S& I-37 & 37 $\pm$ 2\\
RXJ1633.9-2442  &  I-23 & 23 $\pm$ 1\\
RXJ1633.9-2442  &  I-62 & 62 $\pm$ 2\\
\hline
\end{tabular}
\vspace{0.25cm} \\
\noindent (2) Inflection points are labeled with the prefix "I" followed by their location in au. 
\end{table}

\subsection{Substructures from radial profiles}\label{s:profiles}

Using the disc centers and inclinations listed in Table~\ref{t:basic_disc_prop}, we deproject the discs to an inclination of 0.0 deg. The resulting deprojected radial brightness profiles are shown in Fig.~\ref{f:profiles}, including the 3-$\sigma$ errors in the mean calculated as 3$\times$rms / $\sqrt{N}$, where N is the number of beams in the corresponding ellipse. 
Following \citet{huang2018}, for highly inclined systems (ISO-Oph~37 and WSB~82), we only use the azimuthal angles within 20 deg of the semi-major axis orientation in the reprojected coordinates because the substructures are less well resolved along the minor axes.  
We use these deprojected radial profiles to identify and characterize substructures, including gaps, rings, cavities, and inflection points, as discussed in the following subsection.

\subsubsection{Gaps and rings}

Gaps and rings are the most common substructures that have been identified in protoplanetary discs observed at $\sim$5 au resolution in the continuum at (sub)millimeter wavelengths \citep{Andrews2018}.  We find that these structures are also present in 8 of of our 10 targets. 
Following \citet{huang2018}, we label these features with a prefix "B" (``Bright" for rings) or "D" (``Dark" for gaps) followed by a number that indicates their location 
in au.  
We take a conservative approach and only identify gaps and rings that manifest themselves as clear local minima ($I_{min}$) or local maxima ($I_{max}$) in the intensity profile and leave the identification of more subtle features for future work.  
For WSB~82, we derive the radial profile  beyond 1.0$''$ from the image that combines the long-baseline data with the observations at 0.2$''$ resolution. 
Gap and rings are labeled in Fig.~\ref{f:profiles} and listed in Table~\ref{t:gaps}. 
The table also includes a measurement of the depths of the gaps and the widths of the gaps and rings. 
Also following \citet{huang2018}, the depths of the gaps are defined from the ratio of the local maxima and local minima in gap-ring pairs, $I_{min}$/$I_{max}$.  Similarly, the widths of the gaps and the rings are measured at the mean intensity  between these adjacent pairs of gaps and rings, 0.5 $\times$ ($I_{min}$ + $I_{max}$). 
For rings adjacent to a cavity, we measure the width of the ring at 50$\%$ of the peak intensity.
The large cavities seen in EM* SR24~S, DoAr~44, and WSB~82 could also be considered to be wide gaps because they also have unresolved inner discs within the cavities. However, we treat them in a different way because inner cavities represent a distinct type of substructure: discs with cavities have an off-center absolute maximum.

\subsubsection{Discs with large (r $>$ 20 au) dust inner cavities} 
Five objects in our sample (ISO-Oph~2, EM*~SR~24S, DoAr~44, RX-J1633.9-2442, and WSB~82) have well resolved cavities. 
Cavity radii at millimeter wavelengths are typically measured from the star to the location of the absolute maximum in the radial profile, a measurement that we label $R_{Cav.}$ in Table~\ref{t:cav}. 
We find that cavities show radial profiles with different steepness/sharpness (see Fig.~\ref{f:profiles}).
In order to quantify the sharpness of these cavities, 
in Table~\ref{t:cav} we also report the radii at which the profiles reach 10\%, 50$\%$ and 90$\%$ of the peak flux ($R_{Cav.,10\%}$, $R_{Cav.,50\%}$, and $R_{Cav.,90\%}$, respectively). 
For  DoAr~44, WSB~82, and EM* SR 24S, Table~\ref{t:cav} also includes the fluxes of their central beams and the corresponding dust masses for their inner discs. 
These masses have been calculated using the same temperature (20 K) and opacities (2.3 cm$^2$ g$^{-1}$) as in \citet{cieza2019}, but we warn the reader that this might overestimate the dust mass in a warm inner disc and that the real dust masses can be lower by factors of T/20K.
Alternatively, the mass of the inner disc could be 
underestimated if the emission is optically thick. 
For ISO-Oph 2 and RX~J1633.9-2442, we list the 3-$\sigma$ upper limits in the fluxes at the center of the cavities and in the corresponding dust masses.  

\subsubsection{Small (r $<$ 10 au) dust inner cavities}

While the five objects discussed above have  dust cavities with radii larger than 20 au,  three additional targets in our sample have  discs with much smaller inner dust cavities, which are not so obvious in the images shown in  Figure~\ref{f:all_images}. These targets are ISO-Oph 54, the secondary object in ISO-Oph~2, and ISO-Oph 196 (see Figure~\ref{f:small_cav}). The  ODISEA long-baseline data for ISO-Oph 2 system have already been analyzed by \citet{2020ApJ...902L..33G}. The small disc around ISO-Oph 2B has a cavity  with a radius, $R_{Cav.}$, of only 2.2 au 
based on the location of a null in the deprojected visibility profile \citep{hughes2007}.

The deprojected visibility profile for ISO-Oph~2B can be approximated as a simple Bessel function because the disc seems to be a single ring in the image plane. 
However, the ISO-Oph 54 and ISO-Oph 196 discs are much more complex in the image plane (they have multiple concentric rings of different radii and widths) and thus also have more complex visibility profiles.
Therefore, we estimate the sizes of their cavities (also listed in Table~\ref{t:cav}) as half the distance between the intensity peaks seen in the image plane.  

\subsubsection{Inflection points}

In addition to the rings, gaps, and cavities discussed above, we also identify  changes in the slopes of the radial profiles that are abrupt but do not constitute a local minimum. We call them "inflection points" and label them  in Figure~\ref{f:profiles} with the "I" prefix following their location in au. The location of these inflection points are listed in Table~\ref{t:inflections}. 
Like with the gaps and rings, we only list the most conspicuous  inflection points identified by inspection of the radial profiles. 
These inflection points in the brightness profiles could have different origins, including unresolved gaps \citep{huang2018}, changes in the optical depths at snow lines \citep{cieza2016}, or the accumulation of dust at planet-induced pressure bumps (see \citet{pinilla2019} and Section~\ref{s:TD}).

\begin{figure*}
\includegraphics[trim=15mm 20mm 15mm 15mm,clip,width=18.0cm]{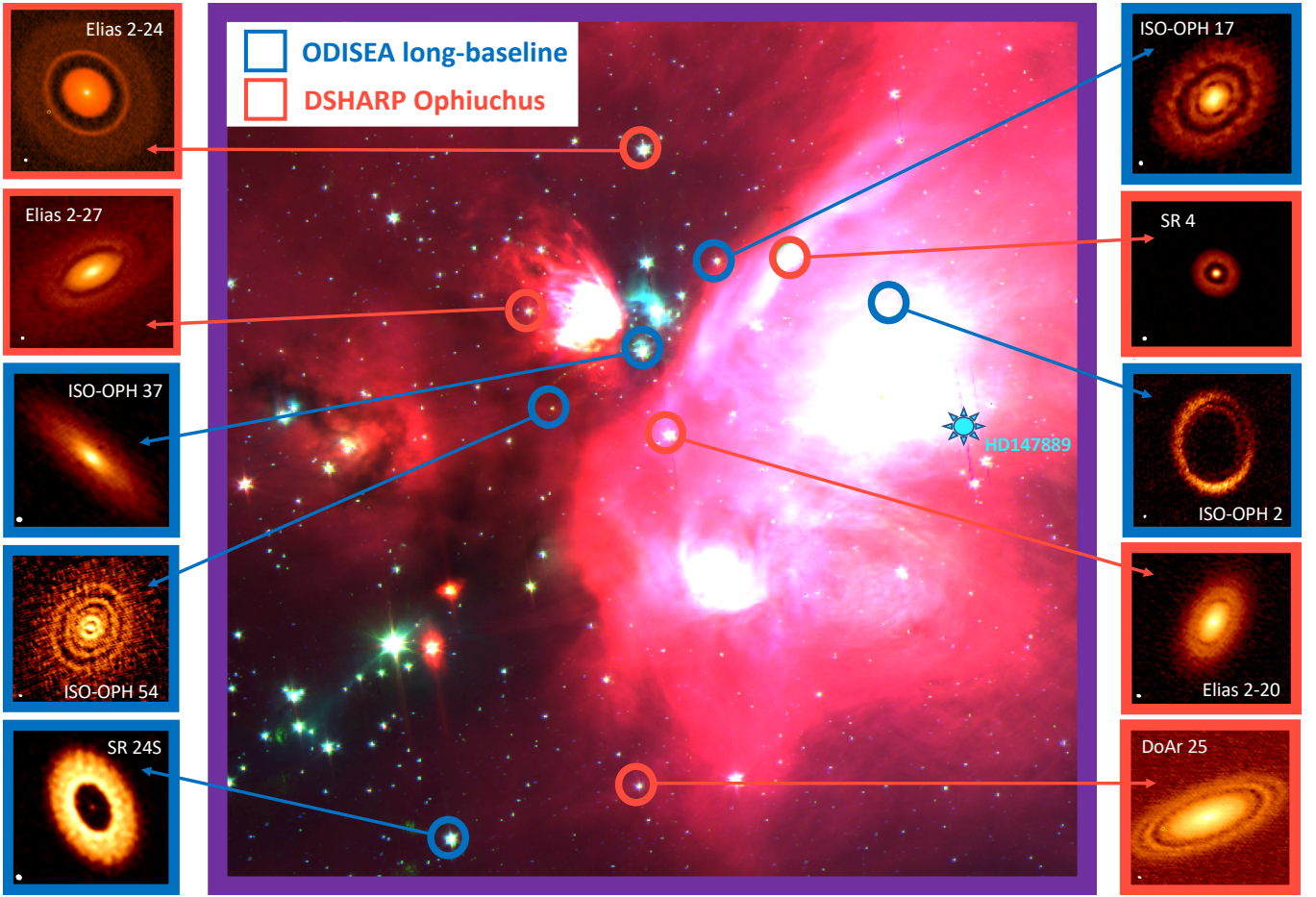}
\caption{Location of the ODISEA (blue frames) and DSHARP (red frames) long-baseline targets in the L1688 star-forming cluster. The locations are overlaid over a 0.6 deg $\times$ 0.6 deg color composite from \emph{Spitzer} 3.6, 4.5, and 
8.0 $\mu$m data. The location of HD~147889, the brightest UV source in the region, is also indicated.}
\label{f:L1688}
\end{figure*}

\begin{figure*}
\includegraphics[trim=40mm 20mm 20mm 20mm,clip,width=19.0cm]{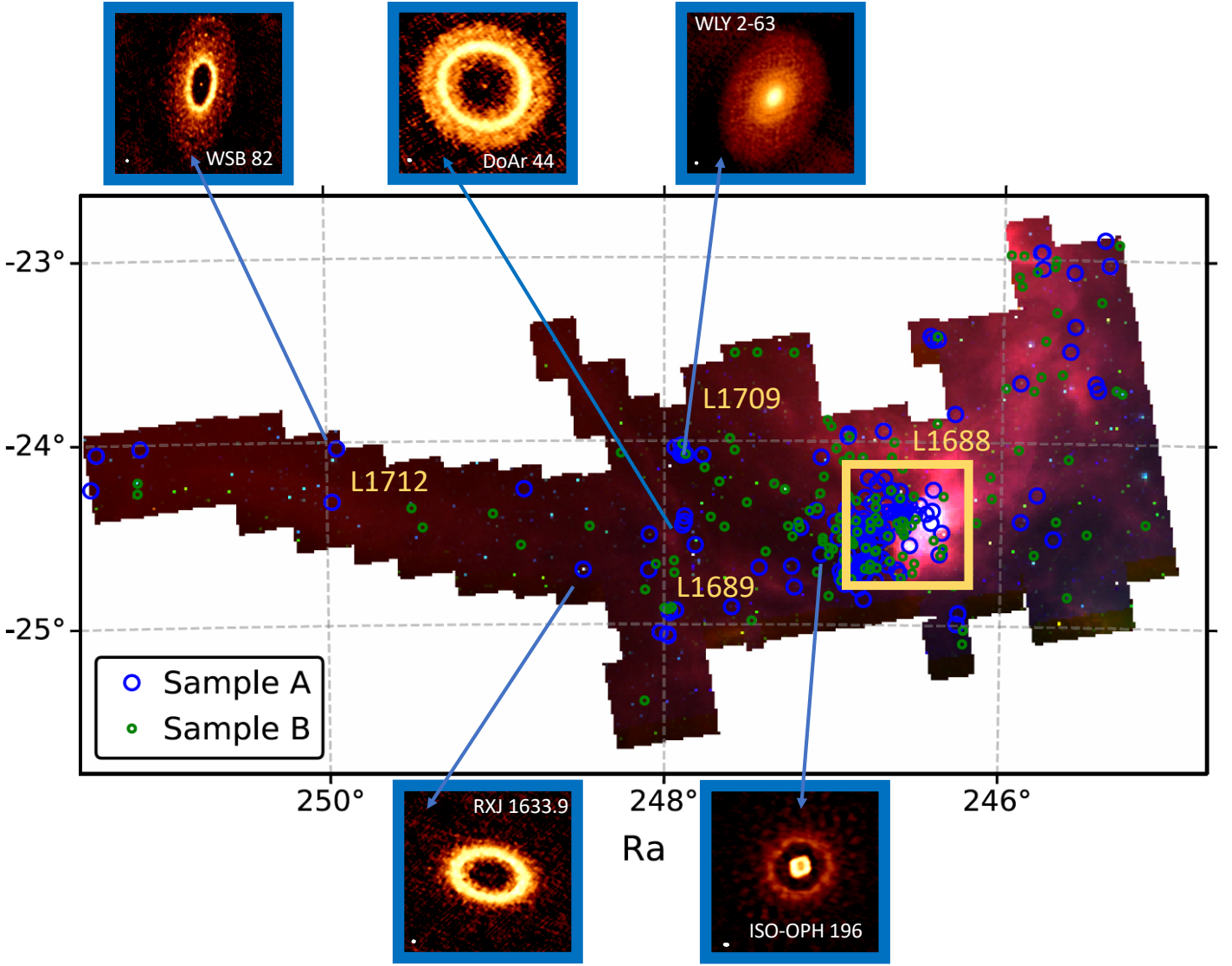}
\caption{Location of the ODISEA long-baseline targets outside of the L1688 star-forming cluster shown in Fig.\ref{f:L1688} (yellow square). Other regions of the cloud are also labeled in yellow. The locations are overlaid over a color composite image from \emph{Spitzer} 3.6, 4.5, and 8.0 $\mu$m data corresponding to the Fig. 2 in \citet{cieza2019}, defining the ODISEA samples first observed at 0.2$''$ (Sample A) and 0.6$''$ (Sample B)  resolutions.  Some of the targets (e.g., WSB 82 and RX J1633.9-2442) are quite isolated, specially when compared to the number density of objects observed in L1688.}
\label{f:c2d_all}
\end{figure*}

\subsection{Caveats and uncertainties}\label{s:caveats}

The substructures observed at 1.3 mm and listed in Tables~\ref{t:gaps},~\ref{t:cav}, and~\ref{t:inflections} are the result of exploratory work at unprecedented resolution for the given targets. However, we emphasize that the characterization of these substructures is subject to important caveats and uncertainties. 
First, the (sub)mm continuum observations are mostly sensitive to dust grains with sizes that are similar to the observing wavelength and that the grain size population is a complex function of radius due to dust evolution (e.g., growth, migration, and accumulation). Therefore,  the location and shape of the features are highly dependent on the wavelength observed. 
Second, all the features are convolved with the synthesized ALMA beam, which can make deep and narrow features look wider and shallower than they really are. 
Third,  the sizes and locations of the features all scale with the distance to the objects, which have uncertainties ranging from $\sim$1$\%$ for targets with accurate Gaia parallaxes to $\sim$20$\%$ for objects without such information. 
Based on the above, the errors in the values in au reported in Tables~\ref{t:gaps},~\ref{t:cav}, and~\ref{t:inflections} correspond to the errors introduced by the  uncertainties in the distance and half the beam size added in quadrature. Similarly, the errors in the depths of the gaps are calculated by propagating the 1-$\sigma$ errors in  the radial profiles at $I_{min}$ and $I_{max}$ to the $I_{min}$/$I_{max}$ ratio. 

Besides the more physical limitations noted above, we also make some methodological assumptions that might have small impacts on the tabulated results.  Namely, we assume that all features in a given disc have the same center, position, angle and inclination. 
This is likely to be the case on most features, but this assumption should be tested in the detail study of individual discs and features.  
We stress, however,  that none of our main conclusions (see Sec.~\ref{s:con}) are dependent on these assumptions or affected by the uncertainties previously mentioned. 

In order investigate how the properties of the substructures we derive depend on the data processing, in Appendix~\ref{s:frank} we use the python module Frankenstein (frank, \citealt{jennings2020}) to produce model images 
(Figs.~\ref{f:frank_1}~and~\ref{f:frank_2}) and a new set of radial profiles.  
Frank recovers  axisymmetric disc structures at sub-beam resolution by fitting the visibilities directly, creating a model and a radial brightness profile of the model  using a Gaussian process.  We find that frank recovers the gaps and rings we list in Table~\ref{t:gaps}, but  since it avoids the resolution loss from the beam convolution used by CLEAN, some of the features identified by frank are deeper and narrower (Fig.~\ref{f:frank_profiles}).  
Frank also identifies several additional local minima and local maxima. Some of them might be real gaps and rings that are not resolved in the CLEANed images, but they should be taken with caution and considered on a case-by-case basis. 
Since the Frank models assume azimuthal symmetry,  the models necessarily fail to identify the azimuthal structures present in the ISO-Oph 2A disc.  Frank also fails to reproduce the cavity we find in the CLEANed images of ISO-Oph 196, but it finds an infection point close to the center of the disc instead.  Further tests on the object reveal that the morphology of the inner disc is dependent on the details of the CLEANing process. Some of the CLEANed images show a cavity while others resemble more a faint inner disc and a narrow gap.  This suggests that the inner disc has low signal-to-noise structures close to the resolution of the observations and further investigation is needed. 

\section{Discussion}\label{s:dis} 

\subsection{Individual objects}

In this section, we review the main properties of each of the ODISEA long-baseline targets  and summarize the main substructures revealed by the observations at 3-5 au resolution.  

\subsubsection{ISO-Oph 54}

ISO-Oph 54 (also known as Oph Emb 22 or GY 91) is a Class~I protostar with  effective temperature of 3300 K  and an stellar luminosity of 1.7 L$_{\odot}$ \citep{doppmann2005} located in the L1688 stellar cluster (\citealt{lynds1962,padgett2008}) of the Ophiuchus molecular cloud (see Fig.~\ref{f:L1688}).  The source is deeply  embedded in a 0.06 M$_{\odot}$ envelope and has a bolometric temperature of 370 K \citep{enoch2009}. From previous ALMA  observations at 0.05$''$  and 0.13$''$ resolution (at 3 mm and  870 $\mu$m, respectively)  three gaps were previously identified at $\sim$10 au, $\sim$40 au, and $\sim$70 au \citep{sheehaneisner2018}.

With an $\alpha_{IR}$ value of 0.45, ISO-Oph 54 is the most embedded source of the ODISEA long-baseline sample based on its IR SED. 
We find that the object has an inner dust cavity $\sim$2.5 au in radius (see Fig.~\ref{f:small_cav}). 
The outer edge of this cavity constitute one of the five rings (local maxima) with peaks at 2.5 au, 21 au, 38 au, 58 au, and 88 au (Fig.~\ref{f:profiles}). We also identify 4 clear local  minima in the brightness profiles (gaps) at 12 au, 32 au, 48 au, and 72. 
The first and last of these gaps roughly correspond to those identified by \citet{sheehaneisner2018}, but our observations resolve the gap they report at 40 au into the 2 gaps and a ring (D32, B38, and D48).  
In addition to these local extremes, the brightness profile of ISO-Oph 54 shows an inflection point at 102 au.
ISO-Oph 54 is a clear demonstration that very young sources (age $\lesssim$ 0.5 Myr) still embedded in their natal envelopes can already show a rich variety of substructures.  

\subsubsection{WLY 2-63}

WLY 2-63 (also known as Oph Emb 17 or IRS 63) is a Flat Spectrum source located in the L1709 region (\citealt{lynds1962,padgett2008}) of the  Ophiuchus cloud. The system has a bolometric temperature of 327 K and is deeply embedded in a 0.07  M$_{\odot}$ envelope falling at an accretion rate of 1.2$\times$10$^{-7}$ M$\odot$ yr$^{-1}$ (Kristensen et al., 2012; Brinch $\&$ Jørgensen, 2013). Based on high-resolution near-IR spectroscopy, Doppmann et al. (2005) estimate an effective temperature of 4200 K and a luminosity of 3.8 L$_{\odot}$ for the central object. 
With a 1.3 mm flux of 335 mJy, WLY 2-63 is the brightest of the ODISEA long-baseline targets. No clear substructures are seen at 0.2$''$ resolution \citep{cieza2019}, but the long-baseline image of the system (Fig.~\ref{f:all_images}) resembles three "concentric discs". 
The deprojected brightness profile (Fig.~\ref{f:profiles}) shows that the inner most region is much brighter than the rest of the disc and shows two inflection points at 21 and 38 au, corresponding the the boundaries of the "concentric discs" seen in the image. 
\citet{2020Natur.586..228S}, recently presented a detailed analysis of the WLY 2-63  disc using ALMA high-resolution data very similar to ours.  They showed the two shallow gaps are revealed at the location of the inflection points when a smooth disc model is subtracted from the data. They also find that the temperature of the disc at the location of the outer gap/inflection point is consistent with the CO snowline.

\subsubsection{ISO-Oph 37}

ISO-Oph 37 (also known as GY 21) is protostar with a Flat Spectrum IR SED, an effective temperature of 3900 K and a luminosity of 0.8 L$_{\odot}$ \citep{doppmann2005}. It is located in the L1688 cluster. 
We find that the disc around ISO-Oph 37 is highly inclined (i $\sim$ 72 deg) but the disc is large (R$_{90\%}$ = 91 au) and very well resolved along the major axis. Still, the radial profile shows no substructures, except for an inflection point at 31 au. 
As with WLY~63, it would be interesting to investigate the temperature structure of the disc and the potential connection of this inflection point to a snow line.  
WLY~63 and ISO-Oph~37 are rare examples of large and  massive discs without clear gaps and rings at 3-5 au resolution and is worth noting that both are very young ($\lesssim$ 1 Myr) embedded (Flat Spectrum) sources based on their IR SEDs.

\subsubsection{ISO-Oph 17}

ISO-Oph 17 (also known as GSS 26) is a Class II source also located in the  L1688 cluster of the Ophiuchus molecular cloud. It shows a "full" SED (without any indication of large gaps or inner hole). As such, it is the typical target included in the DSHARP sample. 
As most objects in DSHARP, ISO-Oph~17 shows concentric gaps and rings, with clear local minima at 20 au and 37 au and the corresponding local maxima at 25 and 47 au. %
To our knowledge, these gaps and rings have not been reported thus far; however,  
the D-37 gap was previously noted as an inflection point based on the 0.2$''$ resolution ODISEA data \citep[see object $\#$30 in][]{cieza2019}.

\subsubsection{DoAr 44}

DoAr~44 (also known as WSB 72 and HBC 268) is located in the L1689 region of the Ophiuchus cloud (\citealt{lynds1962,padgett2008}) and has a pre-transition disc IR SED. 
The large cavity in the system has been imaged before both in the submillimeter \citep{vandermarel2016} and in the near-IR \citep{casassus2018} using  Differential Polarized Imaging (DPI). The DPI observations shows shadows in the outer disc that are likely to be produced by a highly inclined inner disc. 
Our long-baseline observations detect this inner disc, which remains unresolved at $\sim$4 au resolution and has an estimated mass of 0.04 M$_{\oplus}$. 
DoAr~44 exhibits the shallowest cavity edge in the sample with $R_{Cav.10\%}$ = 26 au and $R_{Cav.}$ = 47 au, a narrow peak extending from an inflection point at 39 au to another inflection point at 58 au, followed by a narrow shoulder. 
The combination of these features results in a peculiar 1.3 mm image (see Fig.~\ref{f:all_images}) with a "ring-within-a-ring" morphology.  

\subsubsection{WSB~82}

WSB~82 (also known as IRAS 16367-2356) is a relatively isolated source located in the L1712 region,  in eastern part of the Ophiuchus molecular cloud (see Fig.~\ref{f:c2d_all}).  
With a spectral type of K0, WSB~82 is the hottest star in the ODISEA long-baseline sample. 
Using ALMA, \citet{cox2017} imaged the cavity of WSB~82 for the first time and identified a gap at $\sim$150 au.
Our observations reveal an unresolved inner disc with a mass of $\sim$0.1 M$_{\oplus}$, a narrow ring at the edge of the cavity,  followed by an inflection point (I-80 au) and a narrow shoulder behind this inflection point. 
Unlike DoAr 44, WSB~82 has an extended outer disc with two
pairs of gaps and rings. The local minima are located at 114 and 186 au, with the corresponding local maxima at 123 and 269 au.
The gap identified by \citet{cox2017} at $\sim$150 au corresponds to D-186. The difference in location is partially due to the difference in adopted distance (137 vs 155 pc). 
WSB~82 also has, by far, the largest disc in the sample (R$90\%$ = 256 au).

\subsubsection{ISO-Oph 2}\label{s:iso-oph_2}

ISO-Oph 2 is a wide separation (240 au) binary system in the western edge of the L1688 star-forming cluster (\citealt{lynds1962,padgett2008}). The primary is M0 star \citep{gatti2006}, while the secondary is at the brown dwarf limit \citep{2020ApJ...902L..33G}. %
The system shows a pre-transition disc SED, although it is unclear whether the near-IR excess arises from the primary or the secondary disc. 
The dust cavity around the primary was resolved for the first time by the ODISEA observations at 0.2$''$ resolution \citep{cieza2019}. 

With a flux of 72 mJy at 1.3 mm, ISO-Oph 2A is the faintest disc  of the ODISEA long-baseline targets. 
The system holds several records within the Ophiuchus molecular cloud. Its primary disc has the largest dust cavity in Ophiuchus \citep{cieza2019}, while the secondary is the lowest-mass object ($\sim$0.08 M$\odot$) with a resolved transition disc, which in turn  features the smallest dust cavity resolved thus far in the cloud (r $\sim$2.2 au).
The primary disc shows two closely-packed narrow rings with strong azimuthal brightness asymmetries.  
As shown in Figure~\ref{f:L1688}, the brightest part of the outer ring is in the direction of  HD~147889, a binary system composed of a B2IV-B3IV pair, which is the brightest UV source in the cluster \citep{casassus2008} and might play a role in the asymmetric illumination of the rings around ISO-Oph~2A.
Furthermore, the ISO-Oph 2 system shows a bridge of gas connecting the primary and the secondary discs, which could be an indication that the secondary has recently flown by the primary. 
For a detailed discussion of  this extraordinary system, see \citet{2020ApJ...902L..33G}. 

\subsubsection{ISO-Oph 196}

ISO-Oph 196 (also known as WLY 1-58 or WSB 60) is located in the outer edge of the L1688 cluster, in the direction of the L1689 region. ISO-Oph~196 has a Class II SED, and with an M5 spectral type \citep{manara2014}, it is the coolest star in the sample. Based on ALMA observations at 0.15$''$ resolution, a gap was recently reported extending from from 11 au to 32 au \citep{francisvandermarel2020}.   
Our long-baseline observations reveal an inner dust cavity 8.5 au in radius, a deep gap at 18 au (consistent with the values mentioned above), followed by a ring that is centered at 34 au and  has a width of 16 au. We also identify inflection points  at 23 au,  65 au, and 82 au. We note that the inner dust cavity is subject to the specific caveats discussed at the end of Sect.~\ref{s:caveats}.

\subsubsection{EM* SR 24S}\label{s:SR24}

EM* SR\,24S or SR\,24S is part of a hierarchical triple system, where SR 24S is a single star and SR24N is a binary system. The separation between SR 24S and SR 24N is 5.2’’ \citep{reipurth1993}.
Both  discs were resolved in the scattered light infrared images that show that the primary disc and secondary discs are connected by a spiral arm pattern \citep{mayama2010}. However, ALMA observations at multiple wavelengths for the SR24 system do not reveal any dust emission in SR24N disc, while the SR24S disc shows a ring-like structure peaking at $\sim$0.32’’ (\citealt{pinilla2017,pinilla2019}).
The multi-wavelength analysis from ALMA observations of the ring of SR\,24S supports the scenario of a planet carving a gap and forming the observed cavity and ring-like structure. These observations also revealed an inner disc, which is likely dominated by dust thermal emission instead of free–free emission. The existence of such an inner disc puts an upper limit to the potential planet(s) embedded in this disc of around $\sim$5 M$_{JUP}$ \citep{pinilla2019}. CO emission has been detected in both systems (SR\,24N and SR\,24S, \citealt{andrewswilliams2005,pinilla2017}). 
Our long-baseline observations of SR 24S resolve the ring-like morphology of the outer disc into several components:
a sharp inner edge  ($R_{Cav.50\%}$ = 24 au, $R_{Cav.90\%}$ = 27 au), a relatively narrow peak at 30 au, followed by an inflection point at 37 au, and a disc "shoulder" or "plateau" extending to R$_{90\%}$ = 58 au.
We also detect an unresolved inner disc with an estimated mass of $\sim$0.08 M$_{\oplus}$. 

\subsubsection{RX J1633.9-2442}

RX J1633.9-2442 is young stellar object with a K5 spectral type. It was classified  by \citet{cieza2010} as a "giant planet-forming disc" based on its transition disc SED, low but detectable accretion rate ($\sim$10$^{-10}$ M$_{\odot}$/yr), and high disc mass ($>$10 M$_{JUP}$ assuming a gas to dust mass ratio of 100).  Its dust cavity was first resolved by observation with the Submillimeter Array, from which \citet{cieza2012} estimated the radius of the cavity to be $\sim$25 au. Our ALMA long-baseline observations show that the outer disc is particularly narrow, with R$_{Cav.50\%}$=24, R$_{Cav.}$ = 36 au and R$_{90\%}$ = 53 au.  
The radial brightness profile is very noisy close to the star, but  no inner disc is detected with an upper limit of $\sim$0.02 M$_{\oplus}$ for the central beam. This non-detection is consistent with the lack of near-IR excess in the SED (see Fig.~\ref{f:seds}). Deep aperture masking observations in the near-IR rule out the presence of companions more massive than $\sim$6 M$_{JUP}$ at distances larger than $\sim$3 au from the star \citep{cieza2012}, which suggests that the cavity and the ring-like morphology of the outer disc are due to the dynamical interactions with a planet rather than a brown dwarf or low-mass star. 

\begin{figure*}
\includegraphics[trim=0mm 0mm 0mm 0mm,clip,width=18.0cm]{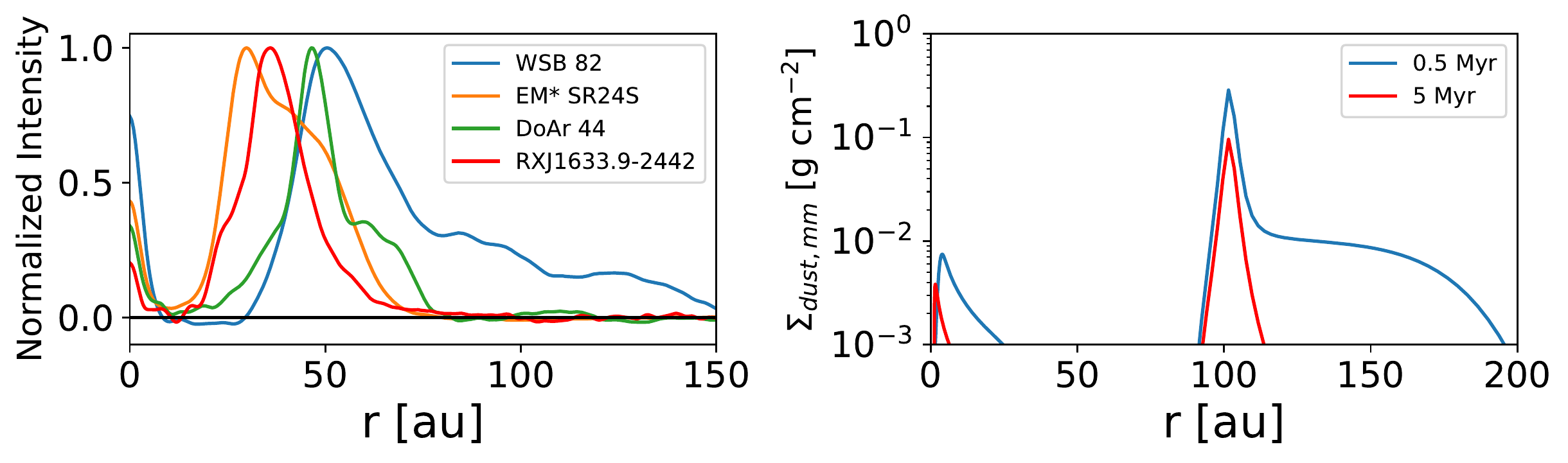}
\caption{The left panel shows the brightness profiles of  WSB~82, EM* SR24, DoAr~44, and RXJ1633.9-2442, all discs with large dust cavities. ISO-Oph~2 is not included because it exhibits two non-axisymmetric rings. 
The profiles \emph{qualitatively}  resemble the predictions of dust evolution models of discs with planet-induced cavities, in which rings become narrower with time  due to the radial migration of the dust in the outer disc and inner discs also start to dissipate.  
As an illustration, the right panel shows the dust density after 0.5 and 5\,Myr from dust evolution models that include a massive planet (1 M$_{JUP}$) at 70\,au to create a large gap \citep{pinilla2019}. In this context, the profile of RXJ1633.9-2442 could be considered more evolved than those of WSB~82, DoAr 44 or EM*~SR~24S. 
}
\label{f:TD_models}
\end{figure*}

\subsection{The origin and evolution of substructures in massive discs}

\subsubsection{The possible origins of gaps and rings} 

Since the first ALMA long baseline observations revealed a stunning sequence of gaps and rings seen in continuum emission in the borderline Class I/II object HL Tau \citep{alma2015}, and the subsequent DSHARP survey revealing that these are common \citep{Andrews2018}, there has been much speculation about their origin. No shortage of mechanisms have been proposed to explain them. Since dust grains emitting at mm-wavelengths likely undergo relatively fast radial drift \citep{Weidenschilling1977}, axisymmetric dust rings can in principle be created by any mechanism that creates local pressure maxima (i.e. dust traps) in the disc. Examples of proposed mechanisms include condensation fronts \citep{Zhang2015}, self-induced dust traps \citep{Gonzalez2015}, magnetic dead zones \citep{flock2015} and secular dust instabilities \citep{takahashi2014,loren-Aguilar2015}. Some of these mechanisms, such as condensation fronts \citep{Zhang2015}, are unlikely to occur in most cases as the rings do not precisely align with the required temperature profile \citep{huang2018,long2018}, although uncertanties remain due to the unkowns in the disc temperature. Other mechanism require the change of the dust properties to change their local opacity, e..g as sintering might be able to do and explain the brightness variations \citep{okuzumi2016}.

Part of the initial confusion arose because of the axisymmetry of the observed rings and gaps, since the expectation from planet-induced structures was that non-axisymmetric structures should be present \citep[e.g.][]{deValBorro2006}. \citet{Dipierro2015} showed that this discrepancy can be resolved because the dust disc is much thinner than the gas disc (e.g. $H$/$R$ $\ll 0.01$ for the dust disc in HL~Tau; see \citealt{Pinte2016}) and relatively decoupled. 
Multiple gaps can also arise from a single planet \citep{Dong2017}.
The low aspect ratio implies that even low mass planets can carve detectable dust gaps (e.g. super-Earths in TW Hya; \citealt{Dong2018,Mentiplay2019}), since the mass threshold required for dust gap opening scales as ($H$/$R$)$^{3}$ \citep{Dipierro2017}. However, such low mass planets would not be expected to carve corresponding gaps in the gas \citep{Dong2015,Dipierro2016}. 
More recently, \citet{Pinte2018} found strong evidence in favour of an embedded planet in the HD163296 disc, in the form of localised deviations from non-Keplerian motion seen in CO channel maps, consistent with signatures from planet-disc interaction predicted by \citet{Perez2015}. 
Subsequent to this, \citet{pinte2019} detected a similar localised velocity `kink' in the transition disc around HD97048, providing direct kinematic evidence for a planet with a mass of 2--3 times that of Jupiter being responsible for carving the dust gap. 

Most recently, \citet{Pinte2020} found tentative evidence for localised deviations from Keplerian motion in 8 of the 18 discs observed in the DSHARP survey, with the strongest evidence in favour of planets being that all of the candidate planets lie either in a dust gap or at the tip of a spiral arm seen in continuum emission. So while the dust has not yet settled on the mechanism behind gap and ring formation, the balance of evidence is tipping in favour of planets being responsible for at least a fraction of the observed rings and gaps. The planet hypothesis was already explored in detail for the DSHARP survey by \citet{Zhang2018}.

The absence of deep rings and gaps in WLY~2-63 and ISO-Oph~37 is also intriguing, suggesting that these discs either have not yet formed planets, or that any planets are not yet sufficiently massive to carve deep gaps in the disc. 
High optical depths might also play a role in the lack of clear substructures. If the continuum optical depth is sufficiently high, even significant drops in surface density could remain undetected.  However, for most sources, this is only likely to occur in the inner disc \citep{huang2018}. Even in the case of the massive disc around WLY 2-63, the optical depth  at 1.3 mm is estimated to approach unity outward of  $\sim$25 au \citep{2020Natur.586..228S}.  
Another possibility is that the mm-emitting dust may be well coupled to the gas in these discs and therefore not yet settled to the mid-plane.  
By contrast,  the gap depths in ISO-Oph~54, ISO-Oph~17 and ISO-Oph~196 suggest the presence of sufficiently massive planets that kinematic follow-up may be able to reveal their presence within these gaps. ISO-Oph~196 appears similar, in terms of the size and location of the gap, to DS Tau for which predictions have been made suggesting that the kinematic detection of the planet with ALMA is possible \citep{ Veronesi2020}.

Besides condensation fronts and planet-disc interaction, there are still other alternatives for the formation of rings and gaps, such as radial variations of disc viscosity which are expected from non-ideal MHD simulations \citep{uribe2011,flock2012,flock2015} and disc winds \citep{suriano2017,suriano2018,suriano2019,riols2019,riols2020}. The observational predictions from such models are still unclear and therefore it remains as a crucial question how we can definitely distinguish a planetary gap from other mechanisms. In case of large gaps or cavities as observed in transition discs, there have been several efforts from the theoretical point of view to distinguish between planets and other alternatives, such as dead zones and/or MHD winds, as discuss in the next section.

\subsubsection{Discs with large inner holes and their connection to disc evolution and planet formation}\label{s:TD}

Since transition objects where first identified as discs with ``inner holes" \citep{strom1993}, they have been considered to be a key clue in the disc evolution and planet formation puzzle.  
However, transition discs are clearly a heterogeneous group of objects,  with a wide range of SED shapes, disc masses,  acretion rates, and multiplicity (\citealt{cieza2010,espaillat2014,pinilla2018,vandermarel2018}).
While some famous transition objects such as CoKu Tau 4 and  HD 142527 are known to be circumbinary discs 
(\citealt{irelandkraus2008,lacour2016,price2018}), most transition objects lack stellar companions that could explain their inner holes \citep{ruizrodriguez2016}, although in several cases observations of strongly non-Keplerian kinematics and spiral arms hint at the presence of relatively massive bodies inside the cavity \citep{Calcino2019,Poblete2020,Calcino2020}.

Non-circumbinary transition discs can be divided into two distinct groups based in their disc masses  and accretion rates, which might indicate different clearing processes \citep{owenclarke2012}.
On the one hand, weak-line T Tauri stars (non-accreting objects) are mostly discless (\citealt{cieza2007,wahhaj2010}), but the ones that do have IR excesses tend to show transition disc SEDs and very low disc masses (\citealt{cieza2008,hardy2015}).
The properties of such objects are most consistent with photoevaporation models and disc dissipation \citep{alexander2014} rather than with planet formation.
On the other hand,  massive and still accreting transition discs around single stars have been considered prime candidates for the formation sites of giant planets  (e.g., \citealt{zhu2012,pinilla2012}).  
All transition discs in our long-baseline sample belong to this latter category and provide an opportunity to further investigate the planet-formation hypothesis. 

We find that the transition objects in  our sample have outer discs with complex radial profiles indicating substructures at 3-5 au scales (see   Fig.~\ref{f:profiles}).  All the profiles peak near the edge of the cavities; however,  ISO-Oph 2 and RXJ1633.9-2442 show only narrow rings, while SR 24S DoAr 44, and WSB~82 show  inflection points and significant "plateaus" beyond  the mm peaks. 
These profiles are strikingly similar to the model  predictions of planet-disc interactions \citep{pinilla2015,pinilla2019},
in which giant planets inside the cavities produce pressure bumps in the outer discs, which in turn efficiently trap mm dust particles, resulting in relatively narrow peaks in the surface density profiles. 
The details of the profiles depend on time, the mass of the planet, the viscosity of the disc, and the distribution of particles sizes, among other parameters. Therefore, constraining the masses of the planets producing observed profiles requires both dedicated modeling and radiative transfer (to compare observed brightness profiles to numerical surface density profiles). 

Just as an illustration, in  Fig.~\ref{f:TD_models} we show the radial profiles of WSB~82, DoAr 44, EM* SR 24S, and RXJ1633.9-2442, (left panel), next to the surface density distribution after 0.5 and 5\,Myr of evolution of the millimeter-sized particles (0.1-10\,mm) from dust evolution models (right panel). The model includes a massive planet (1 M$_{JUP}$) at 70\,au to create a large gap and  the typical set-up from models in e.g.
\citet{pinilla2012,pinilla2015,pinilla2019}.
In short, this model includes the growth and fragmentation of dust particles when there is a large gap formed due to an embedded planet. The disc viscosity is assumed to be $\alpha$=10$^{-3}$ and the threshold for the fragmentation is 10\,ms$^{-1}$. The surface density distribution at 0.5 Myr  shows the "plateau" beyond  the accumulation of the millimeter-sized particles in pressure maximum. This "plateau" is a result of dust that is still growing in the outer parts of the disc and drifting towards pressure maxima. The "plateau" vanishes with time when all the mm-sized particles have finally drift to the pressure bump. The inner dust disc also becomes more compact and it dissipates with time. 
We note that Fig.~\ref{f:TD_models} compares observations in linear scale to model results in logarithmic scale and should only be considered qualitatively, not only for the reasons mentioned above (the lack of any fine-tuning and radiative transfer), but also because models of dust evolution in gas-rich discs tend to greatly overestimate the radial migration of mm-sized grains \citep[e.g., ][]{brauer2007}.
The detailed comparison of models to the observed profiles shown in Fig.~\ref{f:TD_models} might provide  hints on how to overcome the so-called "drift barrier",  which is currently one of the main challenges in the fields of dust evolution and planet formation \citep{testi2014}. 

\citet{pinilla2019} compared their models to multi-frequency ALMA data of EM* SR 24S at 0.1$''$ (11 au) resolution and concluded that the observed brightness profiles were most consistent with a planet-induced inner hole and inconsistent with photoevaporation and dead zone models.  As discussed in Sect.~\ref{s:SR24}, they also concluded that the presence of an inner disc constrains the mass of the planet, even if only one is responsible for the cavity, to be less than 5 M$_{JUP}$.
Our high-resolution observations of transition discs should allow us to place further constrains on the properties of the embedded planets (Gonzalez-Ruilova et al. in prep).  But, what is perhaps even more important,  they provide strong evidence for planet formation being the main driver of substructures in massive (and accreting) discs.

The idea that the cavities of transition discs and the gaps seen in most disc sources observed at high resolution are mostly due to planet formation has been taken with much caution, considering that current core accretion models can not explain the formation of planets at tens of au separations within the age of the targets ($\lesssim$ 1-3 Myr) and that most disc do not show evidence of gravitational instability. 
However, given the mounting evidence connecting transition disc cavities and rings to planets, planet formation is clearly becoming the leading  explanation for the origin of substructures. 
This evidence now includes, not only the dust subtructures as in in Fig~\ref{f:TD_models}, but also independent indications such as the  direct detection of protoplanets in the optical and infrared (\citealt{keppler2018,haffert2019}) and the kinematic detection of planets with ALMA \citep{pinte2019}.   

Since the observations very strongly suggest that giant planets can in fact form within cavities and gaps at tens of au or even at $>$100 au from the host star,  it is worth considering whether most substructures seen in Ophiuchus at 3-5 au resolution are a result of  planet formation processes. Before speculating  about such idea in more detail in Sec.~\ref{s:evol-sequence}, we first discuss the demographic context of the long-baseline sample.

\begin{figure*}
\includegraphics[trim=10mm 30mm 10mm 15mm,clip,width=18.0cm]{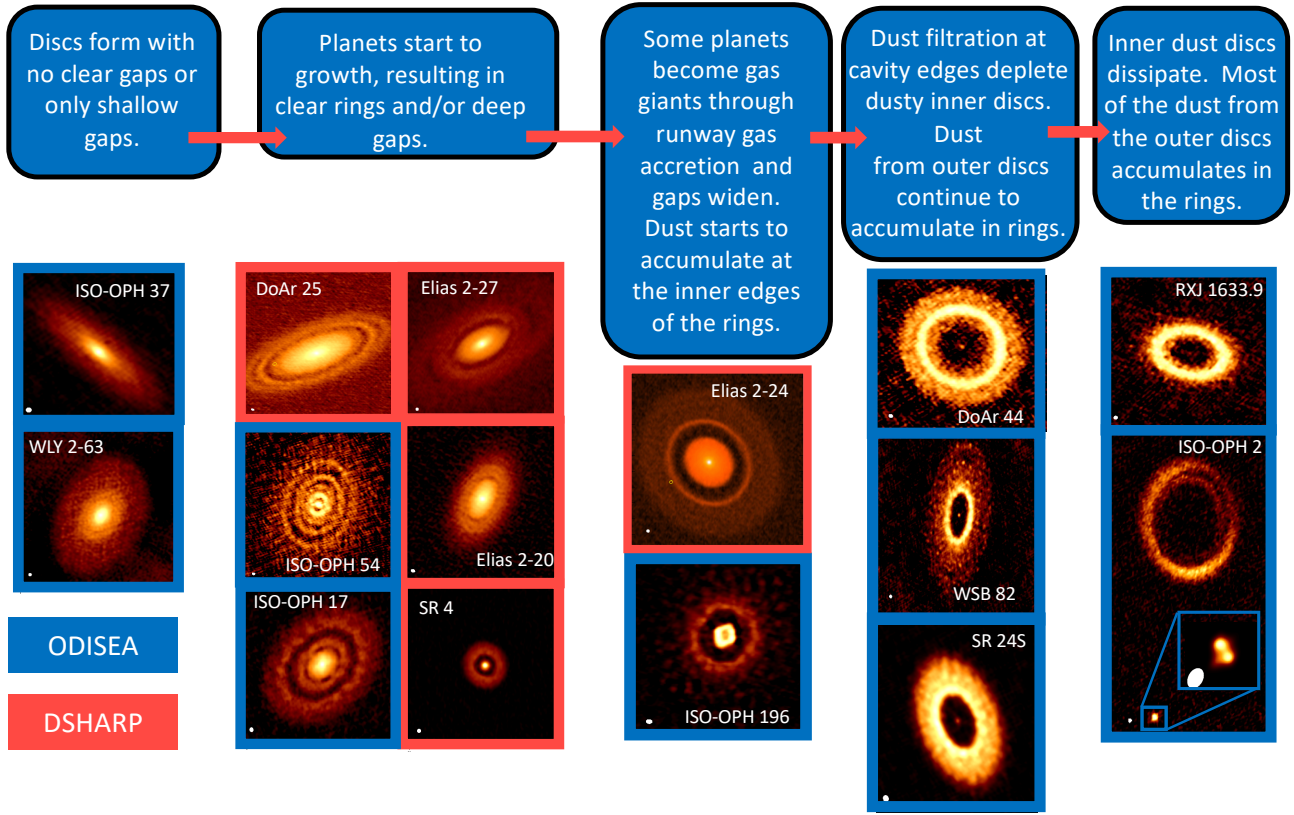}
\caption{Schematic figure of the possible evolution of substructures in massive discs  (M$_{dust}$ $\gtrsim$ 40 M$_{\oplus}$) using the objects in the Ophiuchus long-baseline sample to illustrate different stages.
In the proposed scenario, the progression of the features observed in 1.3 mm continuum at 3-5 au resolution is driven the formation of giant planets though core accretion and dust evolution. By construction, the scenario only applies to systems massive enough to form giant planets. }
\label{f:disc_evo}
\end{figure*}

\subsubsection{Demographic context}

The architectures of planetary systems depend on different  stellar properties, mainly mass, metallicity, and multiplicity (\citealt{winnfabrycky2015}).
Since the metallicity is a very difficult parameter to investigate in pre-main-sequence star, stellar mass and multiplicity become the important variable to consider for the stellar population in Ophiuchus.
Discs masses show a strong correlation with the mass of the host:  
M$_{dust}$ $\propto$ (M$_{\odot}$)$^{1.3-1.9}$ (\citealt{pascucci2016,andrews2020}). There is also at least an  order of magnitude scatter around this correlation for a given stellar mass and age. 
Given our sample selection (the brightest 5$\%$ of the Ophiuchus disc population), our sample is heavily biased toward massive stars (solar or super solar) in the cloud and the most massive discs for a given stellar mass.  
As shown in
Figure~\ref{f:histo_flux}, most of the discs in Ophiuchus are fainter than a few mJy at 1.3 mm and only have a few Earth masses worth of dust. Most of the dusty discs are also smaller than 15 au in radius \citep{cieza2019}. However they still meet the minimum requirements (in terms of mass and size) to form the kind of  the terrestrial planets detected by \emph{Kepler} or \emph{TESS}, which are restricted to orbital separations $<$ 1 au \citep{batalha2013}.

With disc masses of the order of a Minimum Mass Solar Nebula \citep{Weidenschilling1977}, $\gtrsim$10 M$_{JUP}$ (assuming a gas to dust mass ratio of 100), and radii $\gtrsim$30 au, the Ophiuchus long-baseline sample actually contains the few objects in the cloud that are large enough and (currently) have enough mass to form giant planets at large separation. 
We find that 2/3 of these objects are located in the L1688 cluster (Fig.~\ref{f:L1688}), which contains $\sim$50$\%$ of the YSOs in Ophiuchus.  This suggests that the environment of this cluster does not strongly affect the presence of large and massive discs.
However, as shown by \citet{zurlo2020},  very large and massive discs in Ophiuchus are typically only present around single stars or wide separation binaries such as ISO-Oph 2 or EM* SR 24. 
Given the  strong effects binaries have on disc sizes, masses, and lifetimes (\citealt{cieza2009,kraus2012}), the entire ODISEA sample is likely to be biased against medium-separation (1-100 au) binary systems. Instead, such binaries are likely to be part of the vast population of discless  Ophiuchus members recently identified by Gaia \citep{canovas2019}. Gaia has identified over 150 new members, but the census of discless stars remains incomplete given the high extinction in the cloud; therefore, the total population of pre-main-sequence stars in Ophiuchus is likely to be over 500. This implies that the 15 objects in the Ophiuchus long-baseline sample correspond to 5$\%$ of the objects with discs but $\lesssim$ 3$\%$ of the pre-main-sequence population in the cloud.  We therefore conclude that the Ophiuchus long-baseline sample represent a very small and special subset of the objects in the region
and that these objects are the most likely formation sites of giant planets at large separations (r $>$ 5-100) au in the entire molecular cloud if such planets form through core accretion. 

The discs around stars with and without giant planets are expected to evolve in different ways. 
The SEDs of YSOs do show evidence for different evolutionary paths (\citealt{cieza2007,williamscieza2011}). 
While transition discs are very overrepresented in the literature, most of the Class~II objects in molecular clouds have similar SEDs, which can be reproduced by a continuous disc extending down to the dust sublimation radius at $\sim$0.01 au \citep{ribas2017}.
Such SEDs can evolve from strong excesses at all near-IR and mid-IR wavelengths  into SEDs with weaker but detectable excesses at the same wavelengths.  Objects with these latter types of SEDs are know as anemic \citep{lada2006}, homologously depleted \citep{currie2009}, or weak excess \citep{muzerolle2010} discs. 
These objects are consistent with the dissipation of the disc without the formation of large cavities,  except perhaps for the final inside-out clearing of the disc through photoevaporation, a fast  process \citep{alexander2014} that is statistically difficult to observe. 
To match the statistics of exoplanet indicating that 
that M-type stars host an average of $\sim$2.2 $\pm$ 0.3 rocky planets  (R = 1-4 R$_{\oplus}$ and orbital period = 1.5-180 days \citep{gaidos2016}, these numerous and "unremarkable" discs must be their main formation site. 
On the other hand, if massive discs with large dust cavities are associated to the formation of giant planets as discussed in the previous section, it is not surprising that such objects represent a significant fraction of the Ophiuchus long-baseline sample (33$\%$). 
Unfortunately, the incidence of $\sim$ 1 M$_{JUP}$ exoplanets at 10-100 au separations is not yet well constrained \citep{winnfabrycky2015}, and there is currently little overlap  in the mass-separation plane between the population of known exoplanets and the potential population of planets derived from the properties of dust gaps observed in discs \citep[e.g., ][]{Zhang2018} to allow for a direct comparison.

\subsubsection{Towards an evolutionary sequence of substructures in massive discs}\label{s:evol-sequence}

The Ophiuchus long-baseline sample is extraordinarily diverse in terms of SED Class and mm continuum morphologies and therefore provides a unique opportunity to speculate about the evolution of substructures of the most massive discs in a given molecular cloud. Even though each system is different and likely to form a distinct set of planets, most discs might still follow some common evolutionary steps, from formation to dissipation, which we illustrate in Fig.~\ref{f:disc_evo}.
As discussed in Sec.~\ref{s:sed}, the SED provides a good reference for the evolutionary stage of YSOs. While relatively small, the Ophiuchus long-baseline sample covers a wide range of SED classes (I, FS, II/full, II/PTD and II/TD).
We find that only 33$\%$ (1/3) of the embedded (Class I/FS) sources show deep gaps and bright rings (i.e, ISO-Oph 54), while 100$\%$ (12/12) of the Class II objects present such features.  
Massive FU Ori discs like V883 Ori, V2775 Ori and V1647 Ori, which are believed to represent very early stages of disc evolution, also lack the obvious gaps seen in Class II sources (\citealt{cieza2016}; Principe et al. in prep.) but show radial profiles with inflection points that could be due to changes in grain properties at snow lines.
This suggests that the onset of bright rings and deep gaps occurs while discs are still embedded in their envelopes \citep[age $\lesssim$ 1.0 Myr; ][]{evans2009},  but that substructures are not present from the start, with the possible exception of more subtle features that might be related to snow lines (first column in Fig.~\ref{f:disc_evo}). 
The first pressure perturbations could be caused by magnetic effects (e.g., \citealt{flock2015}) combined with wind-driven mass loss (e.g., \citealt{riols2019}). Such perturbations would be enough to cause variations of the radial dust drift and so dust concentration and efficient growth. These locations could serve as the formation sites of planetary embryos. 

By the time envelopes dissipate and the YSOs become Class II sources, the vast majority of massive discs have developed bright rings and/or deep gaps (second column). 
If the gaps are due to the growth of planets, their widths and depths should depend on the mass and location of the planet \citep{long2018}.   Most of the planets inferred from the gaps that have been observed have masses below or close to Neptune \citep{andrews2020}.
In the core accretion model, gas giants form by the rapid runaway accretion of gaseous envelopes onto $\sim$10 M$_{\oplus}$ cores \citep{helled2014}. The runaway gas accretion phase is expected to last of the order of 10$^5$ years for a 1 M$_{JUP}$ planet \citep{marley2007} and will continue for as long as there is material in the feeding zone. 
 As the planets grows from  $\sim$10 M$_{\oplus}$ cores to a 
 $\gtrsim$300 M$_{\oplus}$ giant planets, the gaps are expected to widen.
 At the same time, dust particles migrating in from the outer discs will start to accumulate at the strong pressure bumps produced by the newly-formed massive planets (third row in Fig.~\ref{f:disc_evo}). 
 
When giant planets accrete their envelopes, accretion is diverted onto the planets themselves and the accretion onto the stars is reduced by a factor of $\sim$10 \citep{lubowdangelo2006}, in agreement with demographic results showing that transition discs have low accretion rates for a given disc mass \citep{najita2007}.
If planets become massive enough, they can even halt accretion onto the star  completely \citep{lubow1999}, isolating the inner disc from the outer disc, as with brown dwarf and stellar-mass companions. 
The complete isolation of the outer disc is not seen in the ODISEA long-baseline sample because all the transition discs have detectable accretion (see Table~\ref{t:basic_prop}). 
In any case, when a giant planet is formed, the strong pressure bumps created will efficiently filter mm-sized dust grains (\citealt{zhu2012,pinilla2016}). This will prevent the  replenishment of dust from the outer to the inner disc and the dusty inner disc will start to dissipate as dust grain rapidly migrate in (fourth row in Fig.~\ref{f:disc_evo}). 
Meanwhile, in the absence of additional pressure bumps, dust particles in outer discs will also continue to  radially drift and rings will become narrower. Eventually the dust in the inner disc will dissipate completely and/or grow beyond detectable sizes,  leaving only detectable outer rings even in gas-accreting systems (last row in Fig.~\ref{f:disc_evo}).

The scenario discussed above should be considered a tentative hypothesis at this point, but it represents a coherent picture of the evolution of substructures in discs that can be tested by detailed numerical modeling of each "evolutionary step" (e.g., see Sec.~\ref{s:TD}) and future observations. 
More observations of embedded sources at 3-5 au resolution would be particularly useful to investigate the timescale for the formation of gaps and rings. 
Given the low mass of the planets needed to produce the gaps and cavities observed, the  proposed scenario is still consistent with the non-detection of planets in most transition discs (e.g., \citealt{zurlob2020}).  However, future observations with JWST and the ELTs  should result in more detection of protoplanets within gaps and cavities (e.g., PDS 70 b, PDS 70c), confirming the planetary origin of these substructures. 

The implications of the proposed scenario are far-reaching as it would require planet formation through core accretion to be extremely fast ($\lesssim$1-3 Myr) and efficient at many tens of au from the star, something that cannot be reproduced by current numerical models \citep{benz2014}.  It would also represent an opportunity to use disc substructures to investigate the architecture of nascent planetary systems and a population of planets that is currently very difficult to study by any other technique (planets with masses $\lesssim$ Saturn at tens of au from the host star). 
Since the ALMA images of TW Hydra at 1~au resolution \citep{andrews2016} suggest that rings and gaps are also common in discs less massive than the ones we are considering in Ophiuchus,
similar studies with the future ngVLA would allow extending this novel "planet detection technique" to rocky planets within the snow lines of protoplanetary discs in nearby star-forming regions \citep{2018ApJ...853..110R}.

\subsection{Connection to substructures in debris discs}

The substructures that we found in all massive Class II discs are likely to play an important role in the structure of massive debris discs if these are formed from the most massive Class II systems.
This is because the substructure at tens of au overlays with the typical radii of exoKuiper belts \citep{Matra2018}. This type of debris disc is the extrasolar analogues of the Kuiper belt, although with much higher dust levels sustained over Gyr timescales by destructive collisions of planetesimals at tens of au. Such planetesimal discs have masses $\gtrsim10\ M_\oplus$ \citep{Krivov2020}, consistent with the present
range of solid mass in the form of dust in our Class II
targets. How the substructures in Class II discs relate to those of a young exoKuiper belt is an open question. \cite{Stammler2019} proposed that the bright rings in Class II discs are ideal formation locations for planetesimals via streaming instability, and this could explained the relatively uniform optical depth of these rings in the DHSARP sample. It is tempting to conclude then that a debris disc formed through this process will inherit the dust distribution in those locations where planetesimal formation conditions are met. However, as proposed here the structure of Class II discs is likely to evolve significantly over a few Myr. Therefore the structure of a debris disc might be set instead by the cumulative planetesimal formation history over multiple Class II disc stages. This might explain why debris discs are typically wider than substructure in Class II discs \citep[][Matra et al. in prep]{Matra2018}. Alternatively, planetesimals in debris discs could form over a wide range of radii at the last stages of protoplanetary discs as gas densities drop due to photoevaporation and streaming instability becomes more efficient \citep{Carrera2017}. Both scenarios could lead to exoKuiper belts, although the latter scenario is less constrained by observations.

Thanks to ALMA, in the last few years tens of exoKuiper belts have
been resolved \citep[e.g.][and references therein]{MacGregor2013, Dent2014, Marino2016, Matra2018} and their radial structure revealed. Of that sample only 6 systems have been resolved with a beam small enough to find substructure, 5 of which show evidence of gaps \citep{Marino2018, Marino2019, Marino2020, MacGregor2019, Daley2019}. Such gaps in the dust and planetesimal distribution could have been inherited from the dust distribution in their parent protoplanetary discs. Discs such as the one around ISO-OPH~2 ISO-OPH~54 indeed show gaps at 50-80~au, a range that is similar to the range in which gaps have been found in exoKuiper belts \citep{Marino2020}. Increasing the sample of observed systems with large protoplanetary and debris discs would provide better statistics on the propierties of gaps, which would help to assess if this type of substructures is simply inherited from the Class II stage or established afterwards.

A caveat in this direct comparison is that if the structure in Class
II discs varies significantly over time in a given system, the structure of the product planetesimal disc might not resemble the
instantaneous distribution of dust in a Class II disc. However, if
gaps are indeed carved by planets, the properties of gaps in wide
debris discs might already provide additional constraints for models
that have attempted to predict the evolution of protoplanets
responsible for gaps in protoplanetary discs \citep[e.g.][]{Lodato2019, Ndugu2019}. For example, the location and width of a gap in debris discs could constrained the final mass and radial migration of those planets.

\section{Summary and conclusions}\label{s:con}

We present 1.3 mm continuum ALMA long-baseline observations at 3-5 au resolution of the 10 brightest ODISEA targets not included in the DSHARP ALMA Cycle-4 Large Program.  Our sample includes 
objects with a wide range of SED types: Class I, Flat Spectrum, and Class II. This latter Class can be further divided into full, pre-transition, and transition discs. 
Our main results can be summaries as follows: \\

\noindent 1) By number, gaps and rings are the most common types of substructures. 
We identify a total of 26 narrow rings and gaps  (width/radius $\lesssim$ 1) distributed in 8 sources: 
ISO-Oph~54 (4 rings and 4 gaps); ISO-Oph~17 (2 rings and 2 gaps); DoAr~44 (1 ring); WSB~82 (3 ring and 2 gaps), ISO-Oph~2A (2 rings, 1 gap); ISO-Oph~196 (2 rings and 1 gap), EM*~SR~24S (1 ring), and RXJ1633.9-2442 (1 ring). \\

\noindent 2)  Two discs (WLY 2-63 and ISO-Oph 37) around embedded protostars with Flat Spectrum SEDs lack the clear gaps and rings that are ubiquitous in more evolved sources with Class II SEDs. However, the only Class I object in the sample (ISO-Oph 54), shows a large number of rings and gaps.
Since other embedded sources in the literature also lack clear gaps and rings, 
this suggests that these substructures typically appear in the embedded stage of YSOs (age $\lesssim$ 1.0 Myr), but are not necessarily present from the formation of the disc.  More observations of embedded sources at 3-5 au resolution are needed to confirm this trend. \\

\noindent 3) The five Objects with large ($>$ 20 au) dust cavities (ISO-Oph~2A; EM* SR 24S, DoAr 44, RXJ1633.9-2442, and WSB 82) exhibit different degrees of dust accumulation at the edges of their cavities, in agreement with numerical simulations of pressure bumps produced by giant planets.  Three of the five objects mentioned above (EM*~SR ~24S, DoAr~44, and WSB~82) have small inner dust discs detected by ALMA, which strengthen the evidence for a planetary origin of the cavities and rules out the photoevaporation mechanism.  At least in the case of EM*~SR~24S, analysis of multi-frequency ALMA data has been used to also rule out an inner cavity produced by dead zones.  The other 2 objects with large dust cavities (ISO-Oph~2A  and RXJ1633.9-2442) have outer discs with ring-like morphologies and lack ALMA-detected inner dust discs. These systems are consistent with more "evolved"  planet-induced cavities in which the dust in the inner disc has already drifted towards the star, and  most of the dust in the outer disc has accumulated in narrow rings (i.e., the dust is trapped in a pressure bump). 
Three additional discs (ISO-Oph 2B; ISO-Oph 54, and ISO-Oph 196) have much smaller  (r $\sim$ 2.2-8.5 au)  dust cavities.
\\

\noindent 4) We combine our ODISEA long-baseline sample with the brightest Ophiuchus objects observed by DSHARP to create a flux-limited sample (F$_{1.3mm} >$ 70 mJy) and construct a putative evolutionary sequence in which the substructures observed in massive protoplanetary discs are mostly the result of planet formation and dust evolution. 
The sequence starts with discs without detectable gaps, but these start to appear with the formation of planets massive enough to carve them. Eventually, some of the planets reach the critical mass to undergo runaway gas accretion and become gas giants. These gas giants produce wider gaps and pressure bumps in the outer disc. The pressure bumps efficiently filter the mm-sized dust grain and the  dusty inner discs start to dissipate. Radial drift in the outer disc also accumulate the dust near the edge of the cavity. The process continues until the inner dust discs dissipate in the absence of an inner trap that stop the radial drift of the particles, and all the dust in the outer disc is accumulated in a narrow ring.  \\

\noindent 5) Given the age of the sample ($\lesssim$1-3 Myr), the scenario outlined above requires planet formation to be extremely fast and efficient at many tens of au distances. Once the cores of giant planets are formed, the run away gas accretion process is fast and relatively well understood.  How planet cores massive enough to open detectable gaps at distances as large as $\sim$100 au in the timescales involved remains an open question. However, observational evidence is accumulating in this direction from several independent lines of research (dust continuum imaging, gas kinematics, and direct detection in the IR).  Independently of the details of the planet formation process, dedicated studies of substructure in discs with ALMA might provide an avenue to investigate the demographics of recently-formed ice and gas giants at $\sim$5-100 au distances from the star.

\section*{Acknowledgements}
We thank the anonymous referee for his/her constructive comments and suggestions.
This paper makes use of the following ALMA data: ADS/JAO.ALMA  2018.1.00028.S and 2016.1.00545.S.
ALMA is a partnership of ESO (representing its member states), NSF (USA) and NINS (Japan), together with NRC (Canada), MOST and ASIAA (Taiwan), and KASI (Republic of Korea), in cooperation with the Republic of Chile. The Joint ALMA Observatory is operated by ESO, AUI/NRAO and NAOJ.  The National Radio Astronomy Observatory is a facility of the National Science Foundation operated under cooperative agreement by Associated Universities, Inc. 
This work has made use of data from the European Space Agency (ESA) mission
{\it Gaia} (\url{https://www.cosmos.esa.int/gaia}), processed by the {\it Gaia}
Data Processing and Analysis Consortium (DPAC,
\url{https://www.cosmos.esa.int/web/gaia/dpac/consortium}). Funding for the DPAC
has been provided by national institutions, in particular the institutions
participating in the {\it Gaia} Multilateral Agreement.
 J.P.W. acknowledges support from NSF grant AST-1907486. P.P. acknowledges support provided by the Alexander von Humboldt Foundation in the framework of the Sofja Kovalevskaja Award endowed by the Federal Ministry of Education and Research. 
D.J.P. acknowledges Australian Research Council funding via grants DP180104235 and FT130100034. M.F. acknowledges support by the \emph{European Research Council (ERC)} project under the European Union's Horizon 2020 research and innovation program number 757957.
 \\

\section*{Data availability}
The data underlying this article are available in the ALMA archive at 
\url{https://almascience.eso.org/asax/} under project codes  2018.1.00028.S and 2016.1.00545.S.

%




\bibliographystyle{mnras}
\bibliography{ms} 

\begin{thebibliography}{}
\makeatletter
\relax
\def\mn@urlcharsother{\let\do\@makeother \do\$\do\&\do\#\do\^\do\_\do\%\do\~}
\def\mn@doi{\begingroup\mn@urlcharsother \@ifnextchar [ {\mn@doi@}
  {\mn@doi@[]}}
\def\mn@doi@[#1]#2{\def\@tempa{#1}\ifx\@tempa\@empty \href
  {http://dx.doi.org/#2} {doi:#2}\else \href {http://dx.doi.org/#2} {#1}\fi
  \endgroup}
\def\mn@eprint#1#2{\mn@eprint@#1:#2::\@nil}
\def\mn@eprint@arXiv#1{\href {http://arxiv.org/abs/#1} {{\tt arXiv:#1}}}
\def\mn@eprint@dblp#1{\href {http://dblp.uni-trier.de/rec/bibtex/#1.xml}
  {dblp:#1}}
\def\mn@eprint@#1:#2:#3:#4\@nil{\def\@tempa {#1}\def\@tempb {#2}\def\@tempc
  {#3}\ifx \@tempc \@empty \let \@tempc \@tempb \let \@tempb \@tempa \fi \ifx
  \@tempb \@empty \def\@tempb {arXiv}\fi \@ifundefined
  {mn@eprint@\@tempb}{\@tempb:\@tempc}{\expandafter \expandafter \csname
  mn@eprint@\@tempb\endcsname \expandafter{\@tempc}}}

\bibitem[\protect\citeauthoryear{{ALMA Partnership} et~al.,}{{ALMA Partnership}
  et~al.}{2015}]{alma2015}
{ALMA Partnership} et~al., 2015, \mn@doi [\apjl] {10.1088/2041-8205/808/1/L3},
  \href {https://ui.adsabs.harvard.edu/abs/2015ApJ...808L...3A} {808, L3}

\bibitem[\protect\citeauthoryear{{Alexander}, {Pascucci}, {Andrews}, {Armitage}
   \& {Cieza}}{{Alexander} et~al.}{2014}]{alexander2014}
{Alexander} R.,  {Pascucci} I.,  {Andrews} S.,  {Armitage} P.,   {Cieza} L.,
  2014, in {Beuther} H.,  {Klessen} R.~S.,  {Dullemond} C.~P.,   {Henning} T.,
  eds, Protostars and Planets VI. p.~475 (\mn@eprint {arXiv} {1311.1819}),
  \mn@doi{10.2458/azu_uapress_9780816531240-ch021}

\bibitem[\protect\citeauthoryear{{Allard}, {Homeier}  \& {Freytag}}{{Allard}
  et~al.}{2012}]{allard2012}
{Allard} F.,  {Homeier} D.,   {Freytag} B.,  2012, \mn@doi [Philosophical
  Transactions of the Royal Society of London Series A]
  {10.1098/rsta.2011.0269}, \href
  {https://ui.adsabs.harvard.edu/abs/2012RSPTA.370.2765A} {370, 2765}

\bibitem[\protect\citeauthoryear{{Andrews}}{{Andrews}}{2020}]{andrews2020}
{Andrews} S.~M.,  2020, arXiv e-prints, \href
  {https://ui.adsabs.harvard.edu/abs/2020arXiv200105007A} {p. arXiv:2001.05007}

\bibitem[\protect\citeauthoryear{{Andrews} \& {Williams}}{{Andrews} \&
  {Williams}}{2005}]{andrewswilliams2005}
{Andrews} S.~M.,  {Williams} J.~P.,  2005, \mn@doi [\apjl] {10.1086/427325},
  \href {https://ui.adsabs.harvard.edu/abs/2005ApJ...619L.175A} {619, L175}

\bibitem[\protect\citeauthoryear{{Andrews} et~al.,}{{Andrews}
  et~al.}{2016}]{andrews2016}
{Andrews} S.~M.,  et~al., 2016, \mn@doi [\apjl] {10.3847/2041-8205/820/2/L40},
  \href {https://ui.adsabs.harvard.edu/abs/2016ApJ...820L..40A} {820, L40}

\bibitem[\protect\citeauthoryear{{Andrews} et~al.,}{{Andrews}
  et~al.}{2018}]{Andrews2018}
{Andrews} S.~M.,  et~al., 2018, \mn@doi [\apjl] {10.3847/2041-8213/aaf741},
  \href {https://ui.adsabs.harvard.edu/abs/2018ApJ...869L..41A} {869, L41}

\bibitem[\protect\citeauthoryear{{Baraffe}, {Chabrier}, {Allard}  \&
  {Hauschildt}}{{Baraffe} et~al.}{2002}]{baraffe2002}
{Baraffe} I.,  {Chabrier} G.,  {Allard} F.,   {Hauschildt} P.~H.,  2002,
  \mn@doi [\aap] {10.1051/0004-6361:20011638}, \href
  {https://ui.adsabs.harvard.edu/abs/2002A&A...382..563B} {382, 563}

\bibitem[\protect\citeauthoryear{{Batalha} et~al.,}{{Batalha}
  et~al.}{2013}]{batalha2013}
{Batalha} N.~M.,  et~al., 2013, \mn@doi [\apjs] {10.1088/0067-0049/204/2/24},
  \href {https://ui.adsabs.harvard.edu/abs/2013ApJS..204...24B} {204, 24}

\bibitem[\protect\citeauthoryear{{Beckwith}, {Sargent}, {Chini}  \&
  {Guesten}}{{Beckwith} et~al.}{1990}]{1990AJ.....99..924B}
{Beckwith} S. V.~W.,  {Sargent} A.~I.,  {Chini} R.~S.,   {Guesten} R.,  1990,
  \mn@doi [\aj] {10.1086/115385}, \href
  {https://ui.adsabs.harvard.edu/abs/1990AJ.....99..924B} {99, 924}

\bibitem[\protect\citeauthoryear{{Benz}, {Ida}, {Alibert}, {Lin}  \&
  {Mordasini}}{{Benz} et~al.}{2014}]{benz2014}
{Benz} W.,  {Ida} S.,  {Alibert} Y.,  {Lin} D.,   {Mordasini} C.,  2014, in
  {Beuther} H.,  {Klessen} R.~S.,  {Dullemond} C.~P.,   {Henning} T.,  eds,
  Protostars and Planets VI. p.~691 (\mn@eprint {arXiv} {1402.7086}),
  \mn@doi{10.2458/azu_uapress_9780816531240-ch030}

\bibitem[\protect\citeauthoryear{{Birnstiel} et~al.,}{{Birnstiel}
  et~al.}{2018}]{2018ApJ...869L..45B}
{Birnstiel} T.,  et~al., 2018, \mn@doi [\apjl] {10.3847/2041-8213/aaf743},
  \href {https://ui.adsabs.harvard.edu/abs/2018ApJ...869L..45B} {869, L45}

\bibitem[\protect\citeauthoryear{{Brauer}, {Dullemond}, {Johansen}, {Henning},
  {Klahr}  \& {Natta}}{{Brauer} et~al.}{2007}]{brauer2007}
{Brauer} F.,  {Dullemond} C.~P.,  {Johansen} A.,  {Henning} T.,  {Klahr} H.,
  {Natta} A.,  2007, \mn@doi [\aap] {10.1051/0004-6361:20066865}, \href
  {https://ui.adsabs.harvard.edu/abs/2007A&A...469.1169B} {469, 1169}

\bibitem[\protect\citeauthoryear{{Calcino}, {Price}, {Pinte}, {van der Marel},
  {Ragusa}, {Dipierro}, {Cuello}  \& {Christiaens}}{{Calcino}
  et~al.}{2019}]{Calcino2019}
{Calcino} J.,  {Price} D.~J.,  {Pinte} C.,  {van der Marel} N.,  {Ragusa} E.,
  {Dipierro} G.,  {Cuello} N.,   {Christiaens} V.,  2019, \mn@doi [\mnras]
  {10.1093/mnras/stz2770}, \href
  {https://ui.adsabs.harvard.edu/abs/2019MNRAS.490.2579C} {490, 2579}

\bibitem[\protect\citeauthoryear{{Calcino}, {Christiaens}, {Price}, {Pinte},
  {Davis}, {van der Marel}  \& {Cuello}}{{Calcino} et~al.}{2020}]{Calcino2020}
{Calcino} J.,  {Christiaens} V.,  {Price} D.~J.,  {Pinte} C.,  {Davis} T.~M.,
  {van der Marel} N.,   {Cuello} N.,  2020, \mn@doi [\mnras]
  {10.1093/mnras/staa2468}, \href
  {https://ui.adsabs.harvard.edu/abs/2020MNRAS.498..639C} {498, 639}

\bibitem[\protect\citeauthoryear{{C{\'a}novas} et~al.,}{{C{\'a}novas}
  et~al.}{2019}]{canovas2019}
{C{\'a}novas} H.,  et~al., 2019, \mn@doi [\aap] {10.1051/0004-6361/201935321},
  \href {https://ui.adsabs.harvard.edu/abs/2019A&A...626A..80C} {626, A80}

\bibitem[\protect\citeauthoryear{{Carrera}, {Gorti}, {Johansen}  \&
  {Davies}}{{Carrera} et~al.}{2017}]{Carrera2017}
{Carrera} D.,  {Gorti} U.,  {Johansen} A.,   {Davies} M.~B.,  2017, \mn@doi
  [\apj] {10.3847/1538-4357/aa6932}, \href
  {https://ui.adsabs.harvard.edu/abs/2017ApJ...839...16C} {839, 16}

\bibitem[\protect\citeauthoryear{{Casassus} \& {P{\'e}rez}}{{Casassus} \&
  {P{\'e}rez}}{2019}]{casassusperez2019}
{Casassus} S.,  {P{\'e}rez} S.,  2019, \mn@doi [\apjl]
  {10.3847/2041-8213/ab4425}, \href
  {https://ui.adsabs.harvard.edu/abs/2019ApJ...883L..41C} {883, L41}

\bibitem[\protect\citeauthoryear{{Casassus} et~al.,}{{Casassus}
  et~al.}{2008}]{casassus2008}
{Casassus} S.,  et~al., 2008, \mn@doi [\mnras]
  {10.1111/j.1365-2966.2008.13954.x}, \href
  {https://ui.adsabs.harvard.edu/abs/2008MNRAS.391.1075C} {391, 1075}

\bibitem[\protect\citeauthoryear{{Casassus} et~al.,}{{Casassus}
  et~al.}{2018}]{casassus2018}
{Casassus} S.,  et~al., 2018, \mn@doi [\mnras] {10.1093/mnras/sty894}, \href
  {https://ui.adsabs.harvard.edu/abs/2018MNRAS.477.5104C} {477, 5104}

\bibitem[\protect\citeauthoryear{{Chambers} et~al.,}{{Chambers}
  et~al.}{2016}]{chambers2016}
{Chambers} K.~C.,  et~al., 2016, arXiv e-prints, \href
  {https://ui.adsabs.harvard.edu/abs/2016arXiv161205560C} {p. arXiv:1612.05560}

\bibitem[\protect\citeauthoryear{{Cieza} et~al.,}{{Cieza}
  et~al.}{2007}]{cieza2007}
{Cieza} L.,  et~al., 2007, \mn@doi [\apj] {10.1086/520698}, \href
  {https://ui.adsabs.harvard.edu/abs/2007ApJ...667..308C} {667, 308}

\bibitem[\protect\citeauthoryear{{Cieza}, {Swift}, {Mathews}  \&
  {Williams}}{{Cieza} et~al.}{2008}]{cieza2008}
{Cieza} L.~A.,  {Swift} J.~J.,  {Mathews} G.~S.,   {Williams} J.~P.,  2008,
  \mn@doi [\apjl] {10.1086/592965}, \href
  {https://ui.adsabs.harvard.edu/abs/2008ApJ...686L.115C} {686, L115}

\bibitem[\protect\citeauthoryear{{Cieza} et~al.,}{{Cieza}
  et~al.}{2009}]{cieza2009}
{Cieza} L.~A.,  et~al., 2009, \mn@doi [\apjl] {10.1088/0004-637X/696/1/L84},
  \href {https://ui.adsabs.harvard.edu/abs/2009ApJ...696L..84C} {696, L84}

\bibitem[\protect\citeauthoryear{{Cieza} et~al.,}{{Cieza}
  et~al.}{2010}]{cieza2010}
{Cieza} L.~A.,  et~al., 2010, \mn@doi [\apj] {10.1088/0004-637X/712/2/925},
  \href {https://ui.adsabs.harvard.edu/abs/2010ApJ...712..925C} {712, 925}

\bibitem[\protect\citeauthoryear{{Cieza} et~al.,}{{Cieza}
  et~al.}{2012}]{cieza2012}
{Cieza} L.~A.,  et~al., 2012, \mn@doi [\apj] {10.1088/0004-637X/752/1/75},
  \href {https://ui.adsabs.harvard.edu/abs/2012ApJ...752...75C} {752, 75}

\bibitem[\protect\citeauthoryear{{Cieza} et~al.,}{{Cieza}
  et~al.}{2016}]{cieza2016}
{Cieza} L.~A.,  et~al., 2016, \mn@doi [\nat] {10.1038/nature18612}, \href
  {https://ui.adsabs.harvard.edu/abs/2016Natur.535..258C} {535, 258}

\bibitem[\protect\citeauthoryear{{Cieza} et~al.,}{{Cieza}
  et~al.}{2017}]{cieza2017}
{Cieza} L.~A.,  et~al., 2017, \mn@doi [\apjl] {10.3847/2041-8213/aa9b7b}, \href
  {https://ui.adsabs.harvard.edu/abs/2017ApJ...851L..23C} {851, L23}

\bibitem[\protect\citeauthoryear{{Cieza} et~al.,}{{Cieza}
  et~al.}{2019}]{cieza2019}
{Cieza} L.~A.,  et~al., 2019, \mn@doi [\mnras] {10.1093/mnras/sty2653}, \href
  {https://ui.adsabs.harvard.edu/abs/2019MNRAS.482..698C} {482, 698}

\bibitem[\protect\citeauthoryear{{Cox} et~al.,}{{Cox} et~al.}{2017}]{cox2017}
{Cox} E.~G.,  et~al., 2017, \mn@doi [\apj] {10.3847/1538-4357/aa97e2}, \href
  {https://ui.adsabs.harvard.edu/abs/2017ApJ...851...83C} {851, 83}

\bibitem[\protect\citeauthoryear{{Crida}, {Morbidelli}  \& {Masset}}{{Crida}
  et~al.}{2006}]{crida2006}
{Crida} A.,  {Morbidelli} A.,   {Masset} F.,  2006, \mn@doi [\icarus]
  {10.1016/j.icarus.2005.10.007}, \href
  {https://ui.adsabs.harvard.edu/abs/2006Icar..181..587C} {181, 587}

\bibitem[\protect\citeauthoryear{{Currie}, {Lada}, {Plavchan}, {Robitaille},
  {Irwin}  \& {Kenyon}}{{Currie} et~al.}{2009}]{currie2009}
{Currie} T.,  {Lada} C.~J.,  {Plavchan} P.,  {Robitaille} T.~P.,  {Irwin} J.,
  {Kenyon} S.~J.,  2009, \mn@doi [\apj] {10.1088/0004-637X/698/1/1}, \href
  {https://ui.adsabs.harvard.edu/abs/2009ApJ...698....1C} {698, 1}

\bibitem[\protect\citeauthoryear{{Cutri} et~al.,}{{Cutri}
  et~al.}{2003}]{cutri2003}
{Cutri} R.~M.,  et~al., 2003, {2MASS All Sky Catalog of point sources.}

\bibitem[\protect\citeauthoryear{{Daley} et~al.,}{{Daley}
  et~al.}{2019}]{Daley2019}
{Daley} C.,  et~al., 2019, \mn@doi [\apj] {10.3847/1538-4357/ab1074}, \href
  {https://ui.adsabs.harvard.edu/abs/2019ApJ...875...87D} {875, 87}

\bibitem[\protect\citeauthoryear{{Dent} et~al.,}{{Dent}
  et~al.}{2014}]{Dent2014}
{Dent} W.~R.~F.,  et~al., 2014, \mn@doi [Science] {10.1126/science.1248726},
  \href {http://adsabs.harvard.edu/abs/2014Sci...343.1490D} {343, 1490}

\bibitem[\protect\citeauthoryear{{Dipierro} \& {Laibe}}{{Dipierro} \&
  {Laibe}}{2017}]{Dipierro2017}
{Dipierro} G.,  {Laibe} G.,  2017, \mn@doi [\mnras] {10.1093/mnras/stx977},
  \href {https://ui.adsabs.harvard.edu/abs/2017MNRAS.469.1932D} {469, 1932}

\bibitem[\protect\citeauthoryear{{Dipierro}, {Price}, {Laibe}, {Hirsh},
  {Cerioli}  \& {Lodato}}{{Dipierro} et~al.}{2015}]{Dipierro2015}
{Dipierro} G.,  {Price} D.,  {Laibe} G.,  {Hirsh} K.,  {Cerioli} A.,   {Lodato}
  G.,  2015, \mn@doi [\mnras] {10.1093/mnrasl/slv105}, \href
  {https://ui.adsabs.harvard.edu/abs/2015MNRAS.453L..73D} {453, L73}

\bibitem[\protect\citeauthoryear{{Dipierro}, {Laibe}, {Price}  \&
  {Lodato}}{{Dipierro} et~al.}{2016}]{Dipierro2016}
{Dipierro} G.,  {Laibe} G.,  {Price} D.~J.,   {Lodato} G.,  2016, \mn@doi
  [\mnras] {10.1093/mnrasl/slw032}, \href
  {https://ui.adsabs.harvard.edu/abs/2016MNRAS.459L...1D} {459, L1}

\bibitem[\protect\citeauthoryear{{Dong} \& {Fung}}{{Dong} \&
  {Fung}}{2017}]{dongfung2017}
{Dong} R.,  {Fung} J.,  2017, \mn@doi [\apj] {10.3847/1538-4357/835/2/146},
  \href {https://ui.adsabs.harvard.edu/abs/2017ApJ...835..146D} {835, 146}

\bibitem[\protect\citeauthoryear{{Dong}, {Zhu}  \& {Whitney}}{{Dong}
  et~al.}{2015}]{Dong2015}
{Dong} R.,  {Zhu} Z.,   {Whitney} B.,  2015, \mn@doi [\apj]
  {10.1088/0004-637X/809/1/93}, \href
  {https://ui.adsabs.harvard.edu/abs/2015ApJ...809...93D} {809, 93}

\bibitem[\protect\citeauthoryear{{Dong}, {Li}, {Chiang}  \& {Li}}{{Dong}
  et~al.}{2017}]{Dong2017}
{Dong} R.,  {Li} S.,  {Chiang} E.,   {Li} H.,  2017, \mn@doi [\apj]
  {10.3847/1538-4357/aa72f2}, \href
  {https://ui.adsabs.harvard.edu/abs/2017ApJ...843..127D} {843, 127}

\bibitem[\protect\citeauthoryear{{Dong}, {Li}, {Chiang}  \& {Li}}{{Dong}
  et~al.}{2018}]{Dong2018}
{Dong} R.,  {Li} S.,  {Chiang} E.,   {Li} H.,  2018, \mn@doi [\apj]
  {10.3847/1538-4357/aadadd}, \href
  {https://ui.adsabs.harvard.edu/abs/2018ApJ...866..110D} {866, 110}

\bibitem[\protect\citeauthoryear{{Doppmann}, {Greene}, {Covey}  \&
  {Lada}}{{Doppmann} et~al.}{2005}]{doppmann2005}
{Doppmann} G.~W.,  {Greene} T.~P.,  {Covey} K.~R.,   {Lada} C.~J.,  2005,
  \mn@doi [\aj] {10.1086/431954}, \href
  {https://ui.adsabs.harvard.edu/abs/2005AJ....130.1145D} {130, 1145}

\bibitem[\protect\citeauthoryear{{Duffell} \& {MacFadyen}}{{Duffell} \&
  {MacFadyen}}{2013}]{duffellmacfayden2013}
{Duffell} P.~C.,  {MacFadyen} A.~I.,  2013, \mn@doi [\apj]
  {10.1088/0004-637X/769/1/41}, \href
  {https://ui.adsabs.harvard.edu/abs/2013ApJ...769...41D} {769, 41}

\bibitem[\protect\citeauthoryear{{Dullemond} \& {Penzlin}}{{Dullemond} \&
  {Penzlin}}{2018}]{dullemondpenzlin2018}
{Dullemond} C.~P.,  {Penzlin} A.~B.~T.,  2018, \mn@doi [\aap]
  {10.1051/0004-6361/201731878}, \href
  {https://ui.adsabs.harvard.edu/abs/2018A&A...609A..50D} {609, A50}

\bibitem[\protect\citeauthoryear{{Enoch}, {Evans}, {Sargent}  \&
  {Glenn}}{{Enoch} et~al.}{2009}]{enoch2009}
{Enoch} M.~L.,  {Evans} Neal~J. I.,  {Sargent} A.~I.,   {Glenn} J.,  2009,
  \mn@doi [\apj] {10.1088/0004-637X/692/2/973}, \href
  {https://ui.adsabs.harvard.edu/abs/2009ApJ...692..973E} {692, 973}

\bibitem[\protect\citeauthoryear{{Espaillat} et~al.,}{{Espaillat}
  et~al.}{2014}]{espaillat2014}
{Espaillat} C.,  et~al., 2014, in {Beuther} H.,  {Klessen} R.~S.,  {Dullemond}
  C.~P.,   {Henning} T.,  eds, Protostars and Planets VI. p.~497 (\mn@eprint
  {arXiv} {1402.7103}), \mn@doi{10.2458/azu_uapress_9780816531240-ch022}

\bibitem[\protect\citeauthoryear{{Evans} Neal~J. et~al.,}{{Evans}
  et~al.}{2009}]{evans2009}
{Evans} Neal~J. I.,  et~al., 2009, \mn@doi [\apjs]
  {10.1088/0067-0049/181/2/321}, \href
  {https://ui.adsabs.harvard.edu/abs/2009ApJS..181..321E} {181, 321}

\bibitem[\protect\citeauthoryear{{Fitzpatrick}}{{Fitzpatrick}}{1999}]{fitzpatrick1999}
{Fitzpatrick} E.~L.,  1999, \mn@doi [\pasp] {10.1086/316293}, \href
  {https://ui.adsabs.harvard.edu/abs/1999PASP..111...63F} {111, 63}

\bibitem[\protect\citeauthoryear{{Flock}, {Henning}  \& {Klahr}}{{Flock}
  et~al.}{2012}]{flock2012}
{Flock} M.,  {Henning} T.,   {Klahr} H.,  2012, \mn@doi [\apj]
  {10.1088/0004-637X/761/2/95}, \href
  {https://ui.adsabs.harvard.edu/abs/2012ApJ...761...95F} {761, 95}

\bibitem[\protect\citeauthoryear{{Flock}, {Ruge}, {Dzyurkevich}, {Henning},
  {Klahr}  \& {Wolf}}{{Flock} et~al.}{2015}]{flock2015}
{Flock} M.,  {Ruge} J.~P.,  {Dzyurkevich} N.,  {Henning} T.,  {Klahr} H.,
  {Wolf} S.,  2015, \mn@doi [\aap] {10.1051/0004-6361/201424693}, \href
  {https://ui.adsabs.harvard.edu/abs/2015A&A...574A..68F} {574, A68}

\bibitem[\protect\citeauthoryear{{Francis} \& {van der Marel}}{{Francis} \&
  {van der Marel}}{2020}]{francisvandermarel2020}
{Francis} L.,  {van der Marel} N.,  2020, \mn@doi [\apj]
  {10.3847/1538-4357/ab7b63}, \href
  {https://ui.adsabs.harvard.edu/abs/2020ApJ...892..111F} {892, 111}

\bibitem[\protect\citeauthoryear{{Gaia Collaboration} et~al.,}{{Gaia
  Collaboration} et~al.}{2018}]{GaiaDR2}
{Gaia Collaboration} et~al., 2018, \mn@doi [\aap]
  {10.1051/0004-6361/201833051}, \href
  {https://ui.adsabs.harvard.edu/abs/2018A&A...616A...1G} {616, A1}

\bibitem[\protect\citeauthoryear{{Gaidos}, {Mann}, {Kraus}  \&
  {Ireland}}{{Gaidos} et~al.}{2016}]{gaidos2016}
{Gaidos} E.,  {Mann} A.~W.,  {Kraus} A.~L.,   {Ireland} M.,  2016, \mn@doi
  [\mnras] {10.1093/mnras/stw097}, \href
  {https://ui.adsabs.harvard.edu/abs/2016MNRAS.457.2877G} {457, 2877}

\bibitem[\protect\citeauthoryear{{Gatti}, {Testi}, {Natta}, {Randich}  \&
  {Muzerolle}}{{Gatti} et~al.}{2006}]{gatti2006}
{Gatti} T.,  {Testi} L.,  {Natta} A.,  {Randich} S.,   {Muzerolle} J.,  2006,
  \mn@doi [\aap] {10.1051/0004-6361:20066095}, \href
  {https://ui.adsabs.harvard.edu/abs/2006A&A...460..547G} {460, 547}

\bibitem[\protect\citeauthoryear{{Gonz{\'a}lez-Ruilova}
  et~al.,}{{Gonz{\'a}lez-Ruilova} et~al.}{2020}]{2020ApJ...902L..33G}
{Gonz{\'a}lez-Ruilova} C.,  et~al., 2020, \mn@doi [\apjl]
  {10.3847/2041-8213/abbcce}, \href
  {https://ui.adsabs.harvard.edu/abs/2020ApJ...902L..33G} {902, L33}

\bibitem[\protect\citeauthoryear{{Gonzalez}, {Laibe}, {Maddison}, {Pinte}  \&
  {M{\'e}nard}}{{Gonzalez} et~al.}{2015}]{Gonzalez2015}
{Gonzalez} J.~F.,  {Laibe} G.,  {Maddison} S.~T.,  {Pinte} C.,   {M{\'e}nard}
  F.,  2015, \mn@doi [\mnras] {10.1093/mnrasl/slv120}, \href
  {https://ui.adsabs.harvard.edu/abs/2015MNRAS.454L..36G} {454, L36}

\bibitem[\protect\citeauthoryear{{Haffert}, {Bohn}, {de Boer}, {Snellen},
  {Brinchmann}, {Girard}, {Keller}  \& {Bacon}}{{Haffert}
  et~al.}{2019}]{haffert2019}
{Haffert} S.~Y.,  {Bohn} A.~J.,  {de Boer} J.,  {Snellen} I.~A.~G.,
  {Brinchmann} J.,  {Girard} J.~H.,  {Keller} C.~U.,   {Bacon} R.,  2019,
  \mn@doi [Nature Astronomy] {10.1038/s41550-019-0780-5}, \href
  {https://ui.adsabs.harvard.edu/abs/2019NatAs...3..749H} {3, 749}

\bibitem[\protect\citeauthoryear{{Hardy} et~al.,}{{Hardy}
  et~al.}{2015}]{hardy2015}
{Hardy} A.,  et~al., 2015, \mn@doi [\aap] {10.1051/0004-6361/201526504}, \href
  {https://ui.adsabs.harvard.edu/abs/2015A&A...583A..66H} {583, A66}

\bibitem[\protect\citeauthoryear{{Helled} et~al.,}{{Helled}
  et~al.}{2014}]{helled2014}
{Helled} R.,  et~al., 2014, in {Beuther} H.,  {Klessen} R.~S.,  {Dullemond}
  C.~P.,   {Henning} T.,  eds, Protostars and Planets VI. p.~643 (\mn@eprint
  {arXiv} {1311.1142}), \mn@doi{10.2458/azu_uapress_9780816531240-ch028}

\bibitem[\protect\citeauthoryear{{Huang} et~al.,}{{Huang}
  et~al.}{2018}]{huang2018}
{Huang} J.,  et~al., 2018, \mn@doi [\apjl] {10.3847/2041-8213/aaf740}, \href
  {https://ui.adsabs.harvard.edu/abs/2018ApJ...869L..42H} {869, L42}

\bibitem[\protect\citeauthoryear{{Hughes}, {Wilner}, {Calvet}, {D'Alessio},
  {Claussen}  \& {Hogerheijde}}{{Hughes} et~al.}{2007}]{hughes2007}
{Hughes} A.~M.,  {Wilner} D.~J.,  {Calvet} N.,  {D'Alessio} P.,  {Claussen}
  M.~J.,   {Hogerheijde} M.~R.,  2007, \mn@doi [\apj] {10.1086/518885}, \href
  {https://ui.adsabs.harvard.edu/abs/2007ApJ...664..536H} {664, 536}

\bibitem[\protect\citeauthoryear{{Indebetouw} et~al.,}{{Indebetouw}
  et~al.}{2005}]{indebetouw2005}
{Indebetouw} R.,  et~al., 2005, \mn@doi [\apj] {10.1086/426679}, \href
  {https://ui.adsabs.harvard.edu/abs/2005ApJ...619..931I} {619, 931}

\bibitem[\protect\citeauthoryear{{Ireland} \& {Kraus}}{{Ireland} \&
  {Kraus}}{2008}]{irelandkraus2008}
{Ireland} M.~J.,  {Kraus} A.~L.,  2008, \mn@doi [\apjl] {10.1086/588216}, \href
  {https://ui.adsabs.harvard.edu/abs/2008ApJ...678L..59I} {678, L59}

\bibitem[\protect\citeauthoryear{{Isella}, {Benisty}, {Teague}, {Bae},
  {Keppler}, {Facchini}  \& {P{\'e}rez}}{{Isella} et~al.}{2019}]{isella2019}
{Isella} A.,  {Benisty} M.,  {Teague} R.,  {Bae} J.,  {Keppler} M.,  {Facchini}
  S.,   {P{\'e}rez} L.,  2019, \mn@doi [\apjl] {10.3847/2041-8213/ab2a12},
  \href {https://ui.adsabs.harvard.edu/abs/2019ApJ...879L..25I} {879, L25}

\bibitem[\protect\citeauthoryear{{Jennings}, {Booth}, {Tazzari}, {Rosotti}  \&
  {Clarke}}{{Jennings} et~al.}{2020}]{jennings2020}
{Jennings} J.,  {Booth} R.~A.,  {Tazzari} M.,  {Rosotti} G.~P.,   {Clarke}
  C.~J.,  2020, \mn@doi [\mnras] {10.1093/mnras/staa1365}, \href
  {https://ui.adsabs.harvard.edu/abs/2020MNRAS.495.3209J} {495, 3209}

\bibitem[\protect\citeauthoryear{{Keppler} et~al.,}{{Keppler}
  et~al.}{2018}]{keppler2018}
{Keppler} M.,  et~al., 2018, \mn@doi [\aap] {10.1051/0004-6361/201832957},
  \href {https://ui.adsabs.harvard.edu/abs/2018A&A...617A..44K} {617, A44}

\bibitem[\protect\citeauthoryear{{Kraus}, {Ireland}, {Hillenbrand}  \&
  {Martinache}}{{Kraus} et~al.}{2012}]{kraus2012}
{Kraus} A.~L.,  {Ireland} M.~J.,  {Hillenbrand} L.~A.,   {Martinache} F.,
  2012, \mn@doi [\apj] {10.1088/0004-637X/745/1/19}, \href
  {https://ui.adsabs.harvard.edu/abs/2012ApJ...745...19K} {745, 19}

\bibitem[\protect\citeauthoryear{{Krivov} \& {Wyatt}}{{Krivov} \&
  {Wyatt}}{2020}]{Krivov2020}
{Krivov} A.~V.,  {Wyatt} M.~C.,  2020, \mn@doi [\mnras]
  {10.1093/mnras/staa2385}, \href
  {https://ui.adsabs.harvard.edu/abs/2020MNRAS.tmp.2088K} {}

\bibitem[\protect\citeauthoryear{{Lacour} et~al.,}{{Lacour}
  et~al.}{2016}]{lacour2016}
{Lacour} S.,  et~al., 2016, \mn@doi [\aap] {10.1051/0004-6361/201527863}, \href
  {https://ui.adsabs.harvard.edu/abs/2016A&A...590A..90L} {590, A90}

\bibitem[\protect\citeauthoryear{{Lada}}{{Lada}}{1987}]{lada1987}
{Lada} C.~J.,  1987, in {Peimbert} M.,  {Jugaku} J.,  eds,  IAU Symposium Vol.
  115, Star Forming Regions. p.~1

\bibitem[\protect\citeauthoryear{{Lada} et~al.,}{{Lada}
  et~al.}{2006}]{lada2006}
{Lada} C.~J.,  et~al., 2006, \mn@doi [\aj] {10.1086/499808}, \href
  {https://ui.adsabs.harvard.edu/abs/2006AJ....131.1574L} {131, 1574}

\bibitem[\protect\citeauthoryear{{Lodato} et~al.,}{{Lodato}
  et~al.}{2019}]{Lodato2019}
{Lodato} G.,  et~al., 2019, \mn@doi [\mnras] {10.1093/mnras/stz913}, \href
  {https://ui.adsabs.harvard.edu/abs/2019MNRAS.486..453L} {486, 453}

\bibitem[\protect\citeauthoryear{{Long} et~al.,}{{Long}
  et~al.}{2018}]{long2018}
{Long} F.,  et~al., 2018, \mn@doi [\apj] {10.3847/1538-4357/aae8e1}, \href
  {https://ui.adsabs.harvard.edu/abs/2018ApJ...869...17L} {869, 17}

\bibitem[\protect\citeauthoryear{{Lor{\'e}n-Aguilar} \&
  {Bate}}{{Lor{\'e}n-Aguilar} \& {Bate}}{2015}]{loren-Aguilar2015}
{Lor{\'e}n-Aguilar} P.,  {Bate} M.~R.,  2015, \mn@doi [\mnras]
  {10.1093/mnrasl/slv109}, \href
  {https://ui.adsabs.harvard.edu/abs/2015MNRAS.453L..78L} {453, L78}

\bibitem[\protect\citeauthoryear{{Lubow} \& {D'Angelo}}{{Lubow} \&
  {D'Angelo}}{2006}]{lubowdangelo2006}
{Lubow} S.~H.,  {D'Angelo} G.,  2006, \mn@doi [\apj] {10.1086/500356}, \href
  {https://ui.adsabs.harvard.edu/abs/2006ApJ...641..526L} {641, 526}

\bibitem[\protect\citeauthoryear{{Lubow}, {Seibert}  \& {Artymowicz}}{{Lubow}
  et~al.}{1999}]{lubow1999}
{Lubow} S.~H.,  {Seibert} M.,   {Artymowicz} P.,  1999, \mn@doi [\apj]
  {10.1086/308045}, \href
  {https://ui.adsabs.harvard.edu/abs/1999ApJ...526.1001L} {526, 1001}

\bibitem[\protect\citeauthoryear{{Lynds}}{{Lynds}}{1962}]{lynds1962}
{Lynds} B.~T.,  1962, \mn@doi [\apjs] {10.1086/190072}, \href
  {https://ui.adsabs.harvard.edu/abs/1962ApJS....7....1L} {7, 1}

\bibitem[\protect\citeauthoryear{{MacGregor} et~al.,}{{MacGregor}
  et~al.}{2013}]{MacGregor2013}
{MacGregor} M.~A.,  et~al., 2013, \mn@doi [\apjl]
  {10.1088/2041-8205/762/2/L21}, \href
  {http://adsabs.harvard.edu/abs/2013ApJ...762L..21M} {762, L21}

\bibitem[\protect\citeauthoryear{{MacGregor} et~al.,}{{MacGregor}
  et~al.}{2019}]{MacGregor2019}
{MacGregor} M.~A.,  et~al., 2019, \mn@doi [\apjl] {10.3847/2041-8213/ab21c2},
  \href {https://ui.adsabs.harvard.edu/abs/2019ApJ...877L..32M} {877, L32}

\bibitem[\protect\citeauthoryear{{Manara}, {Testi}, {Natta}, {Rosotti},
  {Benisty}, {Ercolano}  \& {Ricci}}{{Manara} et~al.}{2014}]{manara2014}
{Manara} C.~F.,  {Testi} L.,  {Natta} A.,  {Rosotti} G.,  {Benisty} M.,
  {Ercolano} B.,   {Ricci} L.,  2014, \mn@doi [\aap]
  {10.1051/0004-6361/201323318}, \href
  {https://ui.adsabs.harvard.edu/abs/2014A&A...568A..18M} {568, A18}

\bibitem[\protect\citeauthoryear{{Marino} et~al.,}{{Marino}
  et~al.}{2016}]{Marino2016}
{Marino} S.,  et~al., 2016, \mn@doi [\mnras] {10.1093/mnras/stw1216}, \href
  {http://adsabs.harvard.edu/abs/2016MNRAS.460.2933M} {460, 2933}

\bibitem[\protect\citeauthoryear{{Marino} et~al.,}{{Marino}
  et~al.}{2018}]{Marino2018}
{Marino} S.,  et~al., 2018, \mn@doi [\mnras] {10.1093/mnras/sty1790}, \href
  {http://adsabs.harvard.edu/abs/2018MNRAS.479.5423M} {479, 5423}

\bibitem[\protect\citeauthoryear{{Marino}, {Yelverton}, {Booth}, {Faramaz},
  {Kennedy}, {Matr{\`a}}  \& {Wyatt}}{{Marino} et~al.}{2019}]{Marino2019}
{Marino} S.,  {Yelverton} B.,  {Booth} M.,  {Faramaz} V.,  {Kennedy} G.~M.,
  {Matr{\`a}} L.,   {Wyatt} M.~C.,  2019, \mn@doi [\mnras]
  {10.1093/mnras/stz049}, \href
  {https://ui.adsabs.harvard.edu/abs/2019MNRAS.484.1257M} {484, 1257}

\bibitem[\protect\citeauthoryear{{Marino} et~al.,}{{Marino}
  et~al.}{2020}]{Marino2020}
{Marino} S.,  et~al., 2020, \mn@doi [\mnras] {10.1093/mnras/staa2386}, \href
  {https://ui.adsabs.harvard.edu/abs/2020MNRAS.498.1319M} {498, 1319}

\bibitem[\protect\citeauthoryear{{Marley}, {Fortney}, {Hubickyj}, {Bodenheimer}
   \& {Lissauer}}{{Marley} et~al.}{2007}]{marley2007}
{Marley} M.~S.,  {Fortney} J.~J.,  {Hubickyj} O.,  {Bodenheimer} P.,
  {Lissauer} J.~J.,  2007, \mn@doi [\apj] {10.1086/509759}, \href
  {https://ui.adsabs.harvard.edu/abs/2007ApJ...655..541M} {655, 541}

\bibitem[\protect\citeauthoryear{{Matr{\`a}}, {Marino}, {Kennedy}, {Wyatt},
  {{\"O}berg}  \& {Wilner}}{{Matr{\`a}} et~al.}{2018}]{Matra2018}
{Matr{\`a}} L.,  {Marino} S.,  {Kennedy} G.~M.,  {Wyatt} M.~C.,  {{\"O}berg}
  K.~I.,   {Wilner} D.~J.,  2018, \mn@doi [\apj] {10.3847/1538-4357/aabcc4},
  \href {http://adsabs.harvard.edu/abs/2018ApJ...859...72M} {859, 72}

\bibitem[\protect\citeauthoryear{{Mayama} et~al.,}{{Mayama}
  et~al.}{2010}]{mayama2010}
{Mayama} S.,  et~al., 2010, \mn@doi [Science] {10.1126/science.1179679}, \href
  {https://ui.adsabs.harvard.edu/abs/2010Sci...327..306M} {327, 306}

\bibitem[\protect\citeauthoryear{{McMullin}, {Waters}, {Schiebel}, {Young}  \&
  {Golap}}{{McMullin} et~al.}{2007}]{2007ASPC..376..127M}
{McMullin} J.~P.,  {Waters} B.,  {Schiebel} D.,  {Young} W.,   {Golap} K.,
  2007, in {Shaw} R.~A.,  {Hill} F.,   {Bell} D.~J.,  eds,  Astronomical
  Society of the Pacific Conference Series Vol. 376, Astronomical Data Analysis
  Software and Systems XVI. p.~127

\bibitem[\protect\citeauthoryear{{Mentiplay}, {Price}  \& {Pinte}}{{Mentiplay}
  et~al.}{2019}]{Mentiplay2019}
{Mentiplay} D.,  {Price} D.~J.,   {Pinte} C.,  2019, \mn@doi [\mnras]
  {10.1093/mnrasl/sly209}, \href
  {https://ui.adsabs.harvard.edu/abs/2019MNRAS.484L.130M} {484, L130}

\bibitem[\protect\citeauthoryear{{Miotello}, {van Dishoeck}, {Kama}  \&
  {Bruderer}}{{Miotello} et~al.}{2016}]{miotello2016}
{Miotello} A.,  {van Dishoeck} E.~F.,  {Kama} M.,   {Bruderer} S.,  2016,
  \mn@doi [\aap] {10.1051/0004-6361/201628159}, \href
  {https://ui.adsabs.harvard.edu/abs/2016A&A...594A..85M} {594, A85}

\bibitem[\protect\citeauthoryear{{Muzerolle}, {Allen}, {Megeath},
  {Hern{\'a}ndez}  \& {Gutermuth}}{{Muzerolle} et~al.}{2010}]{muzerolle2010}
{Muzerolle} J.,  {Allen} L.~E.,  {Megeath} S.~T.,  {Hern{\'a}ndez} J.,
  {Gutermuth} R.~A.,  2010, \mn@doi [\apj] {10.1088/0004-637X/708/2/1107},
  \href {https://ui.adsabs.harvard.edu/abs/2010ApJ...708.1107M} {708, 1107}

\bibitem[\protect\citeauthoryear{{Najita}, {Strom}  \& {Muzerolle}}{{Najita}
  et~al.}{2007}]{najita2007}
{Najita} J.~R.,  {Strom} S.~E.,   {Muzerolle} J.,  2007, \mn@doi [\mnras]
  {10.1111/j.1365-2966.2007.11793.x}, \href
  {https://ui.adsabs.harvard.edu/abs/2007MNRAS.378..369N} {378, 369}

\bibitem[\protect\citeauthoryear{{Natta}, {Testi}  \& {Randich}}{{Natta}
  et~al.}{2006}]{natta2006}
{Natta} A.,  {Testi} L.,   {Randich} S.,  2006, \mn@doi [\aap]
  {10.1051/0004-6361:20054706}, \href
  {https://ui.adsabs.harvard.edu/abs/2006A&A...452..245N} {452, 245}

\bibitem[\protect\citeauthoryear{{Ndugu}, {Bitsch}  \& {Jurua}}{{Ndugu}
  et~al.}{2019}]{Ndugu2019}
{Ndugu} N.,  {Bitsch} B.,   {Jurua} E.,  2019, \mn@doi [\mnras]
  {10.1093/mnras/stz1862}, \href
  {https://ui.adsabs.harvard.edu/abs/2019MNRAS.488.3625N} {488, 3625}

\bibitem[\protect\citeauthoryear{{Okuzumi}, {Momose}, {Sirono}, {Kobayashi}  \&
  {Tanaka}}{{Okuzumi} et~al.}{2016}]{okuzumi2016}
{Okuzumi} S.,  {Momose} M.,  {Sirono} S.-i.,  {Kobayashi} H.,   {Tanaka} H.,
  2016, \mn@doi [\apj] {10.3847/0004-637X/821/2/82}, \href
  {https://ui.adsabs.harvard.edu/abs/2016ApJ...821...82O} {821, 82}

\bibitem[\protect\citeauthoryear{{Owen} \& {Clarke}}{{Owen} \&
  {Clarke}}{2012}]{owenclarke2012}
{Owen} J.~E.,  {Clarke} C.~J.,  2012, \mn@doi [\mnras]
  {10.1111/j.1745-3933.2012.01334.x}, \href
  {https://ui.adsabs.harvard.edu/abs/2012MNRAS.426L..96O} {426, L96}

\bibitem[\protect\citeauthoryear{{Padgett} et~al.,}{{Padgett}
  et~al.}{2008}]{padgett2008}
{Padgett} D.~L.,  et~al., 2008, \mn@doi [\apj] {10.1086/523883}, \href
  {https://ui.adsabs.harvard.edu/abs/2008ApJ...672.1013P} {672, 1013}

\bibitem[\protect\citeauthoryear{{Pascucci} et~al.,}{{Pascucci}
  et~al.}{2016}]{pascucci2016}
{Pascucci} I.,  et~al., 2016, \mn@doi [\apj] {10.3847/0004-637X/831/2/125},
  \href {https://ui.adsabs.harvard.edu/abs/2016ApJ...831..125P} {831, 125}

\bibitem[\protect\citeauthoryear{{Pecaut} \& {Mamajek}}{{Pecaut} \&
  {Mamajek}}{2013}]{pecautmamajek2013}
{Pecaut} M.~J.,  {Mamajek} E.~E.,  2013, \mn@doi [\apjs]
  {10.1088/0067-0049/208/1/9}, \href
  {https://ui.adsabs.harvard.edu/abs/2013ApJS..208....9P} {208, 9}

\bibitem[\protect\citeauthoryear{{Perez}, {Dunhill}, {Casassus}, {Roman},
  {Szul{\'a}gyi}, {Flores}, {Marino}  \& {Montesinos}}{{Perez}
  et~al.}{2015}]{Perez2015}
{Perez} S.,  {Dunhill} A.,  {Casassus} S.,  {Roman} P.,  {Szul{\'a}gyi} J.,
  {Flores} C.,  {Marino} S.,   {Montesinos} M.,  2015, \mn@doi [\apjl]
  {10.1088/2041-8205/811/1/L5}, \href
  {https://ui.adsabs.harvard.edu/abs/2015ApJ...811L...5P} {811, L5}

\bibitem[\protect\citeauthoryear{{P{\'e}rez}, {Casassus}, {Baruteau}, {Dong},
  {Hales}  \& {Cieza}}{{P{\'e}rez} et~al.}{2019}]{perez2019}
{P{\'e}rez} S.,  {Casassus} S.,  {Baruteau} C.,  {Dong} R.,  {Hales} A.,
  {Cieza} L.,  2019, \mn@doi [\aj] {10.3847/1538-3881/ab1f88}, \href
  {https://ui.adsabs.harvard.edu/abs/2019AJ....158...15P} {158, 15}

\bibitem[\protect\citeauthoryear{{Pinilla}, {Benisty}  \&
  {Birnstiel}}{{Pinilla} et~al.}{2012}]{pinilla2012}
{Pinilla} P.,  {Benisty} M.,   {Birnstiel} T.,  2012, \mn@doi [\aap]
  {10.1051/0004-6361/201219315}, \href
  {https://ui.adsabs.harvard.edu/abs/2012A&A...545A..81P} {545, A81}

\bibitem[\protect\citeauthoryear{{Pinilla}, {de Juan Ovelar}, {Ataiee},
  {Benisty}, {Birnstiel}, {van Dishoeck}  \& {Min}}{{Pinilla}
  et~al.}{2015}]{pinilla2015}
{Pinilla} P.,  {de Juan Ovelar} M.,  {Ataiee} S.,  {Benisty} M.,  {Birnstiel}
  T.,  {van Dishoeck} E.~F.,   {Min} M.,  2015, \mn@doi [\aap]
  {10.1051/0004-6361/201424679}, \href
  {https://ui.adsabs.harvard.edu/abs/2015A&A...573A...9P} {573, A9}

\bibitem[\protect\citeauthoryear{{Pinilla}, {Klarmann}, {Birnstiel}, {Benisty},
  {Dominik}  \& {Dullemond}}{{Pinilla} et~al.}{2016}]{pinilla2016}
{Pinilla} P.,  {Klarmann} L.,  {Birnstiel} T.,  {Benisty} M.,  {Dominik} C.,
  {Dullemond} C.~P.,  2016, \mn@doi [\aap] {10.1051/0004-6361/201527131}, \href
  {https://ui.adsabs.harvard.edu/abs/2016A&A...585A..35P} {585, A35}

\bibitem[\protect\citeauthoryear{{Pinilla} et~al.,}{{Pinilla}
  et~al.}{2017}]{pinilla2017}
{Pinilla} P.,  et~al., 2017, \mn@doi [\apj] {10.3847/1538-4357/aa6973}, \href
  {https://ui.adsabs.harvard.edu/abs/2017ApJ...839...99P} {839, 99}

\bibitem[\protect\citeauthoryear{{Pinilla} et~al.,}{{Pinilla}
  et~al.}{2018}]{pinilla2018}
{Pinilla} P.,  et~al., 2018, \mn@doi [\apj] {10.3847/1538-4357/aabf94}, \href
  {https://ui.adsabs.harvard.edu/abs/2018ApJ...859...32P} {859, 32}

\bibitem[\protect\citeauthoryear{{Pinilla}, {Benisty}, {Cazzoletti}, {Harsono},
  {P{\'e}rez}  \& {Tazzari}}{{Pinilla} et~al.}{2019}]{pinilla2019}
{Pinilla} P.,  {Benisty} M.,  {Cazzoletti} P.,  {Harsono} D.,  {P{\'e}rez}
  L.~M.,   {Tazzari} M.,  2019, \mn@doi [\apj] {10.3847/1538-4357/ab1cb8},
  \href {https://ui.adsabs.harvard.edu/abs/2019ApJ...878...16P} {878, 16}

\bibitem[\protect\citeauthoryear{{Pinte}, {Dent}, {M{\'e}nard}, {Hales},
  {Hill}, {Cortes}  \& {de Gregorio-Monsalvo}}{{Pinte}
  et~al.}{2016}]{Pinte2016}
{Pinte} C.,  {Dent} W.~R.~F.,  {M{\'e}nard} F.,  {Hales} A.,  {Hill} T.,
  {Cortes} P.,   {de Gregorio-Monsalvo} I.,  2016, \mn@doi [\apj]
  {10.3847/0004-637X/816/1/25}, \href
  {https://ui.adsabs.harvard.edu/abs/2016ApJ...816...25P} {816, 25}

\bibitem[\protect\citeauthoryear{{Pinte} et~al.,}{{Pinte}
  et~al.}{2018}]{Pinte2018}
{Pinte} C.,  et~al., 2018, \mn@doi [\apjl] {10.3847/2041-8213/aac6dc}, \href
  {https://ui.adsabs.harvard.edu/abs/2018ApJ...860L..13P} {860, L13}

\bibitem[\protect\citeauthoryear{{Pinte} et~al.,}{{Pinte}
  et~al.}{2019}]{pinte2019}
{Pinte} C.,  et~al., 2019, \mn@doi [Nature Astronomy]
  {10.1038/s41550-019-0852-6}, \href
  {https://ui.adsabs.harvard.edu/abs/2019NatAs...3.1109P} {3, 1109}

\bibitem[\protect\citeauthoryear{{Pinte} et~al.,}{{Pinte}
  et~al.}{2020}]{Pinte2020}
{Pinte} C.,  et~al., 2020, \mn@doi [\apjl] {10.3847/2041-8213/ab6dda}, \href
  {https://ui.adsabs.harvard.edu/abs/2020ApJ...890L...9P} {890, L9}

\bibitem[\protect\citeauthoryear{{Poblete}, {Calcino}, {Cuello}, {Mac{\'\i}as},
  {Ribas}, {Price}, {Cuadra}  \& {Pinte}}{{Poblete} et~al.}{2020}]{Poblete2020}
{Poblete} P.~P.,  {Calcino} J.,  {Cuello} N.,  {Mac{\'\i}as} E.,  {Ribas}
  {\'A}.,  {Price} D.~J.,  {Cuadra} J.,   {Pinte} C.,  2020, \mn@doi [\mnras]
  {10.1093/mnras/staa1655}, \href
  {https://ui.adsabs.harvard.edu/abs/2020MNRAS.496.2362P} {496, 2362}

\bibitem[\protect\citeauthoryear{{Price} et~al.,}{{Price}
  et~al.}{2018}]{price2018}
{Price} D.~J.,  et~al., 2018, \mn@doi [\mnras] {10.1093/mnras/sty647}, \href
  {https://ui.adsabs.harvard.edu/abs/2018MNRAS.477.1270P} {477, 1270}

\bibitem[\protect\citeauthoryear{{Rebollido} et~al.,}{{Rebollido}
  et~al.}{2015}]{rebollido2015}
{Rebollido} I.,  et~al., 2015, \mn@doi [\aap] {10.1051/0004-6361/201425556},
  \href {https://ui.adsabs.harvard.edu/abs/2015A&A...581A..30R} {581, A30}

\bibitem[\protect\citeauthoryear{{Reipurth} \& {Zinnecker}}{{Reipurth} \&
  {Zinnecker}}{1993}]{reipurth1993}
{Reipurth} B.,  {Zinnecker} H.,  1993, \aap, \href
  {https://ui.adsabs.harvard.edu/abs/1993A&A...278...81R} {278, 81}

\bibitem[\protect\citeauthoryear{{Ribas} et~al.,}{{Ribas}
  et~al.}{2017}]{ribas2017}
{Ribas} {\'A}.,  et~al., 2017, \mn@doi [\apj] {10.3847/1538-4357/aa8e99}, \href
  {https://ui.adsabs.harvard.edu/abs/2017ApJ...849...63R} {849, 63}

\bibitem[\protect\citeauthoryear{{Ricci}, {Testi}, {Natta}  \&
  {Brooks}}{{Ricci} et~al.}{2010}]{ricci2010}
{Ricci} L.,  {Testi} L.,  {Natta} A.,   {Brooks} K.~J.,  2010, \mn@doi [\aap]
  {10.1051/0004-6361/201015039}, \href
  {https://ui.adsabs.harvard.edu/abs/2010A&A...521A..66R} {521, A66}

\bibitem[\protect\citeauthoryear{{Ricci}, {Liu}, {Isella}  \& {Li}}{{Ricci}
  et~al.}{2018}]{2018ApJ...853..110R}
{Ricci} L.,  {Liu} S.-F.,  {Isella} A.,   {Li} H.,  2018, \mn@doi [\apj]
  {10.3847/1538-4357/aaa546}, \href
  {https://ui.adsabs.harvard.edu/abs/2018ApJ...853..110R} {853, 110}

\bibitem[\protect\citeauthoryear{{Rice}, {Armitage}, {Wood}  \&
  {Lodato}}{{Rice} et~al.}{2006}]{rice2006}
{Rice} W.~K.~M.,  {Armitage} P.~J.,  {Wood} K.,   {Lodato} G.,  2006, \mn@doi
  [\mnras] {10.1111/j.1365-2966.2006.11113.x}, \href
  {https://ui.adsabs.harvard.edu/abs/2006MNRAS.373.1619R} {373, 1619}

\bibitem[\protect\citeauthoryear{{Riols} \& {Lesur}}{{Riols} \&
  {Lesur}}{2019}]{riols2019}
{Riols} A.,  {Lesur} G.,  2019, \mn@doi [\aap] {10.1051/0004-6361/201834813},
  \href {https://ui.adsabs.harvard.edu/abs/2019A&A...625A.108R} {625, A108}

\bibitem[\protect\citeauthoryear{{Riols}, {Lesur}  \& {Menard}}{{Riols}
  et~al.}{2020}]{riols2020}
{Riols} A.,  {Lesur} G.,   {Menard} F.,  2020, \mn@doi [\aap]
  {10.1051/0004-6361/201937418}, \href
  {https://ui.adsabs.harvard.edu/abs/2020A&A...639A..95R} {639, A95}

\bibitem[\protect\citeauthoryear{{Rosotti}, {Juhasz}, {Booth}  \&
  {Clarke}}{{Rosotti} et~al.}{2016}]{rosotti2016}
{Rosotti} G.~P.,  {Juhasz} A.,  {Booth} R.~A.,   {Clarke} C.~J.,  2016, \mn@doi
  [\mnras] {10.1093/mnras/stw691}, \href
  {https://ui.adsabs.harvard.edu/abs/2016MNRAS.459.2790R} {459, 2790}

\bibitem[\protect\citeauthoryear{{Ru{\'\i}z-Rodr{\'\i}guez}, {Ireland}, {Cieza}
   \& {Kraus}}{{Ru{\'\i}z-Rodr{\'\i}guez} et~al.}{2016}]{ruizrodriguez2016}
{Ru{\'\i}z-Rodr{\'\i}guez} D.,  {Ireland} M.,  {Cieza} L.,   {Kraus} A.,  2016,
  \mn@doi [\mnras] {10.1093/mnras/stw2297}, \href
  {https://ui.adsabs.harvard.edu/abs/2016MNRAS.463.3829R} {463, 3829}

\bibitem[\protect\citeauthoryear{{Segura-Cox} et~al.,}{{Segura-Cox}
  et~al.}{2020}]{2020Natur.586..228S}
{Segura-Cox} D.~M.,  et~al., 2020, \mn@doi [\nat] {10.1038/s41586-020-2779-6},
  \href {https://ui.adsabs.harvard.edu/abs/2020Natur.586..228S} {586, 228}

\bibitem[\protect\citeauthoryear{{Sheehan} \& {Eisner}}{{Sheehan} \&
  {Eisner}}{2018}]{sheehaneisner2018}
{Sheehan} P.~D.,  {Eisner} J.~A.,  2018, \mn@doi [\apj]
  {10.3847/1538-4357/aaae65}, \href
  {https://ui.adsabs.harvard.edu/abs/2018ApJ...857...18S} {857, 18}

\bibitem[\protect\citeauthoryear{{Soderblom}, {Hillenbrand}, {Jeffries},
  {Mamajek}  \& {Naylor}}{{Soderblom} et~al.}{2014}]{soderblomPPVI}
{Soderblom} D.~R.,  {Hillenbrand} L.~A.,  {Jeffries} R.~D.,  {Mamajek} E.~E.,
  {Naylor} T.,  2014, in {Beuther} H.,  {Klessen} R.~S.,  {Dullemond} C.~P.,
  {Henning} T.,  eds, Protostars and Planets VI. p.~219 (\mn@eprint {arXiv}
  {1311.7024}), \mn@doi{10.2458/azu_uapress_9780816531240-ch010}

\bibitem[\protect\citeauthoryear{{Stammler}, {Dr{\k{a}}{\.z}kowska},
  {Birnstiel}, {Klahr}, {Dullemond}  \& {Andrews}}{{Stammler}
  et~al.}{2019}]{Stammler2019}
{Stammler} S.~M.,  {Dr{\k{a}}{\.z}kowska} J.,  {Birnstiel} T.,  {Klahr} H.,
  {Dullemond} C.~P.,   {Andrews} S.~M.,  2019, \mn@doi [\apjl]
  {10.3847/2041-8213/ab4423}, \href
  {https://ui.adsabs.harvard.edu/abs/2019ApJ...884L...5S} {884, L5}

\bibitem[\protect\citeauthoryear{{Strom}, {Edwards}  \& {Skrutskie}}{{Strom}
  et~al.}{1993}]{strom1993}
{Strom} S.~E.,  {Edwards} S.,   {Skrutskie} M.~F.,  1993, in {Levy} E.~H.,
  {Lunine} J.~I.,  eds, Protostars and Planets III. p.~837

\bibitem[\protect\citeauthoryear{{Suriano}, {Li}, {Krasnopolsky}  \&
  {Shang}}{{Suriano} et~al.}{2017}]{suriano2017}
{Suriano} S.~S.,  {Li} Z.-Y.,  {Krasnopolsky} R.,   {Shang} H.,  2017, \mn@doi
  [\mnras] {10.1093/mnras/stx735}, \href
  {https://ui.adsabs.harvard.edu/abs/2017MNRAS.468.3850S} {468, 3850}

\bibitem[\protect\citeauthoryear{{Suriano}, {Li}, {Krasnopolsky}  \&
  {Shang}}{{Suriano} et~al.}{2018}]{suriano2018}
{Suriano} S.~S.,  {Li} Z.-Y.,  {Krasnopolsky} R.,   {Shang} H.,  2018, \mn@doi
  [\mnras] {10.1093/mnras/sty717}, \href
  {https://ui.adsabs.harvard.edu/abs/2018MNRAS.477.1239S} {477, 1239}

\bibitem[\protect\citeauthoryear{{Suriano}, {Li}, {Krasnopolsky}, {Suzuki}  \&
  {Shang}}{{Suriano} et~al.}{2019}]{suriano2019}
{Suriano} S.~S.,  {Li} Z.-Y.,  {Krasnopolsky} R.,  {Suzuki} T.~K.,   {Shang}
  H.,  2019, \mn@doi [\mnras] {10.1093/mnras/sty3502}, \href
  {https://ui.adsabs.harvard.edu/abs/2019MNRAS.484..107S} {484, 107}

\bibitem[\protect\citeauthoryear{{Takahashi} \& {Inutsuka}}{{Takahashi} \&
  {Inutsuka}}{2014}]{takahashi2014}
{Takahashi} S.~Z.,  {Inutsuka} S.-i.,  2014, \mn@doi [\apj]
  {10.1088/0004-637X/794/1/55}, \href
  {https://ui.adsabs.harvard.edu/abs/2014ApJ...794...55T} {794, 55}

\bibitem[\protect\citeauthoryear{{Teague}, {Bae}, {Bergin}, {Birnstiel}  \&
  {Foreman-Mackey}}{{Teague} et~al.}{2018}]{Teague2018}
{Teague} R.,  {Bae} J.,  {Bergin} E.~A.,  {Birnstiel} T.,   {Foreman-Mackey}
  D.,  2018, \mn@doi [\apjl] {10.3847/2041-8213/aac6d7}, \href
  {https://ui.adsabs.harvard.edu/abs/2018ApJ...860L..12T} {860, L12}

\bibitem[\protect\citeauthoryear{{Testi} et~al.,}{{Testi}
  et~al.}{2014}]{testi2014}
{Testi} L.,  et~al., 2014, in {Beuther} H.,  {Klessen} R.~S.,  {Dullemond}
  C.~P.,   {Henning} T.,  eds, Protostars and Planets VI. p.~339 (\mn@eprint
  {arXiv} {1402.1354}), \mn@doi{10.2458/azu_uapress_9780816531240-ch015}

\bibitem[\protect\citeauthoryear{{Uribe}, {Klahr}, {Flock}  \&
  {Henning}}{{Uribe} et~al.}{2011}]{uribe2011}
{Uribe} A.~L.,  {Klahr} H.,  {Flock} M.,   {Henning} T.,  2011, \mn@doi [\apj]
  {10.1088/0004-637X/736/2/85}, \href
  {https://ui.adsabs.harvard.edu/abs/2011ApJ...736...85U} {736, 85}

\bibitem[\protect\citeauthoryear{{Veronesi} et~al.,}{{Veronesi}
  et~al.}{2020}]{Veronesi2020}
{Veronesi} B.,  et~al., 2020, \mn@doi [\mnras] {10.1093/mnras/staa1278}, \href
  {https://ui.adsabs.harvard.edu/abs/2020MNRAS.495.1913V} {495, 1913}

\bibitem[\protect\citeauthoryear{{Wahhaj} et~al.,}{{Wahhaj}
  et~al.}{2010}]{wahhaj2010}
{Wahhaj} Z.,  et~al., 2010, \mn@doi [\apj] {10.1088/0004-637X/724/2/835}, \href
  {https://ui.adsabs.harvard.edu/abs/2010ApJ...724..835W} {724, 835}

\bibitem[\protect\citeauthoryear{{Weidenschilling}}{{Weidenschilling}}{1977}]{Weidenschilling1977}
{Weidenschilling} S.~J.,  1977, \mn@doi [\mnras] {10.1093/mnras/180.2.57},
  \href {https://ui.adsabs.harvard.edu/abs/1977MNRAS.180...57W} {180, 57}

\bibitem[\protect\citeauthoryear{{Williams} \& {Cieza}}{{Williams} \&
  {Cieza}}{2011}]{williamscieza2011}
{Williams} J.~P.,  {Cieza} L.~A.,  2011, \mn@doi [\araa]
  {10.1146/annurev-astro-081710-102548}, \href
  {https://ui.adsabs.harvard.edu/abs/2011ARA&A..49...67W} {49, 67}

\bibitem[\protect\citeauthoryear{{Williams}, {Cieza}, {Hales}, {Ansdell},
  {Ruiz-Rodriguez}, {Casassus}, {Perez}  \& {Zurlo}}{{Williams}
  et~al.}{2019}]{williams2019}
{Williams} J.~P.,  {Cieza} L.,  {Hales} A.,  {Ansdell} M.,  {Ruiz-Rodriguez}
  D.,  {Casassus} S.,  {Perez} S.,   {Zurlo} A.,  2019, \mn@doi [\apjl]
  {10.3847/2041-8213/ab1338}, \href
  {https://ui.adsabs.harvard.edu/abs/2019ApJ...875L...9W} {875, L9}

\bibitem[\protect\citeauthoryear{{Winn} \& {Fabrycky}}{{Winn} \&
  {Fabrycky}}{2015}]{winnfabrycky2015}
{Winn} J.~N.,  {Fabrycky} D.~C.,  2015, \mn@doi [\araa]
  {10.1146/annurev-astro-082214-122246}, \href
  {https://ui.adsabs.harvard.edu/abs/2015ARA&A..53..409W} {53, 409}

\bibitem[\protect\citeauthoryear{{Wright} et~al.,}{{Wright}
  et~al.}{2010}]{wise2010}
{Wright} E.~L.,  et~al., 2010, \mn@doi [\aj] {10.1088/0004-6256/140/6/1868},
  \href {https://ui.adsabs.harvard.edu/abs/2010AJ....140.1868W} {140, 1868}

\bibitem[\protect\citeauthoryear{{Youdin}}{{Youdin}}{2011}]{youdin2011}
{Youdin} A.~N.,  2011, \mn@doi [\apj] {10.1088/0004-637X/731/2/99}, \href
  {https://ui.adsabs.harvard.edu/abs/2011ApJ...731...99Y} {731, 99}

\bibitem[\protect\citeauthoryear{{Zhang}, {Blake}  \& {Bergin}}{{Zhang}
  et~al.}{2015}]{Zhang2015}
{Zhang} K.,  {Blake} G.~A.,   {Bergin} E.~A.,  2015, \mn@doi [\apjl]
  {10.1088/2041-8205/806/1/L7}, \href
  {https://ui.adsabs.harvard.edu/abs/2015ApJ...806L...7Z} {806, L7}

\bibitem[\protect\citeauthoryear{{Zhang} et~al.,}{{Zhang}
  et~al.}{2018}]{Zhang2018}
{Zhang} S.,  et~al., 2018, \mn@doi [\apjl] {10.3847/2041-8213/aaf744}, \href
  {https://ui.adsabs.harvard.edu/abs/2018ApJ...869L..47Z} {869, L47}

\bibitem[\protect\citeauthoryear{{Zhu}, {Nelson}, {Dong}, {Espaillat}  \&
  {Hartmann}}{{Zhu} et~al.}{2012}]{zhu2012}
{Zhu} Z.,  {Nelson} R.~P.,  {Dong} R.,  {Espaillat} C.,   {Hartmann} L.,  2012,
  \mn@doi [\apj] {10.1088/0004-637X/755/1/6}, \href
  {https://ui.adsabs.harvard.edu/abs/2012ApJ...755....6Z} {755, 6}

\bibitem[\protect\citeauthoryear{{Zurlo} et~al.,}{{Zurlo}
  et~al.}{2020a}]{zurlo2020}
{Zurlo} A.,  et~al., 2020a, \mn@doi [\mnras] {10.1093/mnras/staa1886}, \href
  {https://ui.adsabs.harvard.edu/abs/2020MNRAS.496.5089Z} {496, 5089}

\bibitem[\protect\citeauthoryear{{Zurlo} et~al.,}{{Zurlo}
  et~al.}{2020b}]{zurlob2020}
{Zurlo} A.,  et~al., 2020b, \mn@doi [\aap] {10.1051/0004-6361/201936891}, \href
  {https://ui.adsabs.harvard.edu/abs/2020A&A...633A.119Z} {633, A119}

\bibitem[\protect\citeauthoryear{{de Val-Borro} et~al.,}{{de Val-Borro}
  et~al.}{2006}]{deValBorro2006}
{de Val-Borro} M.,  et~al., 2006, \mn@doi [\mnras]
  {10.1111/j.1365-2966.2006.10488.x}, \href
  {https://ui.adsabs.harvard.edu/abs/2006MNRAS.370..529D} {370, 529}

\bibitem[\protect\citeauthoryear{{van der Marel}, {van Dishoeck}, {Bruderer},
  {Andrews}, {Pontoppidan}, {Herczeg}, {van Kempen}  \& {Miotello}}{{van der
  Marel} et~al.}{2016}]{vandermarel2016}
{van der Marel} N.,  {van Dishoeck} E.~F.,  {Bruderer} S.,  {Andrews} S.~M.,
  {Pontoppidan} K.~M.,  {Herczeg} G.~J.,  {van Kempen} T.,   {Miotello} A.,
  2016, \mn@doi [\aap] {10.1051/0004-6361/201526988}, \href
  {https://ui.adsabs.harvard.edu/abs/2016A&A...585A..58V} {585, A58}

\bibitem[\protect\citeauthoryear{{van der Marel} et~al.,}{{van der Marel}
  et~al.}{2018}]{vandermarel2018}
{van der Marel} N.,  et~al., 2018, \mn@doi [\apj] {10.3847/1538-4357/aaaa6b},
  \href {https://ui.adsabs.harvard.edu/abs/2018ApJ...854..177V} {854, 177}

\makeatother
\end{thebibliography}


\newpage
\appendix\label{s:apendix}
\section{Brightness profile reconstruction using Frankenstein code}\label{s:frank}
\begin{figure*}
\includegraphics[trim=20mm 25mm 20mm 10mm,clip,width=15.0cm]{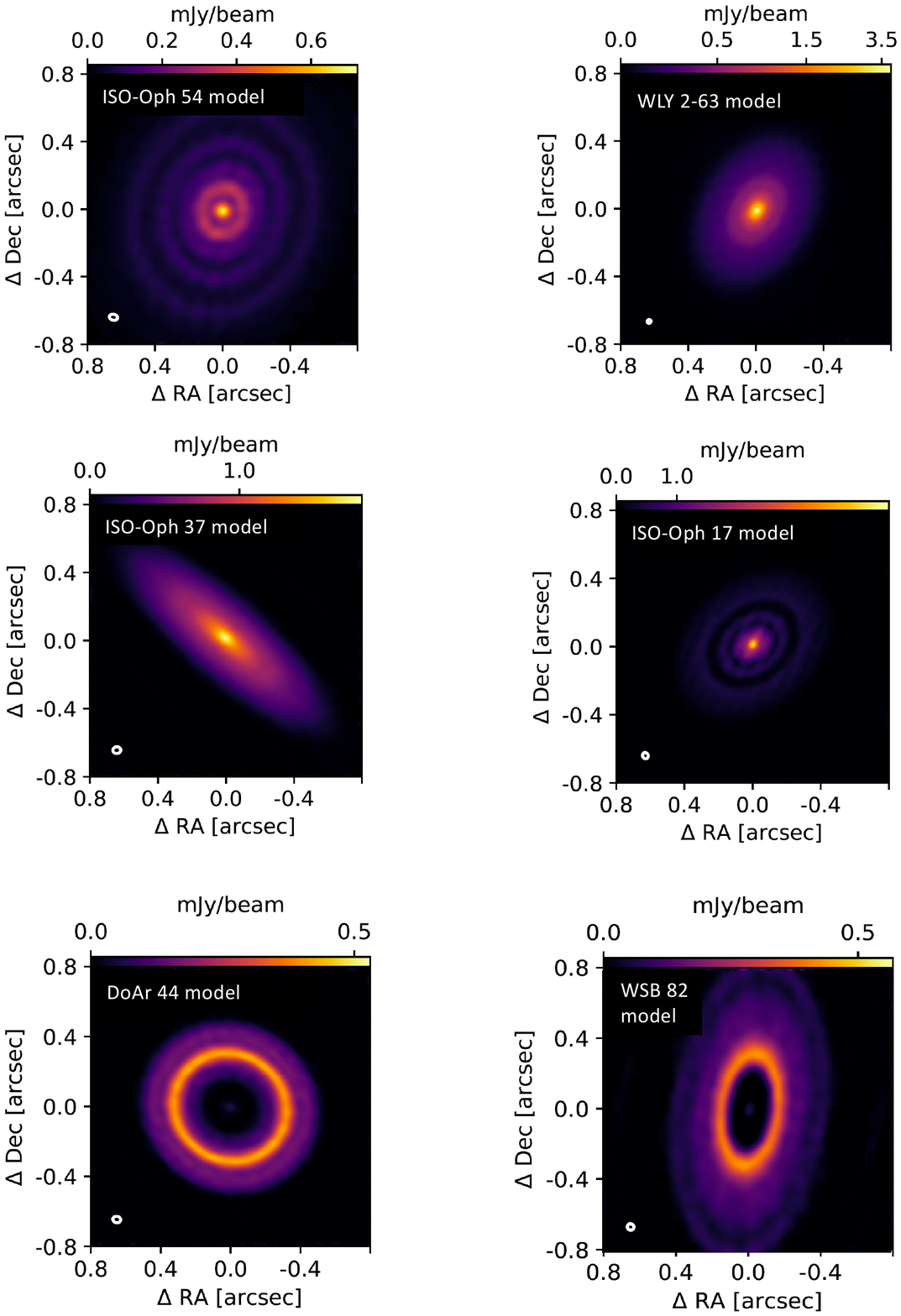}
\caption{Model images of axisymmetric discs produced with the python module Frankenstein (frank, Jennings et al. 2020) for ISO-Oph 54, WLY~2-63, ISO-Oph 37, ISO-Oph 17, DoAr 44, and WSB~82.}
\label{f:frank_1}
\end{figure*}

\begin{figure*}
\includegraphics[trim=20mm 105mm 20mm 0mm,clip,width=15.0cm]{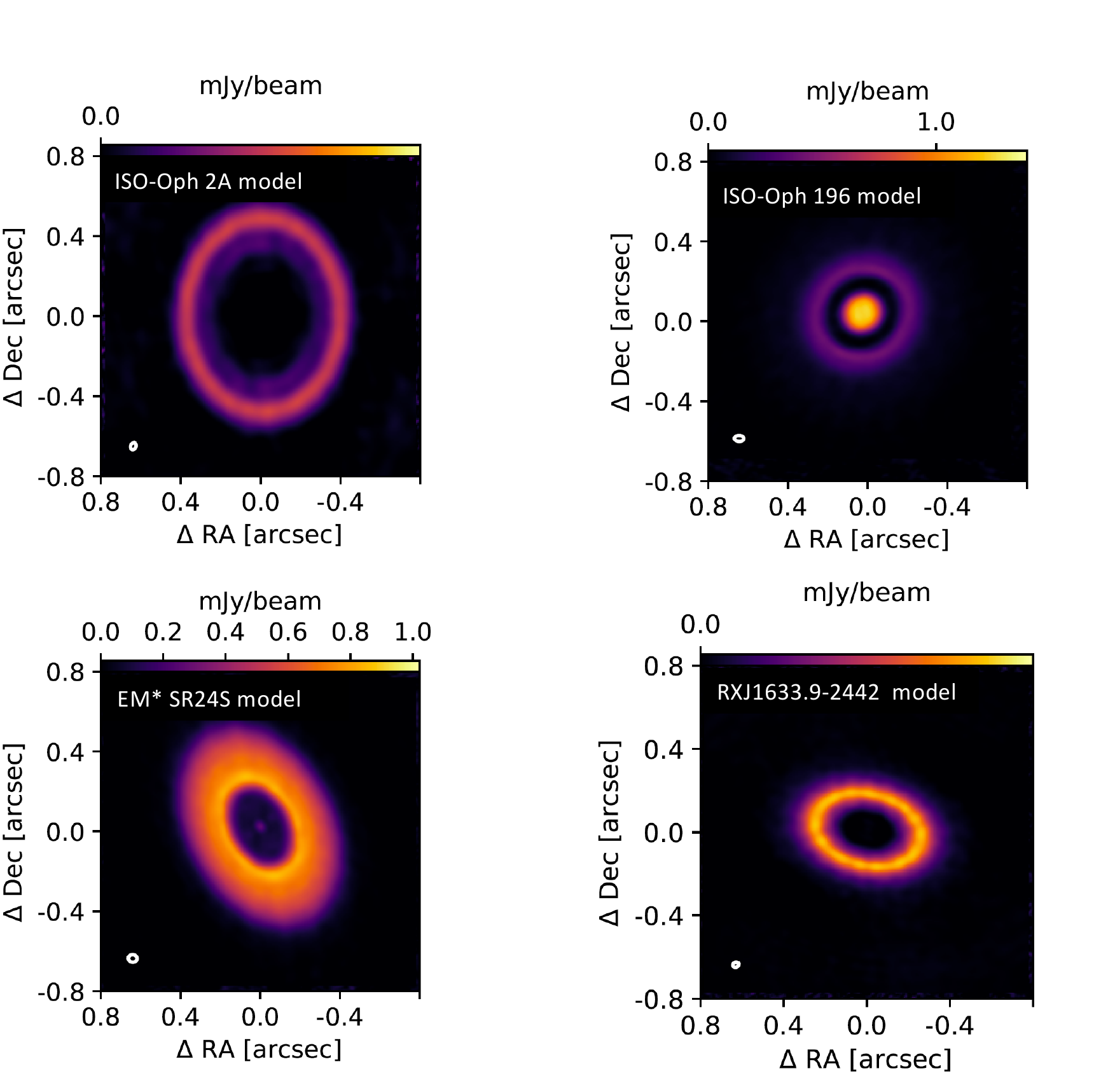}
\caption{Same as Fig.~\ref{f:frank_1}, but for ISO-Oph~2A, ISO-Oph~196, EM* SR2S, and RXJ1633.9-2442.}
\label{f:frank_2}
\end{figure*}

\begin{figure*}
\includegraphics[trim=0mm 95mm 0mm 0mm,clip,width=19.0cm]{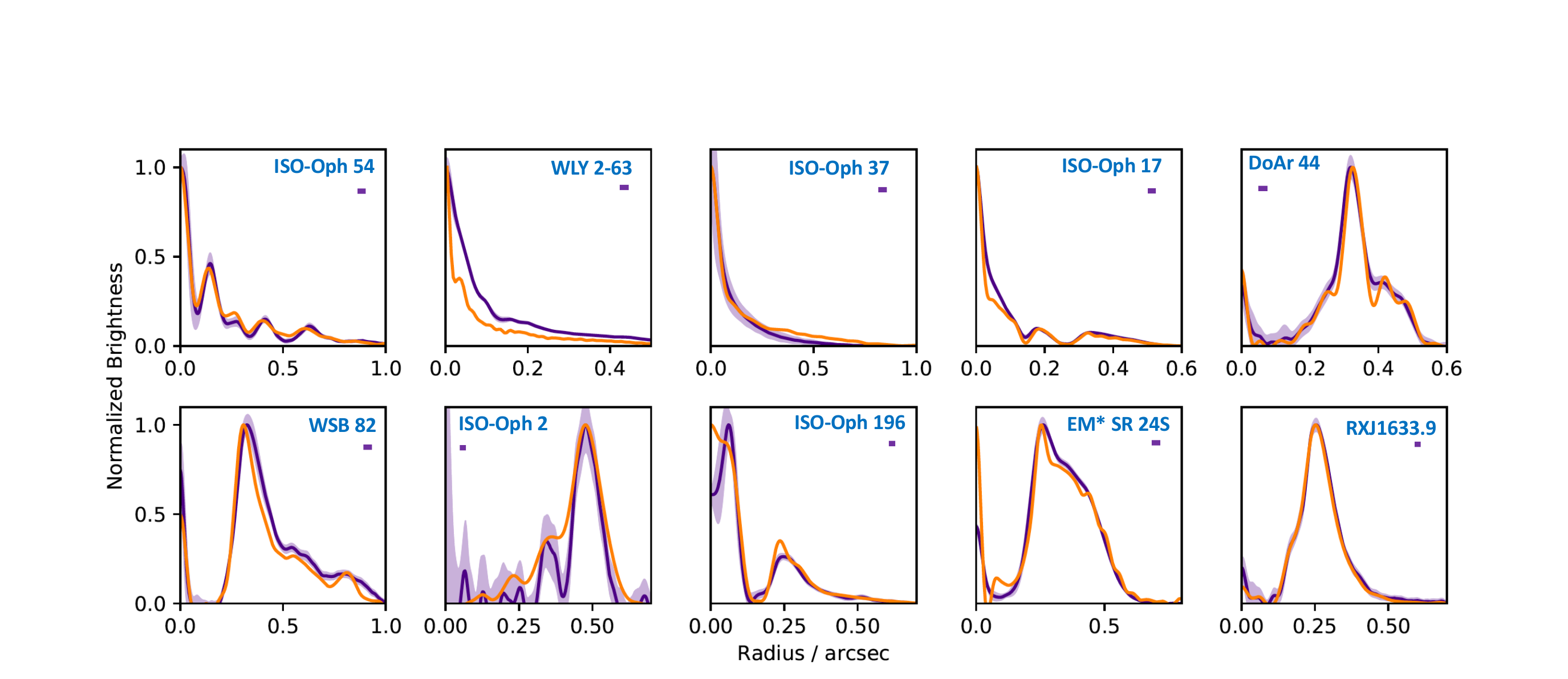}
\caption{Disc brightness profiles corresponding to the model images from frank in Figs~\ref{f:frank_1} and~\ref{f:frank_2} compared to the profiles obtained from the CLEANed images as 
described in Sec.~\ref{s:profiles}.
\textbf{The small bar below the name of each source indicates the size of the beam. With the exceptions of ISO-Oph 2 and ISO-Oph 196 (discussed in Section~\ref{s:caveats}),} the profiles are very similar, but some of the features identified by frank are deeper and narrower. 
Frank also identifies several additional local minima and local maxima (e.g., in WLY 2-63 and DoAr 44) that should be taken with caution.
}
\label{f:frank_profiles}
\end{figure*}


\bsp	
\label{lastpage}
\end{document}